\documentclass[prd,aps,a4paper,superscriptaddress,nofootinbib]{revtex4}
\usepackage{graphicx}
\usepackage{color}
\usepackage{dcolumn}
\usepackage{bm}
\usepackage{slashed}
\usepackage{amsmath}
\usepackage{latexsym}
\usepackage{amssymb}
\usepackage{mathrsfs}
\usepackage{mathtools}
\usepackage{amsfonts}
\usepackage{longtable}
\usepackage{physics}
\usepackage{url}
\usepackage{stackengine}

\allowdisplaybreaks
\begin{document}
\def\SH{\scriptscriptstyle{\mathrm{SH}}}
\def\MH{\scriptscriptstyle{\mathrm{MH}}}
\def\ADM{\scriptscriptstyle{\mathrm{ADM}}}
\def\EOB{\scriptscriptstyle{\mathrm{EOB}}}
\def\tt{\scriptscriptstyle{\mathrm{TT}}}

\newcommand{\fPN}[1]{\tiny{{#1}\mathrm{PN}}}
\newcommand{\ivc}[1]{\frac{1}{c^{#1}}}

\def\pr{p_r}
\def\prt{p_{r_*}}
\def\rdot{\dot{r}}
\def\pnvec{\hat{\pmb{n}}}
\def\pvvec{\pmb{v}}
\def\ppvec{\pmb{p}}

\def\vavg{\bar{v}}
\def\eavg{\bar{e}}
\def\lavg{\bar{l}}
\def\lbdavg{\bar{\lambda}}

\def\vpa{\tilde{v}}
\def\epa{\tilde{e}}
\def\lpa{\tilde{l}}
\def\lbdpa{\tilde{\lambda}}

\def\vphi{v_{\phi}}
\def\xom{x_{\omega}}
\def\vom{v_{\omega}}
\def\et{e_{t}}
\def\betae{\beta_e}

\def\SplusPM{\mathrm{S}^{(+)}}
\def\SminusPM{\mathrm{S}^{(-)}}

\def\ii{\mathrm{i}}
\def\ee{\mathrm{e}}
\def\FinitePart{\mathrm{FP}}

\def\Jint{\mathrm{J}^{(0)}}
\def\Jintd{\mathrm{J}^{(\delta)}}
\def\Jintdd{\mathrm{J}^{(\delta^2)}}
\def\Jintddd{\mathrm{J}^{(\delta^3)}}
\def\Jintln{\mathrm{J}^{(\ln)}}
\def\Jintdln{\mathrm{J}^{(\delta\ln)}}
\newcommand{\fJint}[2]{\mathrm{J}^{(#1)}_{#2}}
\newcommand{\fJintln}[2]{\mathrm{K}^{(#1)}_{#2}}
\newcommand{\fJintlnln}[2]{\mathrm{L}^{(#1)}_{#2}}

\newcommand{\fJints}[3]{{}^{(-)}\mathrm{J}^{(#2,#1)}_{#3}}
\newcommand{\fJintc}[3]{{}^{(+)}\mathrm{J}^{(#2,#1)}_{#3}}

\newcommand{\fcjJint}[2]{\bar{\mathrm{J}}^{(#1)}_{#2}}
\newcommand{\fcjJintln}[2]{\bar{\mathrm{K}}^{(#1)}_{#2}}
\newcommand{\fcjJintlnln}[2]{\bar{\mathrm{L}}^{(#1)}_{#2}}

\def\cK{\mathcal{K}}
\def\cJ{\mathcal{J}}
\def\cL{\mathcal{L}}

\def\hypf{[{}_{2}\mathrm{F}_{1}]}
\def\hypff{[{}_{3}\mathrm{F}_{2}]}

\newcommand{\hypfunc}[4]{[{}_{2}\mathrm{F}_{1}]\bigg(\begin{array}{c}#1,\ #2\\#3\end{array};#4\bigg)}

\newcommand{\DHarmN}[2]{\Delta{\mathrm{H}}_{#2}^{#1}}

\newcommand{\fRLIntegral}[3]{{}_{#2}\pmb{\mathcal{I}}^{#1}_{#3} }
\newcommand{\fRLDeriv}[3]{{}_{#2}\pmb{\mathcal{D}}^{#1}_{#3}}
\def\sgn{\mathrm{sgn}}
\newcommand{\eexp}[1]{\mathrm{e}^{#1}}

\newcommand{\fmodexIp}[2]{\mathop{\mathbb{I}}\limits_{\mathclap{(#1)}}\!_{#2}}
\newcommand{\fmodexJp}[2]{\mathop{\mathbb{J}}\limits_{\mathclap{(#1)}}\!_{#2}}

\newcommand{\fmodeyIp}[2]{\mathop{\tilde{\mathbb{I}}}\limits_{\mathclap{(#1)}}\!_{#2}}
\newcommand{\fmodeyJp}[2]{\mathop{\tilde{\mathbb{J}}}\limits_{\mathclap{(#1)}}\!_{#2}}

\newcommand{\fmodexaIp}[3]{\big[{}_{(#3)}\mathop{\mathbb{I}}\limits_{\mathclap{(#1)}}\!_{#2}\big]}
\newcommand{\fmodexaJp}[3]{\big[{}_{(#3)}\mathop{\mathbb{J}}\limits_{\mathclap{(#1)}}\!_{#2}\big]}

\newcommand{\fmodeyaIp}[3]{\big[{}_{(#3)}\mathop{\tilde{\mathbb{I}}}\limits_{\mathclap{(#1)}}\!_{#2}\big]}
\newcommand{\fmodeyaJp}[3]{\big[{}_{(#3)}\mathop{\tilde{\mathbb{J}}}\limits_{\mathclap{(#1)}}\!_{#2}\big]}

\newcommand{\fmodeI}[3]{\underset{(#1,#2)}{\mathrm{I}_{#3}}}
\newcommand{\fmodeIpn}[4]{\underset{(#1,#2)}{\mathrm{I}_{#3}^{#4}}}

\newcommand{\fmodeIp}[2]{\mathop{\mathrm{I}}\limits_{\mathclap{(#1)}}\!_{#2}}

\newcommand{\fmodeIppn}[3]{\mathop{\mathrm{I}}\limits_{\mathclap{(#1)}}\!_{#2}^{#3}}

\newcommand{\fmodeconjIp}[2]{\mathop{\mathrm{I}}\limits_{\mathclap{(#1)}}\!_{#2}^*}

\newcommand{\fmodeconjIppn}[3]{\mathop{\mathrm{I}}\limits_{\mathclap{(#1)}}\!_{#2}^{*,#3}}

\newcommand{\fmodeJ}[3]{\underset{(#1,#2)}{\mathrm{J}_{#3}}}
\newcommand{\fmodeJpn}[4]{\underset{(#1,#2)}{\mathrm{J}_{#3}^{#4}}}

\newcommand{\fmodeJp}[2]{\mathop{\mathrm{J}}\limits_{\mathclap{(#1)}}\!_{#2}}

\newcommand{\fmodeJppn}[3]{\mathop{\mathrm{J}}\limits_{\mathclap{(#1)}}\!_{#2}^{#3}}

\newcommand{\fmodeconjJp}[2]{\mathop{\mathrm{J}}\limits_{\mathclap{(#1)}}\!_{#2}^*}

\newcommand{\fmodeconjJppn}[3]{\mathop{\mathrm{J}}\limits_{\mathclap{(#1)}}\!_{#2}^{*,#3}}

\newcommand{\fmodemIp}[2]{\big[{}_{(1)}\mathop{\mathrm{I}}\limits_{\mathclap{(#1)}}\!_{#2}\big]}
\newcommand{\fmodemnIp}[3]{\big[{}_{(#3)}\mathop{\mathrm{I}}\limits_{\mathclap{(#1)}}\!_{#2}\big]}
\newcommand{\fmodemIppn}[3]{\big[{}_{(1)}\mathop{\mathrm{I}}\limits_{\mathclap{(#1)}}\!_{#2}^{#3}\big]}
\newcommand{\fmodemnIppn}[4]{\big[{}_{(#4)}\mathop{\mathrm{I}}\limits_{\mathclap{(#1)}}\!_{#2}^{#3}\big]}

\newcommand{\fmodeconjmIp}[2]{\big[{}_{(1)}\mathop{\mathrm{I}}\limits_{\mathclap{(#1)}}\!_{#2}^*\big]}
\newcommand{\fmodeconjmnIp}[3]{\big[{}_{(#3)}\mathop{\mathrm{I}}\limits_{\mathclap{(#1)}}\!_{#2}^*\big]}
\newcommand{\fmodeconjmIppn}[3]{\big[{}_{(1)}\mathop{\mathrm{I}}\limits_{\mathclap{(#1)}}\!_{#2}^{*,#3}\big]}
\newcommand{\fmodeconjmnIppn}[4]{\big[{}_{(#4)}\mathop{\mathrm{I}}\limits_{\mathclap{(#1)}}\!_{#2}^{*,#3}\big]}

\newcommand{\fmodemJp}[2]{\big[{}_{(1)}\mathop{\mathrm{J}}\limits_{\mathclap{(#1)}}\!_{#2}\big]}
\newcommand{\fmodemnJp}[3]{\big[{}_{(#3)}\mathop{\mathrm{J}}\limits_{\mathclap{(#1)}}\!_{#2}\big]}
\newcommand{\fmodemJppn}[3]{\big[{}_{(1)}\mathop{\mathrm{J}}\limits_{\mathclap{(#1)}}\!_{#2}^{#3}\big]}
\newcommand{\fmodemnJppn}[4]{\big[{}_{(#4)}\mathop{\mathrm{J}}\limits_{\mathclap{(#1)}}\!_{#2}^{#3}\big]}

\newcommand{\fmodeconjmJp}[2]{\big[{}_{(1)}\mathop{\mathrm{J}}\limits_{\mathclap{(#1)}}\!_{#2}^*\big]}
\newcommand{\fmodeconjmnJp}[3]{\big[{}_{(#3)}\mathop{\mathrm{J}}\limits_{\mathclap{(#1)}}\!_{#2}^*\big]}
\newcommand{\fmodeconjmJppn}[3]{\big[{}_{(1)}\mathop{\mathrm{J}}\limits_{\mathclap{(#1)}}\!_{#2}^{*,#3}\big]}
\newcommand{\fmodeconjmnJppn}[4]{\big[{}_{(#4)}\mathop{\mathrm{J}}\limits_{\mathclap{(#1)}}\!_{#2}^{*,#3}\big]}

\newcommand{\fmodeM}[3]{\underset{(#1,#2)}{\mathrm{M}_{#3}}}
\newcommand{\fmodeMpn}[4]{\underset{(#1,#2)}{\mathrm{M}_{#3}^{#4}}}
\newcommand{\fmodeMp}[2]{\mathop{\mathrm{M}}\limits_{\mathclap{(#1)}}\!_{#2}}
\newcommand{\fmodemnMp}[3]{\big[{}_{(#3)}\mathop{\mathrm{M}}\limits_{\mathclap{(#1)}}\!_{#2}\big]}
\newcommand{\fmodeconjMp}[2]{\mathop{\mathrm{M}}\limits_{\mathclap{(#1)}}\!_{#2}^*}
\newcommand{\fmodeconjmnMp}[3]{\big[{}_{(#3)}\mathop{\mathrm{M}}\limits_{\mathclap{(#1)}}\!_{#2}^*\big]}

\newcommand{\fmodexMp}[2]{\mathop{\mathbb{M}}\limits_{\mathclap{(#1)}}\!_{#2}}
\newcommand{\fmodexaMp}[3]{\big[{}_{(#3)}\mathop{\mathbb{M}}\limits_{\mathclap{(#1)}}\!_{#2}\big]}

\newcommand{\fmodeyMp}[2]{\mathop{\tilde{\mathbb{M}}}\limits_{\mathclap{(#1)}}\!_{#2}}
\newcommand{\fmodeyaMp}[3]{\big[{}_{(#3)}\mathop{\tilde{\mathbb{M}}}\limits_{\mathclap{(#1)}}\!_{#2}\big]}

\newcommand{\fmodeS}[3]{\underset{(#1,#2)}{\mathrm{S}_{#3}}}
\newcommand{\fmodeSpn}[4]{\underset{(#1,#2)}{\mathrm{S}_{#3}^{#4}}}

\newcommand{\avg}[1]{\left\langle{#1}\right\rangle}

\def\Eflux{\mathcal{F}}
\def\Jflux{\mathcal{G}}

\def\MADM{M_{\ADM}}
\def\ege{\gamma_E}
\def\tail{\mathrm{tail}}
\def\ttail{\tail(\tail)}
\def\tttail{\tail(\tail(\tail))}
\def\tailsq{(\tail)^2}
\def\tailcb{(\tail)^3}

\def\Etflux{\mathcal{F}_{\tail}}
\def\Ettflux{\mathcal{F}_{\ttail}}
\def\Etttflux{\mathcal{F}_{\tttail}}
\def\Etsqflux{\mathcal{F}_{\tailsq}}
\def\Etcbflux{\mathcal{F}_{\tailcb}}

\newcommand{\fEtfluxn}[1]{\mathcal{F}_{\tail(#1) }}
\newcommand{\fEtnflux}[1]{\mathcal{F}_{(\tail)^{#1}}}

\def\Jtflux{\mathcal{G}_{\tail}}
\def\Jttflux{\mathcal{G}_{\ttail}}
\def\Jtttflux{\mathcal{G}_{\tttail}}
\def\Jtsqflux{\mathcal{G}_{\tailsq}}
\def\Jtcbflux{\mathcal{G}_{\tailcb}}

\newcommand{\fJtfluxn}[1]{\mathcal{G}_{\tail(#1) }}
\newcommand{\fJtnflux}[1]{\mathcal{G}_{(\tail)^{#1}}}

\def\Bell{\mathrm{B}}
\def\sigPoly{\mathcal{P}}

\newcommand\fehphi[2]{\varphi_{#1}^{(#2)}}
\newcommand\ftehphi[2]{\tilde{\varphi}_{#1}^{(#2)}}

\newcommand\fehbeta[2]{\beta_{#1}^{(#2)}}
\newcommand\ftehbeta[2]{\tilde{\beta}_{#1}^{(#2)}}

\newcommand\fehgamma[2]{\gamma_{#1}^{(#2)}}
\newcommand\ftehgamma[2]{\tilde{\gamma}_{#1}^{(#2)}}

\newcommand\fehF[2]{F_{#1}^{(#2)}}
\newcommand\ftehF[2]{\tilde{F}_{#1}^{(#2)}}

\newcommand\fehchi[2]{\chi_{#1}^{(#2)}}
\newcommand\ftehchi[2]{\tilde{\chi}_{#1}^{(#2)}}

\newcommand{\feh}[2]{\eta_{#1}^{(#2)}}
\newcommand{\fteh}[2]{\tilde\eta_{#1}^{(#2)}}

\newcommand{\cfinttailU}[1]{\mathfrak{i}^{(\ln)}_{#1}}
\newcommand{\cfinttailV}[1]{\mathfrak{j}^{(\ln)}_{#1}}

\newcommand{\cfgm}[1]{{\bar{\pmb{\gamma}}_{#1}}}

\newcommand{\IContr}[2]{\mathrm{C}^{(#1)}_{#2}}

\newcommand{\cgcoeffs}[2]{\mathrm{CG}^{#1}_{#2}}

\newcommand{\fIehlpn}[2]{\varphi^{(#1)}_{#2}}
\newcommand{\ftIehlpn}[2]{\tilde\varphi^{(#1)}_{#2}}

\newcommand{\fJehlpn}[2]{\gamma^{(#1)}_{#2}}
\newcommand{\ftJehlpn}[2]{\tilde\gamma^{(#1)}_{#2}}

\title{A general Fourier expansion of post-Newtonian binary dynamics based on quasi-Keplerian framework}

\author{Xiaolin Liu}\email[Xiaolin Liu: ]{shallyn.liu@foxmail.com}
\affiliation{Instituto de Física Téorica UAM-CSIC, Universidad Autónoma de Madrid, Cantoblanco 28049 Madrid, Spain}
\author{Zhoujian Cao\footnote{corresponding author}}\email[Zhoujian Cao: ]{zjcao@amt.ac.cn}
\affiliation{Institute for Frontiers in Astronomy and Astrophysics, Beijing Normal University, Beijing 102206, China}
\affiliation{Department of Astronomy, Beijing Normal University, Beijing 100875, China}
\affiliation{School of Fundamental Physics and Mathematical Sciences, Hangzhou Institute for Advanced Study, UCAS, Hangzhou 310024, China}
\affiliation{Institute for Frontiers in Astronomy and Astrophysics, Beijing Normal University, Beijing 102206, China}

\begin{abstract}
We have introduced a new method for computing gravitational-wave emission from non-spinning binaries which systematically unifies the various integrals arising in the Fourier expansions of post-Newtonian dynamics, providing a simple, practical scheme for calculations at arbitrary precision.
Using this approach, we derived the full set of 3PN dynamical quantities and gravitational-wave Fourier modes and have released the corresponding numerical code as open source. Furthermore, when radiation-reaction effects is not included, we found that the tail contribution to the energy and angular momentum fluxes can be resummed into an exceptionally compact expression with the help of the new method. These advances pave the way for more convenient and accurate frequency-domain waveform modeling in the future.
\end{abstract}

\maketitle

\section{Introduction}
Since the first observation in 2015~\cite{GW150914}, gravitational-wave (GW) astronomy has advanced rapidly. 
Joint observing runs by LIGO, Virgo, and KAGRA~\cite{advLIGO, AdvVirgo, KAGRA} have amassed a rich data set~\cite{GWTC_2, GWTC3}, and—using techniques such as matched filtering against theoretical templates—researchers have identified scores of binary-black-hole (BBH), binary-neutron-star (BNS), and even black-hole–neutron-star (NSBH) merger events. 
More recently, pulsar-timing-array (PTA) experiments have reported evidence for a stochastic background produced by super-massive binary-black-hole mergers~\cite{Agazie_2023,Reardon_2023,EPTA2023,Xu_2023}.
Looking ahead, next-generation ground-based detectors like the Einstein Telescope (ET)~\cite{ET_2010} and Cosmic Explorer (CE)~\cite{CosmicExplorer}, together with planned space missions such as LISA~\cite{LISA_1997}, TianQin~\cite{Tianqin}, Taiji~\cite{Taiji}, and DECIGO~\cite{DECIGO}, are expected to uncover many more GW signals.

The dominant sources of GWs in the Universe are generally compact binaries. 
The background generated by systems such as BBHs is thought to prevail over a wide range of frequencies~\cite{landini2025opticalgravitationalwavessignals}. 
Current data-analysis pipelines—whether based on traditional matched filtering or modern machine-learning methods—depend~\cite{George_2018} critically on accurate models of two-body dynamics and GW emission in general relativity. 
Numerical-relativity (NR) simulations have already produced extensive catalogs of binary waveforms, greatly aiding model development~\cite{SXS_2019,CIT_2023,SXS_2025}.
Nevertheless, their high computational cost limits waveform length, a constraint that becomes severe as future detectors push to lower frequencies. 
At the other extreme, the self-force framework treats motion in a known black-hole spacetime as a perturbative expansion in a small mass ratio, yielding precise geodesic orbits and, via perturbation theory, far-zone waveforms~\cite{Mino_1997,Pound_2012}. 
Together, NR and self-force methods offer solutions in the strong-field comparable-mass and large-mass-ratio regimes, respectively.

For weak-field, slow-motion systems, the post-Newtonian–Multipolar-post-Minkowskian (PN–MPM) expansion provides an analytic framework~\cite{blanchet_PNReview}: the field equations are recast as nonlinear wave equations on a flat background, and formal solutions emerge through multipolar expansions. 
PN theory has seen substantial progress, underpinning mainstream waveform models such as effective-one-body (EOB)~\cite{Buonanno_EOB_1999} models, including SEOBNR models~\cite{Pan_SEOBNRv1_2011,seobnrv4,Cao_2017,liu2024effective,EOBv5EHM,EOBv5PHM}, TEOBResum models~\cite{Nagar_2018_TEOB,Akcay_2021,nagar2024TEOBGO} and IMR families~\cite{ajith2007IMR,Ajith_2011,Khan_2016,London_2018,Garc_a_Quir_s_2020,Manna_IMREcc_2025,IMRPhenomEcc}. 
Current ground-based detectors are sensitive to higher-frequency bands, which typically correspond to the late-stage evolution of binaries approaching circular orbits. 
In contrast, future low-frequency detectors will be more concerned with binaries exhibiting non-zero eccentricities. 
Moreover, in recent years, GW events with possible residual eccentricities have been identified~\cite{Abbott_GW190521_2020}, which has driven the development of gravitational wave models that incorporate eccentricity.

However, when we look ahead to future low-frequency detectors, constructing accurate frequency-domain GW models for generic orbits remains a difficult task. 
The standard strategy is to express the post-Newtonian (PN) orbit in a quasi-Keplerian parametrization~\cite{Damour_1985_1PN,Damour_1988_RPA,Schafer_1993_2PN,Memmesheimer_2004_QK} and then compute its evolution using an adiabatic approximation or a multi-scale expansion~\cite{Pound_2008,Miller_2021}. 
Waveform generation, in turn, relies on analytic Fourier-transform techniques—such as the stationary-phase approximation (SPA)~\cite{Yunes_2009} for non-precessing binaries and the shift-uniform approximation (SUA)~\cite{Klein_2013} when spin precession is present. 
Unfortunately, these frequency-domain waveforms all depend on a small-eccentricity expansion, which tends to break down at large eccentricities.

Recent work has shown that by keeping the Bessel functions that naturally appear in the original Fourier modes of the gravitational radiation, one can avoid the accuracy loss at high eccentricity caused by the small-eccentricity expansion~\cite{Morras_2025}. 
This insight suggests that we need to investigate more deeply the structure of the Fourier modes of both the dynamics and the radiative multipole moments within the PN framework. Building on the studies of~\cite{Arun_2008_tail} and~\cite{Morras_2025,Morras_2025_1PN}, we therefore propose a mathematical approach for describing and computing the Fourier modes of the dynamics and waveforms of a PN two-body system in section \ref{sec_generalized_besselj}. In section \ref{sec_rec_pn}, we revisit the theoretical treatment of nonlinear contributions. Finally in section \ref{sec_reexpress_tail}, we will use the new method to show that the tail contribution to the fluxes can be re-summed in a very compact form without considering the radiation-reaction force.
All implementation code is available in the open-source library \texttt{pyPNFourier}.

\section{Theoretical Background and Review}\label{sec_review}
\subsection{Post-Newtonian Dynamical System and the Quasi-Keplerian Parameterization}

The post-Newtonian approximation solves the Einstein field equations perturbatively in the weak-field, slow-motion regime~\cite{Damour_1985,blanchet_PNReview}.
For a two-body system, the equations of motion can be expanded as a power series in $c^{-1}$.
Neglecting spin and radiation-reaction effects, the acceleration is now known through 4PN order~\cite{Bernard_2016,Bernard_2018},
\begin{align}
\pmb{a} = \pmb{a}_N + \ivc{2}\pmb{a}_{\fPN{1}} + \ivc{4}\pmb{a}_{\fPN{2}} + \ivc{6}\pmb{a}_{\fPN{3}} + \ivc{6}\bigl(\pmb{a}_{\fPN{4}} + \pmb{a}_{\tail}\bigr) + \order{c^{-5}},
\end{align}

where the 4PN tail term $\pmb{a}_{\tail}$ represents the back-reaction on the binary from the multipolar GW flux and is a manifestation of the non-linearity of general relativity~\cite{Blanchet_1988_tail,Damour_2014_nonlocal,Bernard_2016}.  
When this non-linear effect is ignored, the system can be described by the two conserved quantities, the energy $E$ and the angular momentum $J$.  
Starting at 2PN order the field integrals become divergent; to cure this one introduces Hadamard finite-part regularization~\cite{Blanchet_2000} and dimensional regularizatioin, which in turn brings in logarithmic terms and an arbitrary length scale $r_0$.  
A modified harmonic (MH) coordinate system is therefore defined to eliminate the logarithms~\cite{Andrade_2001,Blanchet_2002_3PNflux}; we adopt MH coordinates throughout this work.

The PN two-body problem corresponds to a precessing elliptical orbit that can still be described in terms of the familiar orbital elements such as the eccentricity $e$ and semi-major axis $a$.  
This is the essence of the quasi-Keplerian parameterization~\cite{Damour_1985_1PN,Damour_1988_RPA,Schafer_1993_2PN,Memmesheimer_2004_QK}, widely used in contemporary waveform models (e.g. the PN models~\cite{Hinder_2010,Huerta_2014,Moore_taylorf2ecc_2016,Klein_2018}, IMR family~\cite{Manna_IMREcc_2025,IMRPhenomEcc} and the latest EOB model~\cite{EOBv5EHM}).  
The 4PN results of instantaneous part was found in ADM coordinates\cite{Cho_2022}; here we use MH coordinates which is listed in the Appendix~\ref{app_QKMH} and the 4PN transformation among MH, SH and ADM frames are listed in Appendix~\ref{app_4PNct}. Complete results, including EOB coordinates~\cite{Damour_2015_4PNEOB,Bini_2020_5PNEOB,khalil2023_5PNEOB}, can be found in the Supplementary Material `supp\_quasiKeplerialExpansionResults.m'.

We now briefly review the method and the results needed below, following~\cite{Memmesheimer_2004_QK}.  
The radial velocity $\rdot$ can be written as a function of $(E,J,u)$ with $u:=1/r$.  
Solving $\rdot(E,J,u)=0$ yields two roots $u_{\pm}$ corresponding to periastron and apastron, from which we define the semi-major axis and (radial) eccentricity $e_r$
\begin{align}
a_r := \frac{u_+ + u_-}{2u_+u_-}, \qquad
e_r := \frac{u_- - u_+}{u_- + u_+}.
\end{align}
The orbital period $P$ is twice the travel time from $u_-$ to $u_+$.  Since $u$ varies between these bounds we introduce two auxilary angles $\chi,\zeta$
\begin{align}
u = \frac{1 + e_r\cos\zeta}{a_r(1-e_r^2)} 
  = \frac{1}{a_r\big(1-e_r\cos\chi\big)},
\end{align}
where $\zeta$ is the relativistic anomaly and $\chi$ the eccentric anomaly.  
The third angle, called true anomaly $\psi$, is defined by
\begin{align}
\psi := 2\arctan \biggl(
      \Bigl(\frac{1+e_\phi}{1-e_\phi}\Bigr)^{1/2}
      \tan\frac{\chi}{2}\biggr)
    = 2\arctan\biggl(
      \Bigl(\frac{1+e_r}{1-e_r}\Bigr)^{1/2}
      \tan\frac{\zeta}{2}\biggr),
\end{align}
and introducing a new eccentricity $e_\phi$.  
Owing to later naming conflicts we employ a convention that differs from many references.  
The mean anomaly $l:=n(t-t_0)$ with $n:=2\pi/P$, is related to $\chi$ by the Kepler equation; up to 4PN it takes the form
\begin{align}
l &= \chi - e_t\sin\chi + \Big( \ivc{4}g_{4t} + \ivc{6}g_{6t} + \ivc{8}g_{8t} \Big)(\psi-\chi) + \Big(\ivc{4}f_{4t}+\ivc{6}f_{6t}+\ivc{8}f_{8t}\Big)\sin\psi + \Big(\ivc{6}i_{6t}+\ivc{8}i_{8t}\Big)\sin{2\psi} \nonumber\\
& + \Big(\ivc{6}h_{6t}+\ivc{8}h_{8t}\Big)\sin{3\psi} + \ivc{8}j_{8t}\sin{4\psi} + \order{c^{-10}}, \label{eq_kepler}
\end{align}
where the leading-order coefficient of $\sin\chi$ is the time eccentricity $e_t$, the eccentricity used by most models.  
Unless stated otherwise we henceforth drop the subscript and write $e$.

We next introduce
\begin{align}
x := \omega^{2/3},
\qquad
\omega := (1+k)n,
\end{align}
where $k$ is the periastron advance defined via the orbital phase $\phi$ accumulated in one radial period,
\begin{align}
1+k := \frac{1}{2\pi}\biggl(-\int_{u_+}^{u_-}\frac{\dd\phi}{\dd u}\,\dd u\biggr).
\end{align}
For convenience we later set $v:=x^{1/2}$.
With these relations $(E,J)$ can be expressed in terms of $(x,e)$, and the basic dynamical variables $(r,\dot r,\phi,\dot\phi)$ in terms of $(x,e,\chi)$.  Of particular interest is the orbital phase,
\begin{align}
\phi = \lambda + W(v,e;l),
\qquad
\lambda:=(1+k)l\;\;\Rightarrow\;\;\dot\lambda=\omega,
\end{align}
where
\begin{align}
W(v,e;l) = \chi_t - \chi + e\sin\chi + \order{c^{-2}},
\qquad
\chi_t := 2\arctan\biggl(
          \Bigl(\frac{1+e}{1-e}\Bigr)^{1/2}
          \tan\frac{\chi}{2}\biggr).
\end{align}
Defining $\delta\chi:=\chi_t-\chi = 2\arctan\bigl(\betae\sin\chi\big/\bigl(1-\betae\cos\chi\bigr)\bigr)$
with
\begin{align}
\betae := \frac{1-\sqrt{1-e^2}}{e}.
\end{align}
We express $\psi$ in terms of $\chi_t$ in the following sections.
One can perform the needed Fourier transformations following~\cite{Boetzel_2017_PNKeplerEQ}.  
All 4PN quasi-Keplerian parameterization results are collected in the Supplementary Material `supp\_quasiKeplerialExpansionResults.m'.

\subsection{General Method for Gravitational-Wave Calculations} \label{sec_generalized_besselj}
In the PN-MPM framework~\cite{RevModPhys.52.299,blanchet1986radiative}, the GW polarizations $h_{+,\times}$ measured by a distant observer depend on the radiative multipoles $(U_L,V_L)$, which in turn depend on the source multipoles $(I_L,J_L,\dots)$ and non-linear hereditary terms~\cite{Blanchet_1998}.  
All are ultimately functions of the source dynamics, schematically $h(v,e;l)$.  
Within the QK framework one solves for the adiabatic evolution of $(v,e,l)$ under radiation reaction.  
After averaging over one radial period (the adiabatic or orbital average) one obtains $(\vavg,\eavg,\lavg)$ from the energy and angular-momentum fluxes at infinity,
\begin{align}
\avg{\dot{\vavg}} &= \pdv{v}{E}\avg{\Eflux}
                   + \pdv{v}{J}\avg{\Jflux}, \\
\avg{\dot{\eavg}} &= \pdv{e}{E}\avg{\Eflux}
                   + \pdv{e}{J}\avg{\Jflux},
\end{align}
where the orbital average $\avg{\cdot}$ denotes
\begin{align}
\avg{f} := \frac{1}{2\pi}\int_0^{2\pi}f(l)\dd l.
\end{align}
The fluxes $\Eflux$ and $\Jflux$ are computed from the radiative multipoles; their difference from the actual loss of the system energy $E$ and angular $J$ is the Schott term~\cite{Bini_2012}, which vanishes upon orbital averaging for 2.5PN, 3.5PN and 4.5PN and doesn't vanish for 4PN due to the hereditary effect\cite{Trestini_2025_Schott}, leaving some freedom in the explicit form of the radiation-reaction force~\cite{Fumagalli_2025}. The radiation-reaction force of non-spinning binary is known up to 4.5PN~\cite{Iyer_1995,Gopakumar_1997,Leibovich_2023,blanchet_2024_v9RR}.
Multi-scale analysis then yields post-adiabatic corrections~\cite{Pound_2008,Miller_2021}. 
For current 3PN waveform models only the first-order correction $(\vpa,\epa,\lpa)$ is required~\cite{K_nigsd_rffer_2006}, which can be evaluated from $(\vavg,\eavg,\lavg)$ and substituted into $h(v,e;l)$.

Because data analysis is performed mainly in the frequency domain, analytic Fourier approximations are highly desirable—especially for future low-frequency GW detectors. 
In the non-spinning case one expands
\begin{align}
h = \sum_{m,p}\tilde h_{mp}(v,e)\eexp{\ii (m\lambda+pl)},
\end{align}
and applies the stationary-phase approximation (SPA) to each mode~\cite{Yunes_2009}, expanding the phase
\begin{align}
\phi_{mp}(t) = m\lambda+pl-2\pi ft
\end{align}
about the SPA time $t_{mp}$ as
\begin{align}
\phi_{mp}(t)\approx\phi_{mp}(t_{mp})
     + (t-t_{mp})\dot\phi_{mp}(t_{mp})
     + \frac{1}{2}(t-t_{mp})^2\ddot\phi_{mp},
\end{align}
with the condition
\begin{align}
\dot\phi_{mp}=0
\;\;\Longrightarrow\;\;
m\omega(t_{mp})+pn(t_{mp}) = 2\pi f := v_f^{\,3}.
\end{align}
For non-spinning case one could assume the amplitude of the mode grows much slower than phase. However this do not hold for spin-precessing is included, where the Shifted Uniform Asymptotics (SUA) method is applied~\cite{Klein_2018}.

Since post-adiabatic effects enter at $v^5$, one first solves for the averaged part, giving $\vavg^{3} = (m+p)^{-1}v_f^{3}\big(1+\order{\vavg}\big)$, and then substitutes $v_f$ for $\vavg$ in the corrections $(\vpa,\epa,\lpa,\dots)$.

In principle there is no closed-form relation between $\vavg$, $\lavg$, and $\eavg$.  One strategy is to expand $\eavg$ around its initial value $e_0$, but high initial eccentricity requires very high expansion order.  
Padé resummation has been introduced to improve accuracy in that regime~\cite{Tiwari_2020}.
Recent work shows that one can directly integrate numerically with a standard Runge–Kutta scheme~\cite{Morras_2025}, using $\vavg$ rather than $t$ as the independent variable, and then determine $t_{mp}$ by interpolation from the relation $\vavg^{3}=(m+p)^{-1}v_f^{3}\big(1+\order{\vavg}\big)$.  
Given $f$ and $(p,m)$, this yields the averaged orbital frequency $\vavg(t_{mp})$ and all other dynamical quantities.

\section{Fourier series of bounded-binary PN dynamics} \label{sec_rec_pn}
\cite{Arun_2008_tail} showed that, at Newtonian order, the Fourier modes of a binary system’s multipole moments can be expressed in terms of Bessel functions. 
Because the gravitational field is sourced by these radiative multipole moments, its Fourier modes inherit the same structure. 
Recent studies\cite{Morras_2025,Morras_2025_1PN} have demonstrated that this property can be exploited to compute arbitrary Fourier modes rapidly, thereby greatly improving the accuracy of eccentric-orbit models. 
However, at higher PN orders ordinary Bessel functions are no longer sufficient; here we introduce a new family of elliptic integrals that can represent the Fourier modes of all post-Newtonian multipole moments.
\subsection{ Elliptic integrals that appear in the Fourier decomposition of higher-order post-Newtonian dynamics }

The non-spinning binary's dynamics in the PN framework is characterized by a double periodicity. The Fourier series of a dynamical quantity $f$ takes the form
\begin{align}
    f(\lambda,\chi) = \sum_{p,m=-\infty}^{\infty}\hat{f}^{(p,m)}\ee^{\ii(pl+m\lambda)},
\end{align}
where the Fourier coefficient $\hat{f}^{(p,m)}$ is given by
\begin{align}
    \hat{f}^{(p,m)}=\frac{1}{(2\pi)^2}\iint_{-\pi}^{\pi} f(\lambda,\chi)\ee^{-\ii(pl+m\lambda)}\dd{l}\dd{\lambda}.
\end{align}
Here we use $\chi$ is because In most cases the quantity depends on $\chi$ rather than $l$.
And usually, the dependence on $\lambda$ mainly comes from the orbital phase $\phi=\lambda + W(l)$, so we can simplify it via
\begin{align}
    &f(\lambda,\chi) = \sum_{m=-\infty}^{\infty} f^{m}(\chi)\ee^{\ii m\lambda}, \\
    &\hat{f}^{(p,m)} = \frac{1}{2\pi}\int_{-\pi}^{\pi}f^{(m)}(\chi)\ee^{-\ii pl}\dd{l} \nonumber\\
    & = \frac{1}{2\pi}\int_{-\pi}^{\pi}f^{(m)}(\chi)\ee^{-\ii p l}\dv{l}{\chi}\dd{\chi}
\end{align}
One would find that after sorting, up to 3PN order, these Fourier coefficients can be expressed by two kinds of integrals,
\begin{align}
    &\fJint{n}{pqa}(e) := \frac{1}{2\pi}\int_{-\pi}^{\pi}\frac{\ee^{\ii(p\chi-qe\sin\chi)}}{(1-e\cos\chi)^a}(\ii\delta\chi)^n\dd\chi, \\
    &\fJintln{n}{pqa}(e) := \frac{1}{2\pi}\int_{-\pi}^{\pi}\frac{\ee^{\ii(p\chi-qe\sin\chi)}}{(1-e\cos\chi)^a}(\ii\delta\chi)^n\ln(1-e\cos\chi)\dd\chi,
\end{align}
These integrals can be viewed as extensions of Hansen coefficients~\cite{Mik_czi_2015}.
And also can be regarded as the extension of~\cite{Munna_2020,Munna_2022,Trestini_2025_ST}. For example, $\fJint{0}{pq0}\equiv J_p(qe)$.
The factor $(1-e\cos\chi)$ shows in the denominator can be regarded as the derivative with respect to $l$, since
\begin{align}
    \dv{f}{l} = \dv{f}{\chi} \dv{\chi}{l} = \frac{1}{1-e\cos\chi}\dv{f}{\chi} + \order{c^{-4}}.
\end{align}
At 3PN order, we only encounter the term $\fJint{3}{pqa}$ and $\fJintln{0}{pqa}$. 
By observing the symmetry of the integrands we can quickly obtain some properties,
\begin{align}
    &\fJint{n}{(-p)(-q)a}=(-1)^n\fJint{n}{pqa},\ \fJintln{n}{(-p)(-q)a}=(-1)^n\fJintln{n}{pqa}.
\end{align}
And the properties of differentiation,
\begin{align}
    &\pdv{e}\fJint{n}{pqa} = \frac{q}{2}\Big(\fJint{n}{(p-1)qa}-\fJint{n}{(p+1)qa}\Big) + \frac{a}{2}\Big(\fJint{n}{(p-1)q(a+1)}+\fJint{n}{(p+1)q(a+1)}\Big) - \frac{ n}{2\sqrt{1-e^2}}\Big(\fJint{n-1}{(p-1)q(a+1)}-\fJint{n-1}{(p+1)q(a+1)}\Big), \\
    &\pdv{e}\fJintln{n}{pqa} = \frac{q}{2}\Big(\fJintln{n}{(p-1)qa}-\fJintln{n}{(p+1)qa}\Big) + \frac{a}{2}\Big(\fJintln{n}{(p-1)q(a+1)}+\fJintln{n}{(p+1)q(a+1)}\Big) - \frac{ n}{2\sqrt{1-e^2}}\Big(\fJintln{n-1}{(p-1)q(a+1)}-\fJintln{n-1}{(p+1)q(a+1)}\Big) \nonumber\\
    &\qquad\qquad - \frac{1}{2}\Big( \fJintln{n}{(p-1)q(a+1)} + \fJintln{n}{(p+1)q(a+1)} \Big),
\end{align}
And recursion identities
\begin{align}
    &2p\fJint{n}{pqa}=e\bigg[q\Big(\fJint{n}{(p+1)qa}+\fJint{n}{(p-1)qa}\Big)+a\Big(\fJint{n}{(p-1)q(a+1)}-\fJint{n}{(p+1)q(a+1)}\Big)\bigg] + \frac{2n\betae(1-\betae)}{1+\betae^2}\Big(\fJint{n+1}{(p-1)q(a+1)}+\fJint{n+1}{(p+1)q(a+1)}\Big), \\
    &2p\fJintln{n}{pqa}=e\Big(\fJint{n}{(p-1)q(a+1)}-\fJint{n}{(p+1)q(a+1)}\Big)+e\bigg[q\Big(\fJintln{n}{(p+1)qa}+\fJintln{n}{(p-1)qa}\Big)+a\Big(\fJintln{n}{(p-1)q(a+1)}-\fJintln{n}{(p+1)q(a+1)}\Big)\bigg] \nonumber\\
    &\qquad + \frac{2n\betae(1-\betae)}{1+\betae^2}\Big(\fJintln{n+1}{(p-1)q(a+1)}+\fJintln{n+1}{(p+1)q(a+1)}\Big), \\
    &\fJint{n}{pq(a-1)} = \fJint{n}{pqa}-\frac{e}{2}\Big(\fJint{n}{(p+1)qa}+\fJint{n}{(p-1)qa}\Big),\ \fJintln{n}{pq(a-1)} = \fJintln{n}{pqa}-\frac{e}{2}\Big(\fJintln{n}{(p+1)qa}+\fJintln{n}{(p-1)qa}\Big), \\
    &\frac{e}{2}\Big(\fJint{n}{(p+1)qa}-\fJint{n}{(p-1)qa}\Big) = 
        \frac{1}{1-a}\bigg(p\fJint{n}{pq(a-1)} + q\Big(\fJint{n}{pq(a-2)}-\fJint{n}{pq(a-1)}\Big) + n\Big(\sqrt{1-e^2}\fJint{n-1}{pqa} - \fJint{n-1}{pq(a-1)}{}\Big)\bigg), \\
    &\frac{e}{2}\Big(\fJintln{n}{(p+1)qa}-\fJintln{n}{(p-1)qa}\Big) = \frac{1}{1-a}\bigg(p\fJintln{n}{pq(a-1)} + q\Big(\fJintln{n}{pq(a-2)} - \fJintln{n}{pq(a-1)}\Big) + n\Big(\sqrt{1-e^2}\fJintln{n-1}{pqa} - \fJintln{n-1}{pq(a-1)} \Big) \nonumber\\
    &\qquad -\frac{e}{2}\Big(\fJint{n}{(p+1)qa}-\fJint{n}{(p-1)qa}\Big)\bigg)
\end{align}
At some special cases, we have~\cite{PhysRev.136.B1224},
\begin{align}
    &\fJint{0}{00a}= \frac{1}{(1-e^2)^{a/2}}P_{a-1}\Big(\frac{1}{\sqrt{1-e^2}}\Big),\\
    &\fJintln{0}{00a} = \frac{(-1)^{a-1}}{(a-1)!}\Big(\dv[a-1]{Y(y,e)}{y}\Big)_{y=1}, \\
    &\fJintln{0}{p00} = -\frac{\betae^{|p|}}{|p|}.
\end{align}
where $P_N(x)$ is Legendre polynomial and
\begin{align}
    Y(y,e) = \frac{1}{\sqrt{y^2-e^2}}\Big[\ln\Big(\frac{\sqrt{1-e^2}+1}{2}\Big)+2\ln\Big(1+\frac{\sqrt{1-e^2}-1}{y+\sqrt{y^2-e^2}}\Big)\Big].
\end{align}
These integrals generally do not have a closed form, but they can be easily expanded as power series of eccentricity. We reorganize them into the infinite series of Bessel function and $\betae$,
\begin{align}
    &\fJint{0}{pqa} = (1+\betae^2)^a\sum_{k=-\infty}^{\infty}(-1)^kJ_k(qe)\cL_{k+p}^{(a)}, \\
    &\fJint{1}{pqa} = (1+\betae^2)^a\sum_{n=0}^{\infty}\sum_{\delta=\pm 1}\frac{\delta\betae^{n+1}}{n+1}\sum_{k=-\infty}^{\infty}(-1)^kJ_k(qe)\cL_{k+p+\delta(n+1)}^{(a)}, \\
    &\fJint{2}{pqa} = -(1+\betae^2)^a\sum_{n=0}^{\infty}\sum_{s=0}^n\sum_{\delta=\pm 1}\frac{\betae^{n+2}}{(s+1)(n-s+1)}\sum_{k=-\infty}^{\infty}(-1)^kJ_k(qe)\big(\cL_{k+p+\delta(n-2s)}^{(a)} - \cL_{k+p+\delta(n+2)}^{(a)} \big), \\
    &\fJint{3}{pqa} = -(1+\betae^2)^a\sum_{n=0}^{\infty}\sum_{s_1=0}^n\sum_{s_2=0}^{s_1}\sum_{\delta=\pm 1}\frac{\delta\betae^{n+3}}{(s_1-s_2+1)(n-s_1+1)(s_2+1)} \nonumber\\
    &\qquad\qquad\times\sum_{k=-\infty}^{\infty}(-1)^kJ_k(qe)\big(\cL_{k+p+\delta(1+n-2s_2)}^{(a)} + \cL_{k+p+\delta(1+n-2s_1+2s_2)}^{(a)} - \cL_{k+p+\delta(n+3)}^{(a)} - \cL_{k+p+\delta(n-1-2s_1)}^{(a)} \big), \\
    &\fJintln{0}{pqa} = -(1+\betae^2)^a\sum_{k=-\infty}^{\infty}(-1)^kJ_k(qe)\big(\cL_{k+p}^{(a)}\ln(1+\betae^2)+\dv{a}\cL_{k+p}^{(a)}\big)
\end{align}
where $\cL_n^{(a)}$ denotes an extension of Laplace coefficients~\cite{book:celestMechanic},
\begin{align}
    \cL_n^{(a)}(\betae) := \frac{1}{2\pi}\int_{-\pi}^{\pi}\frac{\eexp{\ii ny}}{(1+\betae^2-2\betae\cos{y})^a}\dd{y},
\end{align}
which is also equivalent to
\begin{align}
    \cL_n^{(a)}(\betae) = \Big(\frac{e}{2\betae}\Big)^a\fJint{0}{n0a}(e).
\end{align}
It can be written in the form of a hypergeometric function. Since the coefficients of this hypergeometric function are non-negative, it can be further expanded into a finite polynomial of $\betae$ when $a>0$,
\begin{align}
    &\cL_{n}^{(a)}=\frac{(a)_{|n|}}{|n|!}\betae^{|n|}(1-\betae^2)^{1-2a}\hypfunc{1-a}{|n|+1-a}{|n|+1}{\betae^2} \nonumber\\
    &=(a)_{|n|}\betae^{|n|}(1-\betae^2)^{1-2a}\sum_{m=0}^{a-1}\frac{(-1)^m(|n|+1-a)_m}{(|n|+m)!}\binom{a-1}{m}\betae^{2m},
\end{align}
where the general hypergeometric function is defined by~\cite{hypergeometric_function}
\begin{align}
    [{}_{p}\mathrm{F}_{q}]\bigg(\begin{array}{c}a_1,a_2,...a_p\\b_1,b_2,...b_q\end{array};y\bigg) = \sum_{m=0}^{\infty}\frac{(a_1)_m(a_2)_m...(a_p)_m}{(b_1)_m(b_2)_m...(b_q)_m}\frac{x^m}{m!}.
\end{align}
And $(a)_n:=\Gamma(a+n)/\Gamma(a)$ is Pochhammer symbol.
$\dv{a}\cL_n^{(a)}$ is the formal derivative with respect to $a$, which takes a polynomial series,
\begin{align}
    &\dv{a}\cL_n^{(a)} := |n|!\sum_{m=0}^\infty \frac{1}{(|n|+m)!m!}\betae^{2m}\dv{a}\Big[(1-a)_m(|n|+1-a)_m\Big],
\end{align}
where the Pochhammers' derivative with respective to $a$ is
\begin{align}
    &\dv{a}(1-a)_m = 
    \left\{
    \begin{aligned}
    &(1-a)_m\DHarmN{a}{a-m} & m\le a-1 \\
    &(-1)^a\Gamma(a)\Gamma(m+1-a) & m\ge a
    \end{aligned}
    \right. \\
    &\dv{a}(|n|+1-a) = 
    \left\{
    \begin{aligned}
    &(|n|+1-a)_m\DHarmN{|n|+1-a}{|n|+1+m-a} & a\le |n|, \\
    &(|n|+1-a)_m\DHarmN{a-|n|}{a-m-|n|} & a\ge |n|+1\ \text{and}\ m\le a-|n|-1, \\
    &(-1)^{a-|n|}\Gamma(a-|n|)\Gamma(m+|n|+1-a) & a\ge |n|+1\ \text{and}\ m\ge a-|n|
    \end{aligned}
    \right.
\end{align}
$\DHarmN{n}{m}$ denotes the difference between the $(n-1)$th and $(m-1)$th Harmonic number. The poles produced by the di-Gamma function $\psi(x):=\Gamma'(x)/\Gamma(x)$ divided by the Gamma function are removed by the following limit,
\begin{align}
    \lim_{n\to\mathbb{Z}^+} \frac{\psi(1-n)}{\Gamma(1-n)} = (-1)^n(n-1)!.
\end{align}
Specifically, when $a=0$,
\begin{align}
    \dv{a}\eval{\cL_n^{(a)}}_{a=0} = 
    \begin{cases}
        \beta^{|n|}/|n| & n \neq 0 \\
        0 & n=0.
    \end{cases}
\end{align}
Here we recall the Kepteyn series of standard Bessel functions~\cite{book:BesselFunction}
\begin{align}
    &\frac{1}{1-e} = \sum_{p=-\infty}^\infty \fJint{0}{pp0}, \\
    &\frac{1}{\sqrt{1-e^2}} = \sum_{p=-\infty}^\infty \big(\fJint{0}{pp0}\big)^2.
\end{align}
These identities can be generalized to
\begin{align}
    &\frac{\delta_{n0}}{(1-e)^{a+1}} = \sum_{p=-\infty}^\infty \fJint{n}{ppa},\quad \frac{\delta_{n0}\ln(1-e)}{(1-e)^{a+1}} = \sum_{p=-\infty}^\infty \fJintln{n}{ppa}, \\
    &\fJint{c+d}{m0(a+b)} = \sum_{p=-\infty}^\infty \fJint{c}{(p+m)qa} \fJint{d}{pqb},\quad \fJintln{c+d}{m0(a+b)} = \sum_{p=-\infty}^\infty \fJint{c}{(p+m)qa}\fJintln{d}{pqb},\quad \fJintlnln{c+d}{m0(a+b)} = \sum_{p=-\infty}^\infty  \fJintln{c}{(p+m)qa} \fJintln{d}{pqb},
\end{align}
where $\fJintlnln{n}{pqa}$ is defined via
\begin{align}
    \fJintlnln{n}{pqa} := \frac{1}{2\pi}\int_{-\pi}^{\pi}\frac{\eexp{\ii(p\chi-qe\sin\chi)}}{(1-e\cos\chi)^a}\ln^2\big(1-e\cos\chi\big)(\ii\delta\chi)^n\dd{\chi}.
\end{align}
The proof of the first identity is easy,
\begin{align}
    &\sum_{p=-\infty}^{\infty}\fJint{n}{ppa} = \frac{1}{2\pi}\int_{-\pi}^{\pi}(\ii\delta\chi)^n(1-e\cos\chi)^{-a}\sum_{p=-\infty}^{\infty} \eexp{\ii(p\chi-pe\sin\chi)}\dd\chi \nonumber\\
    &\qquad = \int_{-\pi}^{\pi}(\ii\delta\chi)^n(1-e\cos\chi)^{-a} \delta\big(p(\chi-e\sin\chi)\big)\dd\chi = \frac{\delta_{n0}}{(1-e)^{a+1}},
\end{align}
where we use the Poisson summation formula
\begin{align}
    \sum_{p=-\infty}^{\infty}\eexp{\ii px} = 2\pi\sum_{k=-\infty}^{\infty}\delta(x-2\pi k).
\end{align}
The proof of the second identity is similar,
\begin{align}
    &\sum_{p=-\infty}^{\infty}\fJint{c}{(p+m)qa}\fJint{d}{pqb} = \frac{1}{(2\pi)^2}\iint_{-\pi}^{\pi}\frac{\eexp{\ii \big[mx-q(\sin{x}-\sin{y})\big]}}{(1-e\cos{x})^a(1-e\cos{y})^b}\ii^{c+d}\delta\chi(x)^{c}\delta\chi(y)^{d}\sum_{p=-\infty}^{\infty} \eexp{\ii p(x-y)}\dd{x}\dd{y}, \nonumber\\
    &=\frac{1}{2\pi}\int_{-\pi}^{\pi}\frac{\eexp{\ii mx}\dd{x}}{(1-e\cos{x})^{a+b}}\big(\ii\delta\chi(x+y)\big)^{c+d} = \fJint{c+d}{m0(a+b)},
\end{align}
where we use $\sum_p\eexp{\ii p(x-y)}=2\pi\delta(x-y)$. Note that this identity is independent of $q$. For other cases involving logarithms, just add the logarithmic term to the integral.
At this point, we can easily expand these integrals to any power of eccentricity. 
Because $J_{p}(qe)\sim (qe/2)^p/p!$ and, for $e\in(0,1)$, there is $\betae <e$, the series expansions of these integrals converge very rapidly.  
In practical calculations one usually determines the dynamical variables first and then evaluates the waveform.  
Hence, for a fixed eccentricity, the coefficients $\cL$, the Bessel functions $J_p(qe)$, and even the generalized coefficients $\fJint{c}{pqa}$ themselves are repeatedly required.
By caching these coefficients the computational cost can be reduced substantially.
All of these integrals, together with the waveform routines discussed later, are packaged in the publicly available module \texttt{pyPNFourier}.

\subsection{Results}
\subsubsection{Solution of PN-Kepler equation}
Here we present the Fourier-series solution of the 4PN Kepler equation~(\ref{eq_kepler}).  We emphasize that the Kepler equation considered here does not include the contribution from the 4PN non-local effects.

First, we replace the orbital energy $E$ and angular momentum $J$ in the equation by $v$ and $e$, and substitute the true anomaly $\psi$ with $\chi_t$.  We then compute the Fourier series of $\chi-l$,
\begin{align}
\chi - l \;=\; \sum_{p=1}^{\infty} X_{p}\,\sin p l .
\end{align}
The Fourier coefficients $X_p$ have the structure
\begin{align}
&X_p = X_p^{0} 
          + v^4X_p^{2} 
          + v^6X_p^{3} 
          + v^8X_p^{4} + \order{c^{-9}}, \\
&X_p^{0} = \frac{2}{p}J_p(e p) \\
&X_p^{2} =  \frac{1}{\sqrt{1-e^2}}\Bigg[(15-6\nu)\big( J_p(eq) + \fJint{1}{pp0} \big) - \frac{3}{2}e(5-2\nu)\big(2J_{p-1}(ep) + 2\fJintc{1}{1}{pp0} + 2p\fJints{1}{1}{pp0} - ep\fJints{2}{1}{pp0}\big) \nonumber\\
&\  - \frac{1}{4}e\sqrt{1-e^2}(60-39\nu+\nu^2)\fJints{1}{0}{pp0}\Bigg], \\
&X_p^{3} = \frac{1}{(1-e^2)^{3/2}}\Bigg\{ -\nu p(1-e^2)^{5/2}\bigg(\frac{13 \nu ^2}{48}-\frac{73 \nu }{48}+\frac{23}{48}\bigg)\fJint{0}{pp1} + \bigg[ 21 \nu ^2+\bigg(\frac{41 \pi ^2}{32}-\frac{427}{3}\bigg) \nu +60 \nonumber\\
&\  - (1-e^2)(30-29\nu+11\nu^2) + \nu p(1-e^2)^{3/2}\bigg( \frac{\nu ^2}{12}+\frac{131 \nu }{12}-\frac{505}{12} \bigg) \bigg]J_p(ep) + \bigg[10 \nu ^2+\bigg(\frac{41 \pi ^2}{32}-\frac{340}{3}\bigg) \nu +30 \nonumber\\
&\  + e^2(30-29\nu+11\nu^2)\bigg]\fJint{1}{pp0} + e\bigg[ e\bigg(\frac{13 \nu ^3}{24}-\frac{347 \nu ^2}{24}+\frac{86 \nu }{3}-30\bigg) + \sqrt{1-e^2}\bigg(-\frac{13 \nu ^3}{24}-\frac{181 \nu ^2}{24} \nonumber\\
&\  +\bigg(\frac{12413}{210}-\frac{41 \pi ^2}{16}\bigg) \nu +50\bigg) \bigg]\fJints{1}{0}{pp1} + e\bigg[ 5 \nu ^2+\bigg(\frac{41 \pi^2}{64}-\frac{170}{3}\bigg) \nu +15 + e^2 \bigg(\frac{11 \nu ^2}{2}-\frac{29 \nu }{2}+15\bigg) \bigg]\big(ep\fJints{2}{1}{pp0} \nonumber\\
&\  - 2J_{p-1}(ep) -2 \fJintc{1}{1}{pp0} - 2p\fJints{1}{1}{pp0} \big) \Bigg\} \\
&X_p^{4} = \frac{1}{(1-e^2)^{5/2}}\Bigg\{ \frac{9}{8}p(1-e^2)^{3/2}(5-2\nu)^2\big( 4\fJint{2}{pp0} - 4e\fJintc{1}{2}{pp0} - 2ep\fJints{1}{2}{pp0} + e^2\fJints{2}{2}{pp0} \big) \nonumber\\
&\  + \frac{3}{8}(1-e^2)^2(-300+315\nu-83\nu^2+2\nu^3)\big(ep\fJints{1}{1}{pp1}-\fJint{1}{pp1}\big) - \frac{3}{16}\nu(1-e^2)(75-35\nu+2\nu^2)\big(e^2p^2\fJintc{2}{1}{pp0} \nonumber\\
&\  - 2(1-e^2)\fJint{1}{pp2}\big) + \Bigg[ \nu p^2(1-e^2)^2(2-e^2)\bigg(\frac{3 \nu ^2}{8}-\frac{105 \nu }{16}+\frac{225}{16}\bigg) -\frac{3}{2} \sqrt{1-e^2} \big(5 (1-e^2)-3\big) (5-2 \nu)^2 -\frac{321 \nu ^3}{4} \nonumber\\
&\  +\bigg(\frac{8977}{8}-\frac{615 \pi ^2}{32}\bigg) \nu ^2+\bigg(\frac{47503 \pi ^2}{1024}-\frac{207253}{120}\bigg) \nu +\frac{2949}{8} + (1-e^2)\Bigg(\frac{171 \nu ^3}{2}+\frac{1}{96} \bigg(615 \pi ^2-64552\bigg) \nu ^2 \nonumber\\
&\  +\bigg(\frac{37567}{36}-\frac{53281 \pi ^2}{3072}\bigg) \nu -\frac{2055}{8}\Bigg) + (1-e^2)^2\bigg( -\frac{143 \nu ^3}{12}+\frac{403 \nu ^2}{24}-\frac{37 \nu }{8}+15 \bigg) \Bigg]\fJint{1}{pp0} \nonumber\\
&\  + \nu p(1-e^2)^{5/2}\bigg(\bigg(\frac{103 e^2}{384}-\frac{107}{192}\bigg) \nu ^3+\bigg(-\frac{35 e^2}{64}-\frac{45}{32}\bigg) \nu ^2+\bigg(\frac{85 e^2}{128}-\frac{41 \pi ^2}{128}-\frac{92303}{4032}\bigg) \nu +\frac{349 e^2}{384}-\frac{903 \pi ^2}{1024} \nonumber\\
&\  -\frac{7111577}{201600}\bigg)\fJint{0}{pp1} + \nu p(1-e^2)^{7/2}\bigg(\frac{5 \nu ^3}{64}-\frac{187 \nu ^2}{32}+\frac{20269 \nu \
}{576}-\frac{1447}{144}\bigg)\fJint{0}{pp2} + \nu p(1-e^2)^{9/2}\bigg(\frac{25 \nu ^3}{256}+\frac{27 \nu ^2}{128}-\frac{1639 \nu }{256} \nonumber\\
&\  -\frac{31}{256}\bigg)\fJint{0}{pp3} + e\Bigg( \frac{321 \nu ^3}{4}+\frac{1}{32} \bigg(615 \pi ^2-35908\bigg) \nu ^2+\bigg(\frac{207253}{120}-\frac{47503 \pi ^2}{1024}\bigg) \nu -\frac{2949}{8} + e^2\Bigg(-\frac{171 \nu ^3}{2} \nonumber\\
&\  -\frac{1}{96} \bigg(615 \pi ^2-64552\bigg) \nu ^2+\bigg(\frac{53281 \pi ^2}{3072}-\frac{37567}{36}\bigg) \nu +\frac{2055}{8}\Bigg) + e^4\bigg( \frac{143 \nu ^3}{12}-\frac{403 \nu ^2}{24}+\frac{37 \nu }{8}-15 \bigg) \nonumber\\
&\quad + \frac{3}{2}\sqrt{1-e^2} \big(2 (1-e^2)-3\big) (5-2 \nu )^2-\frac{1}{32}\nu^2p^2(1-e^2)\Bigg)J_{p-1}(ep) + \Bigg[ p(1-e^2)^{3/2}\Bigg( -\frac{45 \nu ^3}{4}+\bigg(\frac{135 \pi ^2}{8}-\frac{274069}{2520}\bigg) \nu ^2 \nonumber\\
&\  +\bigg(\frac{1730437}{4200}-\frac{34135 \pi ^2}{3072}\bigg) \nu -\frac{1695}{8} + \nu\sqrt{1-e^2}\bigg( -\frac{9 \nu ^2}{4}+\frac{315 \nu }{8}-\frac{675}{8} \bigg) + \nu(1-e^2)\bigg( \frac{5 \nu ^3}{192}+\frac{639 \nu ^2}{32} \nonumber\\
&\  -\frac{5699 \nu }{192}-\frac{1135}{96} \bigg) \Bigg) + \frac{1}{32}\nu^2p^2(1-e^2)^{5/2}(15-\nu)^2 + \frac{3}{2}\sqrt{1-e^2}(5-2\nu)^2\big( 1+2e^2 \big) - \frac{321 \nu ^3}{4}+\bigg(\frac{8977}{8}-\frac{615 \pi ^2}{32}\bigg) \nu ^2 \nonumber\\
&\  +\bigg(\frac{47503 \pi ^2}{1024}-\frac{207253}{120}\bigg) \nu +\frac{2949}{8} + (1-e^2)\Bigg( \frac{171 \nu ^3}{2}+\bigg(\frac{205 \pi ^2}{32}-\frac{8069}{12}\bigg) \nu ^2+\bigg(\frac{37567}{36}-\frac{53281 \pi ^2}{3072}\bigg) \nu -\frac{2055}{8} \Bigg) \nonumber\\
&\  + (1-e^2)^2\bigg(-\frac{143 \nu ^3}{12}+\frac{403 \nu ^2}{24}-\frac{37 \nu }{8}+15\bigg) \Bigg]J_p(ep) + e\Bigg[-\frac{3}{2}\sqrt{1-e^2}(1+2e^2)(5-2\nu)^2 + \frac{20 \nu ^3}{3} \nonumber\\
&\  +\bigg(\frac{205 \pi ^2}{16}-\frac{933}{2}\bigg) \nu ^2+\bigg(\frac{123877}{180}-\frac{22307 \pi ^2}{768}\bigg) \nu -\frac{507}{4} + e^2\Bigg(\frac{185 \nu ^3}{3}+\bigg(\frac{205 \pi ^2}{32}-\frac{3833}{6}\bigg) \nu ^2+\bigg(\frac{18617}{18}-\frac{53281 \pi ^2}{3072}\bigg) \nu \nonumber\\
&\   -\frac{1815}{8}\Bigg) + e^4\bigg(\frac{143 \nu ^3}{12}-\frac{403 \nu ^2}{24}+\frac{37 \nu }{8}-15\bigg) \Bigg]\fJintc{1}{1}{pp0} + ep\Bigg[ \frac{3}{2}\sqrt{1-e^2}(2-5e^2)(5-2\nu)^2 + \frac{49 \nu ^3}{6}+\bigg(\frac{205 \pi ^2}{16}-\frac{1971}{4}\bigg) \nu ^2 \nonumber\\
&\  +\bigg(\frac{67001}{90}-\frac{22307 \pi ^2}{768}\bigg) \nu -\frac{507}{4} + e^2\Bigg(\frac{176 \nu ^3}{3}+\bigg(\frac{205 \pi ^2}{32}-\frac{1759}{3}\bigg) \nu ^2+\bigg(\frac{8296}{9}-\frac{53281 \pi ^2}{3072}\bigg) \nu -\frac{1815}{8}\Bigg) \nonumber\\
&\  + e^4\bigg( \frac{161 \nu ^3}{12}-\frac{1033 \nu ^2}{24}+\frac{487 \nu }{8}-15 \bigg) \Bigg]\fJints{1}{1}{pp0} + e\Bigg[ \sqrt{1-e^2}\bigg[ \frac{141 \nu ^3}{2}+\bigg(-\frac{619691}{1260}-\frac{335 \pi ^2}{16}\bigg) \nu ^2+\bigg(\frac{2584979}{6300}+\frac{99 \pi ^2}{64}\bigg) \nu \nonumber\\
&\  +107 + (1-e^2)\bigg( -\frac{1123 \nu ^3}{12}+\bigg(\frac{69581}{120}-\frac{697 \pi ^2}{64}\bigg) \nu ^2+\bigg(\frac{67951 \pi ^2}{3072}-\frac{10111811}{12600}\bigg) \nu +\frac{135}{8} \bigg) + (1-e^2)^2\bigg( -\frac{283 \nu ^4}{576} \nonumber\\
&\  +\frac{1231 \nu ^3}{96}-\frac{2141 \nu ^2}{192}+\frac{241 \nu }{192}-15 \bigg) \bigg] + (1-e^2)^2\bigg(-\frac{\nu ^3}{2}+\frac{83 \nu ^2}{4}-\frac{315 \nu }{4}+75\bigg) -\frac{1}{64}\nu^2p^2(15-\nu)^2\Bigg]\fJints{1}{0}{pp1} \nonumber\\
&\ + e^2p\Bigg[\frac{3}{4}\sqrt{1-e^2}(1+2e^2)(5-2\nu)^2 + -\frac{321 \nu ^3}{8}+\bigg(\frac{8977}{16}-\frac{615 \pi ^2}{64}\bigg) \nu ^2+\bigg(\frac{47503 \pi ^2}{2048}-\frac{207253}{240}\bigg) \nu +\frac{2949}{16} \nonumber\\
&\ + (1-e^2)\Bigg( \frac{171 \nu ^3}{4}+\bigg(\frac{205 \pi ^2}{64}-\frac{8069}{24}\bigg) \nu ^2+\bigg(\frac{37567}{72}-\frac{53281 \pi ^2}{6144}\bigg) \nu -\frac{2055}{16} \Bigg) + (1-e^2)^2\bigg(-\frac{143 \nu ^3}{24}+\frac{403 \nu ^2}{48} \nonumber\\
&\ -\frac{37 \nu }{16}+\frac{15}{2}\bigg) \Bigg]\fJints{2}{1}{pp0} \Bigg\},
\end{align}
where we introduce shorthands
\begin{align}
{}^{(\pm)}\mathrm{J}^{(b,c)}_{pqa} := \frac{1}{2}\Big(\fJint{b}{(p+c)qa} \pm \fJint{b}{(p-c)qa}\Big), \\
\end{align}
By expanding with small eccentricity, we compared the 3PN results with those of Eq.(A8) in \cite{Boetzel_2017_PNKeplerEQ}. Here we show the expansion of coefficients $X^2_p,X^3_p$ for the case $p=5$ as a reference,
\begin{align}
&X^2_5 = e^5\bigg( \frac{523 \nu ^2}{1024}+\frac{1167 \nu }{5120}-\frac{5049}{256} \bigg) + e^7\bigg( -\frac{50195 \nu ^2}{73728}+\frac{542539 \nu }{122880}+\frac{44521}{3072} \bigg) + e^9\bigg( \frac{473695 \nu ^2}{1376256}-\frac{16009117 \nu }{6881280}-\frac{2439751}{344064} \bigg) \nonumber\\
&\quad+ \order{e^{11}},\\
&X^3_5 = e^5 \bigg( -\frac{2481 \nu ^3}{1024}+\frac{12281 \nu ^2}{256}+\Big(\frac{17571199}{215040}+\frac{9553 \pi^2}{10240}\Big) \nu -\frac{25969}{128} \bigg) + e^7 \bigg( \frac{495725 \nu ^3}{110592}-\frac{33517009 \nu^2}{368640} \nonumber\\
&\quad +\Big(\frac{715241323}{3096576}-\frac{3887989 \pi ^2}{2949120}\Big) \nu +\frac{20815}{4608} \bigg) + e^9\bigg( -\frac{12241405 \nu ^3}{4128768}+\frac{271099897 \nu^2}{10321920}+\left(\frac{5228167549}{96337920}-\frac{2643475 \pi ^2}{16515072}\right) \nu \nonumber\\
&\quad -\frac{46872317}{516096} \bigg) + \order{e^{11}}.
\end{align}
One can check that they are completely consistent with the coefficient $A_s$ in Eq.(A8) of \cite{Boetzel_2017_PNKeplerEQ}.

\subsubsection{Multipole moments}
There are two sets of radiative-type multipole moments $U_L,V_{L}$~\cite{Blanchet_1998}, where $L=i_1...i_l$ denotes the (STF) multi-indices. 
These radiative-type moments is determined by six source-type moments $(I_L,J_L,W_L,X_L,Y_L,Z_L)$ or two kinds of canonical source-type multipole moments $M_L,S_L$,
\begin{align}
    &U_L(t_r) = M_L^{(\ell)}(t_r) + \frac{2\MADM}{c^3}\int_0^{\infty}\dd\tau M_{L}^{(\ell+2)}(t_r-\tau)\Big[\ln\Big(\frac{\tau}{2r_0}\Big)+\kappa_\ell\Big] + \order{c^{-5}}, \\
    &V_L(t_r) = S_L^{(\ell)}(t_r) + \frac{2\MADM}{c^3}\int_0^{\infty}\dd\tau S_{L}^{(\ell+2)}(t_r-\tau)\Big[\ln\Big(\frac{\tau}{2r_0}\Big)+\pi_\ell\Big] + \order{c^{-5}},
\end{align}
where $t_r$ denotes retarded-time and $\MADM$ is ADM mass.
At same PN-level, the number of the indices of $V$ is 1 less than $U$. At present, we have a clear understanding of the form of 3PN in the MH frame~\cite{Arun_2008,Mishra_2015}, and recently the 4PN results were published~\cite{Marchand_2020}. 
In this work, we are concerned with the form of 3PN in the MH frame~\cite{Blanchet_2008}. Following~\cite{Arun_2008}, the structure of the Fourier decomposition of $M_L, S_{L}$ reads
\begin{align}
    &M_L = \sum_{m=-\ell}^\ell \sum_{p=-\infty}^{\infty} \fmodeM{p}{m}{L}\eexp{\ii(m\lambda+pl)}, \\
    &S_{L} = \sum_{m=-\ell}^{\ell}\sum_{p=-\infty}^{\infty} \fmodeS{p}{m}{L}\eexp{\ii(m\lambda+pl)}.
\end{align}
To expand these moments, we use polar coordinates $(r,\rdot,\phi,\dot{\phi})$, then
\begin{align}
    &(x,y,z) = r(\cos\phi, \sin\phi, 0), \\
    &(v_x,v_y,v_z) = (\rdot\cos\phi-r\dot{\phi}\sin\phi, \rdot\sin\phi+r\dot{\phi}\cos\phi, 0).
\end{align}
In this coordinates, the STF matrices can be represented by $(r,\dot{r},\dot{\phi})$. 
Substituting the results of the quasi-Keplerian expansion and then performing Fourier integration, we obtain the Fourier modes of these moments. 
Here we show the 2PN results of $M_{xx}$ as an example, the results of full 3PN order and all $\ell\leq8$ moments are showed in Supplementary Material `supp\_MomFourierCoefficientsResults.m',
\begin{align}
    &\fmodeM{p}{2}{xx} = \frac{\nu}{ v^4}\Big( \fmodeMpn{p}{2}{xx}{N} + v^2\fmodeMpn{p}{2}{xx}{\fPN{1}} + v^4\fmodeMpn{p}{2}{xx}{\fPN{2}} + v^5\fmodeMpn{p}{2}{xx}{\fPN{2.5}} + v^6\fmodeMpn{p}{2}{xx}{\fPN{3}} + \order{c^{-7}}\Big), \\
    &\fmodeMpn{p-2}{2}{xx}{N} = \frac{1}{e^2p^2}\Bigg[e\sqrt{1-e^2}\big(-1+\sqrt{1-e^2}p\big)J_{p-1}\big(ep\big) + \frac{1}{2}\big(-1+\sqrt{1-e^2}\big)\big(-1+\sqrt{1-e^2}-2(1-e^2)p\big)J_{p}(ep)\Bigg], \\
    &\fmodeMpn{p-2}{2}{xx}{\fPN{1}} = \frac{1}{e^2(1-e^2)p^3} \Bigg\{ \frac{3}{8}e^2p^3\Big[5e(2+e^2)\fJintc{1}{1}{pp0} - 2(2+e^2)\fJintc{2}{1}{pp0} + e(2-e^2)\fJintc{3}{1}{pp0}\Big] \nonumber\\
    &\quad + \frac{3}{4}e^2\sqrt{1-e^2}p^3\Big[ 2(1+e^2)\fJints{2}{1}{pp0} - e\fJints{3}{1}{pp0} - 5e\fJints{1}{1}{pp0} \Big] - \frac{15}{4}e^4p^3\fJint{1}{pp0} + e\Bigg[ \sqrt{1-e^2}\bigg[18-p\nu \nonumber\\
    &\quad - p^2(1-e^2)\bigg(\frac{23}{42}+\frac{4\nu}{21}\bigg) + p^3(1-e^2)^2\bigg(\frac{1}{21}-\frac{\nu}{7}\bigg)\bigg] - \frac{15}{2}p + (1-e^2)\bigg(\frac{27}{14} \nu  p^2+\frac{3}{14} (-10 p-91) p\bigg) \nonumber\\
    &\quad - (1-e^2)^2p^2\bigg(\frac{23}{42} + \frac{4\nu}{21}\bigg) \Bigg]J_{p+1}\big(ep\big) + \big(1-\sqrt{1-e^2}\big)\Bigg[9 + \frac{11}{2}p + (1-e^2)\bigg(\frac{1}{84} \nu  p (-162 p-67)+\frac{1}{84} p (180 p+1539)\bigg) \nonumber\\
    &\quad + (1-e^2)^2p^2\bigg( \frac{11}{21} + \frac{11\nu}{42} \bigg) + \sqrt{1-e^2}\bigg[ \nu  p-\frac{13 p}{2}-9 + (1-e^2)\bigg(\frac{1}{84} \nu  p (6 p-17) +\frac{1}{84} p (57-758 p)\bigg) \nonumber\\
    &\quad + (1-e^2)^2p^3\bigg(-\frac{1}{21} + \frac{\nu}{7}\bigg) \bigg] \Bigg]J_{p}\big(ep\big) \Bigg\}, \\
    &\fmodeMpn{p-2}{2}{xx}{\fPN{2}} = \frac{1}{e^2p^4(1-e^2)^2}\Bigg\{ e^4p^4\Bigg[ (1-e^2) \bigg(\frac{249}{16}-\frac{17 \nu }{4}\bigg)+\bigg(1-e^2\bigg)^{3/2} \bigg(\frac{75 p}{16}-\frac{15 \nu  p}{8}\bigg)-\frac{45 \sqrt{1-e^2}}{4}+\frac{135 \nu }{8}-\frac{765}{16} \Bigg]\fJint{1}{pp0} \nonumber\\
    &\quad + \frac{45 e^4 p^4}{4}\fJint{2}{pp0} + (1-e^2)^4p^4\bigg(\frac{613 \nu ^2}{3024}-\frac{7195 \nu }{3024}+\frac{4513}{1512}\bigg)\fJint{0}{pp1} + (1-e^2)^5p^4\bigg(-\frac{613 \nu ^2}{3024}+\frac{2011 \nu }{3024}+\frac{131}{1512}\bigg)\fJint{0}{pp2} \nonumber\\
    &\quad + e\Bigg[ \bigg(\frac{\nu ^2}{16}-\frac{15 \nu }{16}\bigg) p^5 (1-e^2)^4+\bigg(\frac{9 \nu }{14}-\frac{97}{28}\bigg) p^4 (1-e^2)^3+p^3 \bigg(\bigg(-\frac{53 \nu ^2}{168}+\frac{655 \nu }{504}+\frac{2}{7}\bigg) (1-e^2)^3 \nonumber\\
    &\quad +\bigg(\frac{197 \nu ^2}{126}+\frac{344 \nu }{63}+\frac{7375}{504}\bigg) (1-e^2)^2+\bigg(-\frac{149 \nu }{14}-\frac{207}{14}\bigg) (1-e^2)\bigg)+p^2 \bigg(\frac{135 \nu }{4}+\bigg(\frac{4579}{56}-\frac{531 \nu }{14}\bigg) (1-e^2)^2 \nonumber\\
    &\quad +\bigg(\frac{515 \nu }{28}-\frac{180}{7}\bigg) (1-e^2)-\frac{765}{8}\bigg)+p \big(252 (1-e^2)+144\big) + \sqrt{1-e^2}\bigg[p^4 \bigg(\bigg(\frac{73 \nu ^2}{378}-\frac{19 \nu }{54}+\frac{149}{1512}\bigg) (1-e^2)^3 \nonumber\\
    &\quad +\bigg(\frac{32 \nu ^2}{189}+\frac{437 \nu }{378}+\frac{8989}{756}\bigg) (1-e^2)^2\bigg)+p^3 \bigg(\bigg(\frac{397 \nu }{28}-\frac{1891}{56}\bigg) (1-e^2)^2+\bigg(\frac{347}{8}-\frac{63 \nu }{4}\bigg) (1-e^2)\bigg)+p^2 \bigg(\frac{\nu ^2}{2}+\frac{17 \nu }{2} \nonumber\\
    &\quad +\bigg(-\frac{5 \nu ^2}{24}-\frac{15 \nu }{56}+\frac{85}{168}\bigg) (1-e^2)^2+\bigg(-\frac{37 \nu ^2}{21}-\frac{19 \nu }{6}-\frac{2955}{28}\bigg) (1-e^2)-\frac{361}{4}\bigg)+p \bigg(-63 \nu \nonumber\\
    &\quad +\bigg(\frac{348 \nu }{7}-\frac{1737}{14}\bigg) (1-e^2)+\frac{315}{2}\bigg)-216\bigg] \Bigg]J_{p-1}(ep) + \Bigg[ \bigg(\frac{15 \nu }{16}-\frac{\nu ^2}{16}\bigg) p^5 (1-e^2)^4 +p^4 \bigg(\bigg(\frac{31 \nu ^2}{252}-\frac{41 \nu }{84} \nonumber\\
    &\quad+\frac{143}{1008}\bigg) (1-e^2)^4+\bigg(\frac{181 \nu ^2}{756}+\frac{1787 \nu }{756}+\frac{37013}{3024}\bigg) (1-e^2)^3\bigg)+p^3 \bigg(\bigg(\frac{53 \nu ^2}{168}+\frac{6653 \nu }{504}-\frac{501}{14}\bigg) (1-e^2)^3 \nonumber\\
    &\quad +\bigg(-\frac{197 \nu ^2}{126}-\frac{2713 \nu }{126}+\frac{15359}{504}\bigg) (1-e^2)^2+\bigg(\frac{149 \nu }{14}+\frac{207}{14}\bigg) (1-e^2)\bigg)+p^2 \bigg(-\frac{99 \nu }{4}+\bigg(\frac{3 \nu ^2}{112}+\frac{37 \nu }{48}-\frac{155}{336}\bigg) (1-e^2)^3 \nonumber\\
    &\quad +\bigg(-\frac{503 \nu ^2}{336}+\frac{4385 \nu }{112}-\frac{9419}{48}\bigg) (1-e^2)^2+\bigg(-\frac{681 \nu }{28}-\frac{253}{7}\bigg) (1-e^2)+\frac{617}{8}\bigg)+p \bigg(-\frac{81 \nu }{2} \nonumber\\
    &\quad +\bigg(\frac{111 \nu }{7}-\frac{729}{28}\bigg) (1-e^2)^2+\bigg(\frac{159 \nu }{14}-\frac{2151}{7}\bigg) (1-e^2)-\frac{117}{4}\bigg)-108 (1-e^2)-108 + \sqrt{1-e^2}\bigg[\bigg(-\frac{53 \nu ^2}{378}-\frac{103 \nu }{378} \nonumber\\
    &\quad +\frac{131}{1512}\bigg) p^5 (1-e^2)^4+p^4 \bigg(\bigg(-\frac{73 \nu ^2}{378}+\frac{188 \nu }{189}-\frac{5387}{1512}\bigg) (1-e^2)^3+\bigg(-\frac{32 \nu ^2}{189}-\frac{437 \nu }{378}-\frac{8989}{756}\bigg) (1-e^2)^2\bigg) \nonumber\\
    &\quad +p^3 \bigg(\bigg(\frac{149 \nu ^2}{504}-\frac{317 \nu }{168}+\frac{295}{504}\bigg) (1-e^2)^3+\bigg(\frac{19 \nu ^2}{42}-\frac{185 \nu }{63}+\frac{443}{8}\bigg) (1-e^2)^2+\bigg(\frac{\nu ^2}{2}+\frac{211 \nu }{84}-\frac{10993}{168}\bigg) (1-e^2)\bigg) \nonumber\\
    &\quad +p^2 \bigg(-\frac{\nu ^2}{2}+\frac{67 \nu }{2}+\bigg(\frac{5 \nu ^2}{24}-\frac{239 \nu }{8}+\frac{1799}{24}\bigg) (1-e^2)^2+\bigg(\frac{37 \nu ^2}{21}+\frac{116 \nu }{21}+\frac{191}{2}\bigg) (1-e^2)-\frac{59}{4}\bigg)+p \bigg(63 \nu \nonumber\\
    &\quad +\bigg(\frac{4005}{14}-\frac{348 \nu }{7}\bigg) (1-e^2)+\frac{153}{2}\bigg)+216\bigg] \Bigg]J_p(ep) + e^3\Bigg[ p^4 \bigg(-\frac{405 \nu }{16}+\bigg(\frac{1055}{224}-\frac{65 \nu }{56}\bigg) (1-e^2)^2 \nonumber\\
    &\quad +\bigg(\frac{1983 \nu }{112}-\frac{1581}{28}\bigg) (1-e^2)+\frac{2295}{32}\bigg) + \sqrt{1-e^2}\bigg[ p^5 \bigg(\bigg(\frac{75}{32}-\frac{15 \nu }{16}\bigg) (1-e^2)^2+\bigg(\frac{45 \nu }{16}-\frac{225}{32}\bigg) (1-e^2)\bigg) \nonumber\\
    &\quad +p^4 \bigg(9-\frac{9 (1-e^2)}{2}\bigg) \bigg] \Bigg]\fJintc{1}{1}{pp0} - \frac{45}{8}e^3p^4(1+2e^2)\fJintc{1}{2}{pp0} + e^2\Bigg[ p^4 \bigg(\frac{81 \nu }{8}+\bigg(\frac{55 \nu }{28}-\frac{883}{112}\bigg) (1-e^2)^2\nonumber\\
    &\quad +\bigg(\frac{471}{14}-\frac{591 \nu }{56}\bigg) (1-e^2)-\frac{459}{16}\bigg) + \sqrt{1-e^2}\bigg[p^5 \bigg(\bigg(\frac{3 \nu }{8}-\frac{15}{16}\bigg) (1-e^2)^2+\bigg(\frac{45}{16}-\frac{9 \nu }{8}\bigg) (1-e^2)\bigg) \nonumber\\
    &\quad +p^4 \big(9-9 (1-e^2)\big)\bigg] \Bigg]\fJintc{2}{1}{pp0} + \frac{9}{4}e^2p^4(1+2e^2)\fJintc{2}{2}{pp0} + e^2\Bigg[ p^4 \bigg(-\frac{27 e \nu }{16} +(1-e^2)^2 \bigg(\frac{37 e \nu }{56}-\frac{243 e}{224}\bigg) \nonumber\\
    &\quad +(1-e^2) \bigg(\frac{53 e \nu }{112}-\frac{129 e}{56}\bigg)+\frac{153 e}{32}\bigg) + \sqrt{1-e^2}\bigg[p^5 \bigg((1-e^2)^2 \bigg(\frac{3 e \nu }{16}-\frac{15 e}{32}\bigg)+(1-e^2) \bigg(\frac{3 e \nu }{16}-\frac{15 e}{32}\bigg)\bigg) \nonumber\\
    &\quad +p^4 \bigg(\frac{9 e (1-e^2)}{2}-9 e\bigg)\bigg] \Bigg]\fJintc{3}{1}{pp0} + \frac{9}{8}e^3p^4( -2+e^2 )\fJintc{3}{2}{pp0} + \frac{9}{4}e^4p^4\sqrt{1-e^2}\fJintc{4}{1}{pp0} + e^3\Bigg[ \bigg(\frac{75}{16}-\frac{15 \nu }{8}\bigg) p^5 (1-e^2)^2 \nonumber\\
    &\quad +p^4 \bigg(\frac{63}{4}-\frac{81 (1-e^2)}{4}\bigg) + \sqrt{1-e^2}p^4\bigg[ \frac{105 \nu }{8}+\bigg(\frac{1439}{112}-\frac{61 \nu }{14}\bigg) (1-e^2)-\frac{525}{16} \bigg] \Bigg]\fJints{1}{1}{pp0} + \frac{45}{4}e^3p^4\sqrt{1-e^2}\fJints{1}{2}{pp0} \nonumber\\
    &\quad + ep^4(1-e^2)^{7/2}\bigg(-\frac{43}{14}+\frac{12}{7}\nu\bigg)\fJints{1}{0}{pp1} + e^2\Bigg[p^5 \bigg(\bigg(\frac{15}{8}-\frac{3 \nu }{4}\bigg) (1-e^2)^3+\bigg(\frac{3 \nu }{2}-\frac{15}{4}\bigg) (1-e^2)^2\bigg) \nonumber\\
    &\quad +p^4 \bigg(-\frac{27 (1-e^2)^2}{4}+\frac{45 (1-e^2)}{2}-\frac{63}{4}\bigg) + \sqrt{1-e^2}p^4\bigg[-\frac{21 \nu }{2}+\bigg(\frac{223}{56}-\frac{11 \nu }{7}\bigg) (1-e^2)^2 \nonumber\\
    &\quad +\bigg(\frac{295 \nu }{28}-\frac{1529}{56}\bigg) (1-e^2)+\frac{105}{4}\bigg]\Bigg]\fJints{2}{1}{pp0} - \frac{9}{2}e^2p^4\sqrt{1-e^2}(1+e^2)\fJints{2}{2}{pp0} + e^3p^4\Bigg[ \bigg(\frac{15}{16}-\frac{3 \nu }{8}\bigg) p (1-e^2)^2-\frac{9 (1-e^2)}{4} \nonumber\\
    &\quad +\frac{27}{4} + \sqrt{1-e^2}\bigg[\frac{21 \nu }{8}+\bigg(\frac{579}{112}-\frac{29 \nu }{14}\bigg) (1-e^2)-\frac{105}{16}\bigg] \Bigg] \fJints{3}{1}{pp0} + \frac{9}{4}e^3p^4\sqrt{1-e^2}\fJints{3}{2}{pp0} + \frac{9}{8}e^4p^4(-2+e^2)\fJints{4}{1}{pp0} \Bigg\},
\end{align}
where the 1PN results are consistent with Eq.(68)-(70) of~\cite{Loutrel_2017}.
$\fJintln{n}{pqa}$ would appear in the 3PN terms of the Fourier coefficients of the dynamical multipole moments.

\subsubsection{Gravitational waveform}
The GWs emitted to null infinity can be expressed in transverse-traceless frame $h_{ij}^{\tt}$ constituted by two polarizations $h_+,h_\times$. The perturbation matrix $h_{ij}^{\tt}$ can be uniquily decomposed in to radiative-type multipole moments~\cite{RevModPhys.52.299}
\begin{align}
    h_{ij}^{\tt} = \frac{4}{c^2R}P_{ijab}^{\tt}(\pmb{N})\sum_{l=2}^\infty \frac{1}{c^ll!}\bigg[ N_{L-2}U_{ab(L-2)}-\frac{2\ell}{c(\ell+1)}N_{c(L-2)}\epsilon_{cd(a}V_{b)d(L-2)} \bigg] + \order{R^{-2}},
\end{align}
where $\pmb{N} = (\sin\theta\cos\phi,\sin\theta\sin\phi,\cos\theta)$ is the direction from the source to the observer, $R$ denotes the (luminosity) distance of the source, and $P_{ijab}^{\tt}$ is transverse-traceless projector.
The GW polarization can be decomposed by spin-weighted -2 spherical harmonics~\cite{Blanchet_2008}
\begin{align}
    h_+-\ii h_\times = \sum_{l=2}^\infty \sum_{m=-\ell}^\ell h_{\ell m} [{}_{-2}Y_{\ell m}](\theta,\phi).
\end{align}
The spherical harmonic modes are given by
\begin{align}
    h_{\ell m} = -\frac{1}{\sqrt{2}R c^{\ell +2}}\Big(U_{\ell m}-\frac{\ii}{c}V_{\ell m}\Big),
\end{align}
where the spherical modes of the radiative moments $U_{lm},V_{lm}$ are
\begin{align}
    &U_{\ell m} = \frac{4}{\ell!}\sqrt{\frac{(\ell+1)(\ell+2)}{2\ell(\ell-1)}}\alpha^{\ell m}_LU_L, \\
    &V_{\ell m} = -\frac{8}{\ell!}\sqrt{\frac{\ell(\ell+2)}{2(\ell+1)(\ell-2)}}\alpha^{\ell m}_L V_L.
\end{align}
The coefficient $\alpha^{\ell m}_L$ denotes the connection between STF tensor $N^{\avg{L}}$ and spherical harmonics $Y_{lm}$~\cite{Ledesma_2020},
\begin{align}
    &N^{\avg{L}} = \sum_{m=-\ell}^\ell \alpha^L_{\ell m}(\pmb{N})Y_{\ell m}(\pmb{N}), \\
    &Y_{\ell m}(\pmb{N}) = \frac{(2\ell +1)!!}{4\pi \ell !}\alpha^{L*}_{\ell m} (\pmb{N})N^{\avg{L}} ,\\
    &\alpha_L^{\ell m} := \int\dd\pmb{N} N^{\avg{L}}Y^*_{\ell m}(\pmb{N}).
\end{align}
The GW waveform includes instantaneous part, tail part, and memory part, which correspond to the three components of the radiation multipole moment respectively. For higher PN orders exceeding 3PN, there will also be a tail-memory interaction part. 
Currently, we only consider the instantaneous part and the tail part within 3PN.
When calculating the tail part, one would encounter integrals involving the power of logarithm. To evaluate such integral, one can use
\begin{align}
    \eval{\dv[s]{\Gamma(z)}{z}}_{z=1} = \int_0^\infty \dd\tau e^{-\tau}\ln^s\tau.
\end{align}
Therefore,
\begin{align}
    &\int_0^\infty \eexp{-\ii y \tau}\ln^s\tau \dd\tau = \eval{\pdv[s]{z}\int_0^\infty t^{z-1}\eexp{-(\varepsilon+\ii y)\tau}\dd\tau}_{z=1,\varepsilon=0} = \eval{\pdv[s]{z}\Big[(\varepsilon+\ii y)^{-z}\Gamma(z)\Big]}_{z=1,\varepsilon=0}\nonumber \\
    &=\frac{1}{\ii y}\sum_{m=0}^s (-1)^m\binom{s}{m}\cfgm{s-m} \Big(\ln|y| + \sgn(y)\frac{\pi}{2}\Big)^m, \label{eq_logintegral}
\end{align}
where $\cfgm{s}:=\eval{\dv[s]{z}\Gamma(z)}_{z=1}$. This formula can recover~\cite{Boetzel_2019}.
Substituting the Fourier expansion of $M_L,S_L$ and integrating, one could express the Fourier modes in terms of $\fJint{n}{pqa}$. Here we show 2PN results of $h_{22}$ for example, the results of full 3PN order and all $\ell\leq 8$ spherical modes can be found in Supplementary Material `supp\_MomFourierCoefficientsResults.m',
\begin{align}
    &h_{22} = \sqrt{\frac{\pi}{5}}\frac{4\nu v^2}{c^4R}\eexp{-\ii 2\lambda}\sum_{p=-\infty}^\infty \tilde{H}_{2(-2)p}\eexp{\ii p l}, \\
    &\tilde{H}_{2(-2)p} = \tilde{H}^{N}_{2(-2)p} + \ivc{2}v^2\tilde{H}_{2(-2)p}^{\fPN{1}} + \ivc{3}v^3\tilde{H}_{2(-2)p}^{\tail,\fPN{1.5}} + \ivc{4}v^4\tilde{H}_{2(-2)p}^{\fPN{2}} + 
    \ivc{5}v^5\Big( \tilde{H}_{2(-2)p}^{\fPN{2.5}} + \tilde{H}_{2(-2)p}^{\tail,\fPN{2.5}}\Big) \nonumber\\
    &\qquad + 
    \ivc{6}v^6\Big( \tilde{H}_{2(-2)p}^{\fPN{3}} + \tilde{H}_{2(-2)p}^{\tail,\fPN{3}} \Big) + \order{c^{-7}}, \\
    &\tilde{H}_{2(-2)(p+2)}^{N} = \frac{1}{e^2}\Bigg[ (-2+2e^2)\fJint{0}{pp1} + (2-4e^2+2e^4)\fJint{0}{pp2} + 2e\sqrt{1-e^2}J_{p-1}(e p) \nonumber\\
    &\quad + \bigg[ e^2-2 - 2\sqrt{1-e^2}(1+p-e^2p) \bigg]J_p(e p) \Bigg] \\
    &\tilde{H}_{2(-2)(p+2)}^{\fPN{1}} = \frac{1}{e^2p(1-e^2)} \Bigg\{ -12p^2(1-e^2)^{3/2}\fJint{1}{pp0} - 12p(1-e^2)\fJint{1}{pp1} + p(1-e^2)\Bigg[ e^2 \bigg(-\frac{8 \nu }{21}-\frac{23}{21}\bigg)-\frac{73 \nu }{21}+\frac{113}{21} \nonumber\\
    &\quad + \sqrt{1-e^2}\bigg[p \bigg(e^2 \bigg(\frac{2 \nu }{7}+\frac{124}{21}\bigg)-\frac{2 \nu }{7}-\frac{124}{21}\bigg)-12\bigg] \Bigg]\fJint{0}{pp1} + p(1-e^2)\Bigg[ e^4 \bigg(-\frac{20 \nu }{21}-\frac{19}{21}\bigg)+e^2 \bigg(\frac{884}{21}-\frac{41 \nu }{21}\bigg) \nonumber\\
    &\quad +\frac{61 \nu }{21}-\frac{865}{21} + \sqrt{1-e^2}\bigg[ -12 e^2+p \bigg(e^4 \bigg(\frac{2 \nu }{7}-\frac{2}{21}\bigg)+e^2 \bigg(\frac{4}{21}-\frac{4 \nu }{7}\bigg)+\frac{2 \nu }{7}-\frac{2}{21}\bigg)+12 \bigg] \Bigg]\fJint{0}{pp2} + 12p(1-e^2)^2\fJint{1}{pp2} \nonumber\\
    &\quad + p(1-e^2)\bigg[e^4 \bigg(\frac{10 \nu }{7}+\frac{242}{21}\bigg)+e^2 \bigg(-\frac{20 \nu }{7}-\frac{484}{21}\bigg)+\frac{10 \nu }{7}+\frac{242}{21}\bigg]\fJint{0}{pp3} + p(1-e^2)\Bigg[e^6 \bigg(\frac{6 \nu }{7}-\frac{2}{7}\bigg)+e^4 \bigg(\frac{6}{7}-\frac{18 \nu }{7}\bigg) \nonumber\\
    &\quad +e^2 \bigg(\frac{18 \nu }{7}-\frac{6}{7}\bigg)-\frac{6 \nu }{7}+\frac{2}{7}\Bigg]\fJint{0}{pp4} + e\Bigg[ p(12-6e^2) + \sqrt{1-e^2}\Bigg[ p \bigg(e^2 \bigg(\frac{37}{7}-\frac{25 \nu }{21}\bigg)+\frac{67 \nu }{21}-\frac{121}{7}\bigg)+12 \Bigg] \Bigg]J_{p-1}(e p) \nonumber\\
    &\quad + \Bigg[6 e^2+\bigg(-12 e^4+24 e^2-12\bigg) p^2+p \bigg(e^4 \bigg(\frac{17 \nu }{42}-\frac{19}{14}\bigg)+e^2 \bigg(\frac{39 \nu }{14}-\frac{279}{14}\bigg)-\frac{67 \nu }{21}+\frac{121}{7}\bigg)-12 \nonumber\\
    &\quad + \sqrt{1-e^2}\bigg[p \bigg(e^2 \bigg(\frac{25 \nu }{21}-\frac{37}{7}\bigg)-\frac{67 \nu }{21}+\frac{121}{7}\bigg)+p^2 \bigg(e^4 \bigg(\frac{11 \nu }{21}+\frac{22}{21}\bigg)+e^2 \bigg(\frac{62 \nu }{21}-\frac{129}{7}\bigg)-\frac{73 \nu }{21}+\frac{365}{21}\bigg)-12\bigg]\Bigg]J_p(e p) \nonumber\\
    &\quad + ep(-12+6e^2)\fJintc{1}{1}{pp0} - 12ep\sqrt{1-e^2}\fJints{1}{1}{pp0} - 6ep\bigg[ 2(1-e^2) + \sqrt{1-e^2}(2-e^2) \bigg]\fJints{1}{0}{pp1} \Bigg\} \\
    &\tilde{H}_{2(-2)(p+2)}^{\tail,\fPN{1.5}} = \frac{\ii }{e^2p}\Bigg(3\ln\bigg(\frac{x}{x_0'} \bigg)+ 2\ln\bigg(\frac{\ii p}{2}\bigg) \Bigg)\Bigg[ p\sqrt{1-e^2}\fJint{0}{pp1} + \bigg[2-e^2 + p(1-e^2)^{3/2}\bigg]\fJint{0}{pp2} + \bigg[ 1-\frac{7}{2}e^2 \nonumber\\
    &\quad -p(1-e^2)^{3/2} \bigg]\fJint{0}{pp3} + \bigg[ -\frac{21 e^4}{2}+\frac{67 e^2}{2}-23 + 3p(1-e^2)^{5/2} \bigg]\fJint{0}{pp4} + 35(1-e^2)^2\fJint{0}{pp5} + 15(1-e^2)^3\fJint{0}{pp6} \Bigg] \\
    &\tilde{H}_{2(-2)(p+2)}^{\fPN{2}} = \frac{1}{e^2p^2(1-e^2)^2}\Bigg\{ p^2\Bigg[ p^2 \bigg(6 \nu  (1-e^2)^3-15 (1-e^2)^3\bigg)-72 p (1-e^2)^2+72 (1-e^2) \nonumber\\
    &\quad+ \sqrt{1-e^2}\Bigg(p \bigg(\bigg(\frac{160 e^2}{7}+\frac{50}{7}\bigg) \nu  (1-e^2)+\bigg(\frac{352}{7}-\frac{401 e^2}{7}\bigg) (1-e^2)\bigg)+36 (1-e^2)+36\Bigg) \Bigg]\fJint{1}{pp0} - 36 p^3(1-e^2)^{3/2}\fJint{2}{pp0} \nonumber\\
    &\quad + p^2(1-e^2)\Bigg[ \frac{227 \nu }{7}+\bigg(\frac{\nu ^2}{8}-\frac{15 \nu }{8}\bigg) p^2 (1-e^2)^3+\bigg(-\frac{30 \nu }{7}-\frac{242}{7}\bigg) p (1-e^2)^2+\bigg(\frac{37 \nu ^2}{42}-\frac{589 \nu }{63}+\frac{97}{14}\bigg) (1-e^2)^2 \nonumber\\
    &\quad +\bigg(-\frac{725 \nu ^2}{252}-\frac{6829 \nu }{252}+\frac{14459}{252}\bigg) (1-e^2)+\frac{279}{7} + \sqrt{1-e^2}\bigg[18 \nu +p \bigg(\bigg(\frac{73 \nu ^2}{189}-\frac{673 \nu }{189}+\frac{7547}{756}\bigg) (1-e^2)^2 \nonumber\\
    &\quad +\bigg(\frac{64 \nu ^2}{189}+\frac{3272 \nu }{189}+\frac{53}{189}\bigg) (1-e^2)\bigg)+\bigg(\frac{785}{7}-\frac{192 \nu }{7}\bigg) (1-e^2)-13\bigg] \Bigg]\fJint{0}{pp1} + p^2 (1-e^2) \Bigg[ \frac{216 \nu }{7} \nonumber\\
    &\quad + \bigg(\frac{403}{7}-\frac{166 \nu }{7}\bigg) (1-e^2)-\frac{555}{7} + \sqrt{1-e^2}\bigg[\bigg(\frac{30 \nu }{7}-\frac{353}{7}\bigg) p (1-e^2)-72\bigg] \Bigg] \fJint{1}{pp1} -36 p^2 (1-e^2)\fJint{2}{pp1} \nonumber\\
    &\quad +  p^2 (1-e^2)^2\Bigg[-\frac{3 \nu ^2}{8}+\frac{4043 \nu }{56}+\bigg(\frac{12 \nu }{7}-\frac{4}{7}\bigg) p (1-e^2)^2+\bigg(-\frac{19 \nu ^2}{72}+\frac{1271 \nu }{504}+\frac{41}{36}\bigg) (1-e^2)^2+\bigg(\frac{772 \nu ^2}{189} \nonumber\\
    &\quad -\frac{5140 \nu }{189}+\frac{28838}{189}\bigg) (1-e^2)-\frac{1923}{14} + \sqrt{1-e^2}\bigg[ -\frac{36 \nu }{7}+p \bigg(\bigg(\frac{11 \nu ^2}{63}+\frac{113 \nu }{63}-\frac{137}{252}\bigg) (1-e^2)^2 \nonumber\\
    &\quad +\bigg(-\frac{64 \nu ^2}{189}-\frac{5 \nu }{189}-\frac{13435}{378}\bigg) (1-e^2)\bigg)+\bigg(\frac{138 \nu }{7}-\frac{424}{7}\bigg) (1-e^2)-\frac{1724}{7} \bigg]\Bigg]\fJint{0}{pp2} + p^2 (1-e^2)^2\Bigg[ -\frac{216 \nu }{7} \nonumber\\
    &\quad +\bigg(\frac{142 \nu }{7}-\frac{395}{7}\bigg) (1-e^2)-\frac{957}{7} + \sqrt{1-e^2}\bigg[\bigg(\frac{101}{7}-\frac{30 \nu }{7}\bigg) p (1-e^2)+72\bigg] \Bigg]\fJint{1}{pp2} + 36p^2 (1-e^2)^2 \fJint{2}{pp2} \nonumber\\
    &\quad + p^2 (1-e^2)^3 \Bigg[ -\frac{1283 \nu ^2}{378}-\frac{14767 \nu }{378}+\bigg(-\frac{\nu ^2}{21}+\frac{1238 \nu }{63}-\frac{659}{24}\bigg) (1-e^2)-\frac{19819}{216} + \sqrt{1-e^2}\bigg[ \bigg(-\frac{265 \nu ^2}{189}-\frac{353 \nu }{189} \nonumber\\
    &\quad +\frac{2707}{756}\bigg) p (1-e^2)+98 \bigg] \Bigg] \fJint{0}{pp3} + p^2 (1-e^2)^3 \bigg( \frac{60 \nu }{7}+\frac{484}{7} \bigg)\fJint{1}{pp3} +p^2 (1-e^2)^4 \Bigg[\frac{2 \nu ^2}{189}-\frac{3986 \nu }{189}+\bigg(-\frac{331 \nu ^2}{126} \nonumber\\
    &\quad-\frac{1193 \nu }{126}+\frac{211}{72}\bigg) (1-e^2)+\frac{16001}{216} + \sqrt{1-e^2}\bigg[ -\frac{36 \nu }{7}+\bigg(\frac{53 \nu ^2}{63}+\frac{103 \nu }{63}-\frac{131}{252}\bigg) p (1-e^2)+\frac{12}{7} \bigg]\Bigg] \fJint{0}{pp4} \nonumber\\
    &\quad + p^2 (1-e^2)^4 \bigg(\frac{12}{7}-\frac{36 \nu }{7}\bigg) \fJint{1}{pp4} + p^2 (1-e^2)^5\bigg(\frac{238 \nu ^2}{27}+\frac{5792 \nu }{189}-\frac{12793}{756}\bigg)\fJint{0}{pp5} + p^2 (1-e^2)^6\bigg(-\frac{265 \nu ^2}{63} \nonumber\\
    &\quad-\frac{515 \nu }{63}+\frac{655}{252}\bigg)\fJint{0}{pp6} + e\Bigg[ p^3 \bigg(\bigg(\frac{\nu ^2}{8}-\frac{15 \nu }{8}\bigg) (1-e^2)^3+\bigg(\frac{\nu ^2}{8}-\frac{15 \nu }{8}\bigg) (1-e^2)^2\bigg)+p^2 \bigg(-27 \nu \nonumber\\
    &\quad +\bigg(\frac{88 \nu }{7}-\frac{313}{14}\bigg) (1-e^2)^2+\bigg(\frac{95 \nu }{7}-\frac{1119}{7}\bigg) (1-e^2)+\frac{81}{2}\bigg)+p (54 (1-e^2)+54) + \sqrt{1-e^2}\bigg[ p^2 \bigg(-\nu ^2+25 \nu \nonumber\\
    &\quad+\bigg(\frac{5 \nu ^2}{12}+\frac{15 \nu }{28}-\frac{85}{84}\bigg) (1-e^2)^2+\bigg(\frac{74 \nu ^2}{21}-\frac{563 \nu }{21}+\frac{585}{14}\bigg) (1-e^2)-\frac{101}{2}\bigg)+p \bigg(-42 \nu +\bigg(\frac{232 \nu }{7}-\frac{579}{7}\bigg) (1-e^2)+33\bigg) \nonumber\\
    &\quad+72 \bigg] \Bigg]J_{p-1}(e p) + \Bigg[ \bigg(\frac{15 \nu }{4}-\frac{\nu ^2}{4}\bigg) p^4 (1-e^2)^3+p^3 \bigg(\bigg(-\frac{\nu ^2}{8}-\frac{1143 \nu }{56}+\frac{430}{7}\bigg) (1-e^2)^3 \nonumber\\
    &\quad+\bigg(-\frac{\nu ^2}{8}+\frac{825 \nu }{56}+\frac{768}{7}\bigg) (1-e^2)^2\bigg)+p^2 \bigg(18 \nu +\bigg(\frac{3 \nu ^2}{56}+\frac{37 \nu }{24}-\frac{155}{168}\bigg) (1-e^2)^3+\bigg(-\frac{503 \nu ^2}{168}+\frac{1881 \nu }{56} \nonumber\\
    &\quad-\frac{12203}{168}\bigg) (1-e^2)^2+\bigg(\frac{1927}{7}-\frac{419 \nu }{7}\bigg) (1-e^2)-28\bigg)+p \bigg(27 \nu +\bigg(\frac{243}{14}-\frac{74 \nu }{7}\bigg) (1-e^2)^2+\bigg(\frac{132}{7}-\frac{53 \nu }{7}\bigg) (1-e^2) \nonumber\\
    &\quad-\frac{189}{2}\bigg)-36 (1-e^2)-36 + \sqrt{1-e^2}\bigg[p^3 \bigg(\bigg(-\frac{149 \nu ^2}{252}+\frac{317 \nu }{84}-\frac{295}{252}\bigg) (1-e^2)^3+\bigg(-\frac{19 \nu ^2}{21}-\frac{709 \nu }{126}-\frac{99}{2}\bigg) (1-e^2)^2 \nonumber\\
    &\quad+\bigg(-\nu ^2-\frac{74 \nu }{21}+\frac{4415}{42}\bigg) (1-e^2)\bigg)+p^2 \bigg(\nu ^2-25 \nu +\bigg(-\frac{5 \nu ^2}{12}-\frac{247 \nu }{28}+\frac{1801}{84}\bigg) (1-e^2)^2+\bigg(-\frac{74 \nu ^2}{21}+\frac{605 \nu }{21}\nonumber\\
    &\quad+\frac{409}{14}\bigg) (1-e^2)+\frac{245}{2}\bigg)+p \bigg(42 \nu +\bigg(\frac{327}{7}-\frac{232 \nu }{7}\bigg) (1-e^2)-105\bigg)-72\bigg] \Bigg]J_p(e p) + e p^2 (1-e^2)\Bigg[ -\frac{53 \nu }{7}\nonumber\\
    &\quad+\bigg(\frac{243}{14}-\frac{74 \nu }{7}\bigg) (1-e^2)+\frac{27 \nu -\frac{81}{2}}{(1-e^2)}+\frac{510}{7} + p\sqrt{1-e^2}\bigg( 3 \nu +\bigg(3 \nu -\frac{15}{2}\bigg) (1-e^2)-\frac{15}{2} \bigg) \Bigg]\fJintc{1}{1}{pp0} \nonumber\\
    &\quad-18 e p^2(2-e^2) \fJintc{1}{2}{pp0} + 36e^2p^2\sqrt{1-e^2}\fJintc{2}{1}{pp0} + e p^2\Bigg[ (6 \nu -15) p (1-e^2)^2 + \sqrt{1-e^2}\bigg[42 \nu +\bigg(\frac{579}{7}-\frac{232 \nu }{7}\bigg) (1-e^2)\nonumber\\
    &\quad-33\bigg] \Bigg]\fJints{1}{1}{pp0} - 36e p^2\fJints{1}{2}{pp0} + e p^2\Bigg[ p \bigg(\bigg(\frac{\nu ^2}{8}-\frac{15 \nu }{8}\bigg) (1-e^2)^3+\bigg(\frac{\nu ^2}{8}-\frac{15 \nu }{8}\bigg) (1-e^2)^2\bigg)+\bigg(\frac{403}{7}-\frac{166 \nu }{7}\bigg) (1-e^2)^2 \nonumber\\
    &\quad+\bigg(\frac{216 \nu }{7}-\frac{555}{7}\bigg) (1-e^2) + \sqrt{1-e^2}\bigg[27 \nu +\bigg(\frac{243}{14}-\frac{74 \nu }{7}\bigg) (1-e^2)^2+\bigg(\frac{6}{7}-\frac{53 \nu }{7}\bigg) (1-e^2)-\frac{81}{2}\bigg] \Bigg]\fJints{1}{0}{pp1} \nonumber\\
    &\quad- 36 e p^2\Big[2(1-e^2) + \sqrt{1-e^2}(2-e^2)\Big]\fJints{1}{1}{pp1} - 18e^2 p^2(2-e^2)\fJints{2}{1}{pp0} \Bigg\}
\end{align}
We compared all 3PN $\ell\leq8$ waveform modes with the results of small eccentricity expansion in \cite{Mishra_2015} and \cite{Boetzel_2019}, and they were completely consistent.

\subsection{Comparison with post-circular expansion}
Building on the preceding results, one can compute GWs from binary systems with arbitrary eccentricity directly by following the procedure outlined in Section~\ref{sec_review}. 
\cite{Morras_2025} shows that as the eccentricity increases, the number of Fourier modes required for a high-fidelity waveform rises steeply; when the eccentricity approaches $0.9$, hundreds of modes are needed. 
Although we have introduced methods and tools for evaluating high-order PN contributions, waveform generation remains time-consuming at large eccentricities, so there is still substantial scope for optimization. 
In the remainder of this section we present several illustrative examples to demonstrate why the new method is indispensable in the high-eccentricity regime.

\begin{figure*}[t]
\begin{tabular}{c}
\includegraphics[width=1\textwidth]{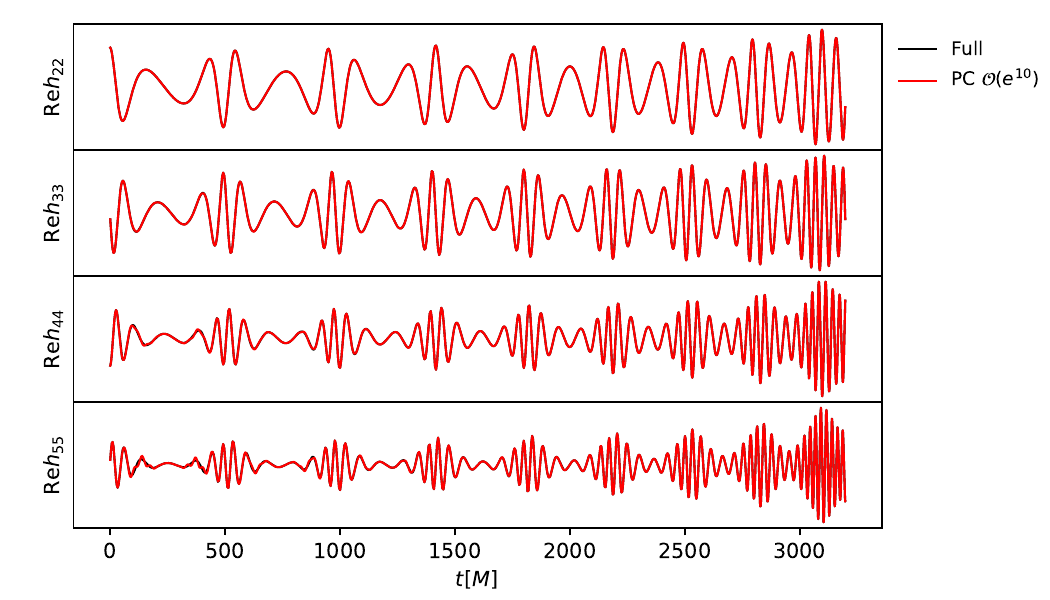}
\end{tabular}
\caption{Comparison of the leading-mode waveforms $h_{\ell\ell}$ (first four multipoles up to 3PN) computed with the Fourier-mode expansion and the post-circular expansion. In this example the binary has mass ratio $\nu=0.22$, initial frequency $v_0=0.25$, and the evolution ends at $v=1/\sqrt{6}$. The initial eccentricity is $e_0=0.3$. The black curve corresponds to the Fourier-mode approach presented in this paper, obtained by summing $H_{\ell m}=\sum_p H_{\ell(-m)p}\eexp{\ii p l}$ until the required accuracy is reached; the red curve shows the post-circular waveform truncated at $\order{e^{10}}$.}
\label{fig_CASE1}
\end{figure*}

First, we compute the 3PN orbital dynamical variables $(\bar{v},\bar{e},\bar{l},\bar{\lambda},\bar\chi)$ together with the post-adiabatic corrections $(\tilde{v},\tilde{e},\tilde{l},\tilde{\lambda})$ as described in Section~\ref{sec_review}. 
We then obtain the 3PN waveform by summing $H_{\ell m}=\sum_p H_{\ell(-m)p},\eexp{\ii p l}$ until the desired accuracy is achieved. 
For comparison, we also use the familiar post-circular expansion truncated at $\order{e^{10}}$.

\begin{figure*}[t]
\begin{tabular}{c}
\includegraphics[width=1\textwidth]{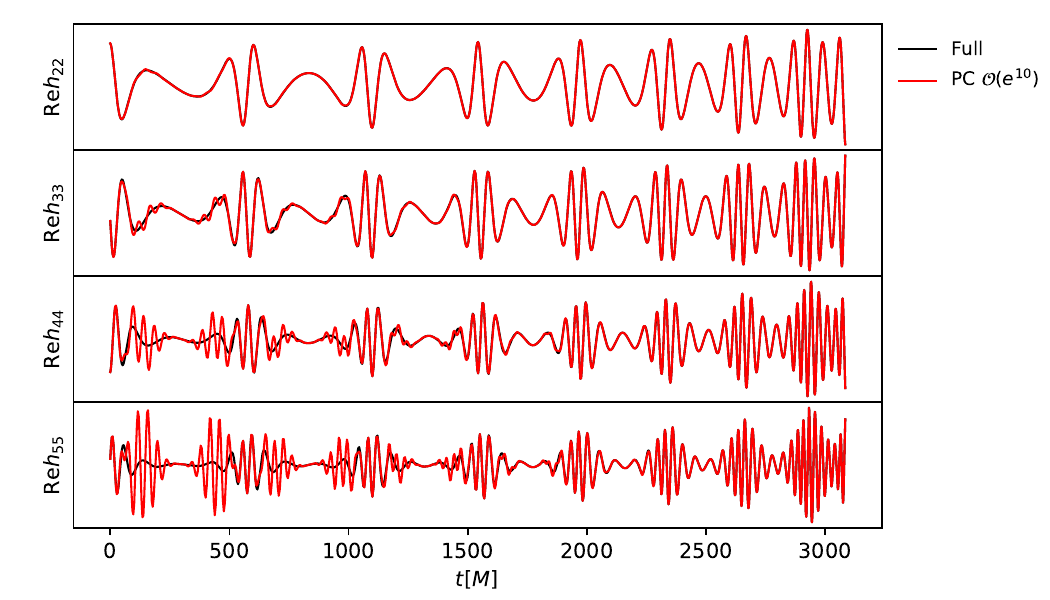}
\end{tabular}
\caption{Comparison of the leading-mode waveforms $h_{\ell\ell}$ (first four multipoles up to 3PN) computed with the Fourier-mode expansion and the post-circular expansion. In this example the binary has mass ratio $\nu=0.22$, initial frequency $v_0=0.25$, and the evolution ends at $v=1/\sqrt{6}$. The initial eccentricity is $e_0=0.4$. The black curve corresponds to the Fourier-mode approach presented in this paper, obtained by summing $H_{\ell m}=\sum_p H_{\ell(-m)p}\eexp{\ii p l}$ until the required accuracy is reached; the red curve shows the post-circular waveform truncated at $\order{e^{10}}$.}
\label{fig_CASE2}
\end{figure*}

\begin{figure*}[t]
\begin{tabular}{c}
\includegraphics[width=1\textwidth]{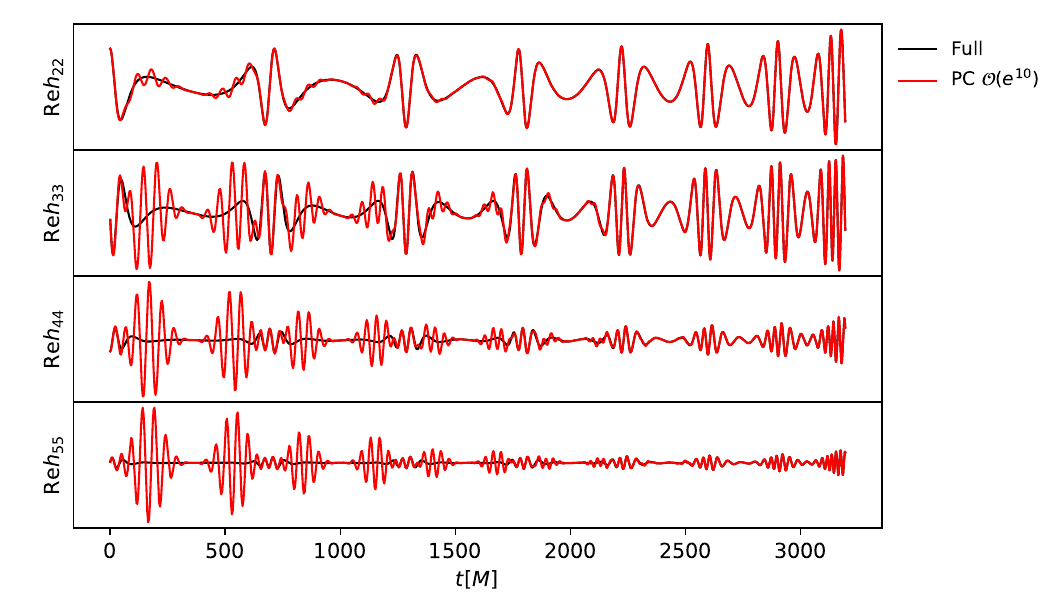}
\end{tabular}
\caption{Comparison of the leading-mode waveforms $h_{\ell\ell}$ (first four multipoles up to 3PN) computed with the Fourier-mode expansion and the post-circular expansion. In this example the binary has mass ratio $\nu=0.22$, initial frequency $v_0=0.25$, and the evolution ends at $v=1/\sqrt{6}$. The initial eccentricity is $e_0=0.5$. The black curve corresponds to the Fourier-mode approach presented in this paper, obtained by summing $H_{\ell m}=\sum_p H_{\ell(-m)p}\eexp{\ii p l}$ until the required accuracy is reached; the red curve shows the post-circular waveform truncated at $\order{e^{10}}$.}
\label{fig_CASE3}
\end{figure*}

Figures~(\ref{fig_CASE1})-(\ref{fig_CASE3}) plot the leading harmonic $h_{\ell\ell}$ of the first four multipoles for three representative cases with symmetric mass ratio $\nu=0.22$, initial frequency $v_0=0.25$, and initial eccentricities of $0.3$, $0.4$, and $0.5$.

Figure~(\ref{fig_CASE1}) indicates that for $e\sim0.3$ the post-circular expansion up to $\order{e^{10}}$ is reasonably reliable. However, its accuracy degrades rapidly as the eccentricity grows. At $e\sim0.4$, higher multipoles with $\ell>3$ already break down; by $e\sim0.5$, errors during the early, highly eccentric inspiral exceed the waveform amplitude itself. These findings make it clear that for moderately eccentric orbits the post-circular approach struggles to describe high-order multipole waveforms accurately. Developing faster and more precise frequency-domain techniques for waveform generation therefore remains a pressing challenge.

\section{Re-expression of tail effects in GW fluxes} \label{sec_reexpress_tail}
The GW energy flux is consist of instantaneous part, which is directly determined by the dynamics of the source, and hereditary part, which is caused by non-linear interactions. 
The reason for the nonlinear effect is the interaction between the radiative moments and the source-type moments or themselves. 
Formally these hereditary part are expressed as integrals. 
These genetic terms can be divided into a tail part and a memory part. The former originates from the interaction related to the radiative moments, while the latter originates from the change of the radiative moment. Sometimes the memory part is also classified as the instantaneous part~\cite{Arun_2008}, while the energy flux would becomes instantaneous after taking a time derivative and the angular momentum would be hereditary~\cite{Trestini_2025_ST}.

Generally, the energy and angular momentum fluxes of GW read~\cite{RevModPhys.52.299}
\begin{align}
    &\Eflux = \sum_{\ell \ge 2} \frac{1}{c^{2\ell+1}} \Big[a_\ell U^{(1)}_LU^{(1)}_L + \frac{1}{c^2}b_\ell V^{(1)}_LV^{(1)}_L\Big], \\
    &\Jflux^i = \epsilon_{iab}\sum_{\ell \ge 2} \frac{1}{c^{2\ell+1}} \Big[\tilde{a}_\ell U_{a(L-1)}U^{(1)}_{b(L-1)} + \frac{1}{c^2}\tilde{b}_\ell V_{a(L-1)}V^{(1)}_{b(L-1)}\Big],
\end{align}
where
\begin{align}
    &a_\ell = \frac{(\ell+1)(\ell+2)}{\ell(\ell-1)\ell!(2\ell+1)!!},\ b_\ell = \frac{4\ell(\ell+2)}{(\ell-1)(\ell+1)!(2\ell+1)!!}, \\
    &\tilde{a}_\ell = \frac{(\ell+1)(\ell+2)}{(\ell-1)\ell!(2\ell+1)!!},\ \tilde{b}_\ell=\frac{4\ell^2(\ell+2)}{(\ell-1)(\ell+1)!(2\ell+1)!!}
\end{align}
Up to 4PN, the radiative moments $U_L$ takes the form~\cite{Trestini_2023,Blanchet_2023},
\begin{align}
&U_L = M_L^{(\ell)} + \ivc{3}U_L^{\tail} + \ivc{5}\Big( U_L^{\mathrm{mem},\fPN{2.5}} + U_L^{M^2,\fPN{2.5}} \Big) + \ivc{6} U_L^{\tail(\tail)} + \ivc{7}\Big( U_L^{\mathrm{mem},\fPN{3.5}} + U_L^{M^2,\fPN{3.5}} \Big) \nonumber\\
&\quad + \ivc{8} U_L^{\tail(\mathrm{mem})} + \order{c^{-9}}
\end{align}
And similar for $V_L$.
The 4.5PN tail-of-tail-of-tail contribution to canonical moments were known~\cite{Marchand_2016}. 
All of these tail terms involve infinite integrals over logarithms. 
The usual procedure is to insert the Fourier expansion of the multipole moments and then average over the orbit. 
In this way, the contribution of the tails to the flux can be written in terms of several eccentricity-enhancement functions. 
In this section, we follow the method of~\cite{Arun_2008_tail} and, together with the new J-integrals introduced in this paper, express these eccentricity-enhancement functions as sums of these J-integrals.
We introduce
\begin{align}
&\mathcal{M}^{\Eflux}_\ell(0) = M_L^{(\ell+1)}\int_0^\infty \dd\tau M_L^{(\ell+3)}(t_r-\tau)\Big(\ln\tau + \kappa^{(\ell,0)}_0\Big), \\
&\mathcal{M}^{\Eflux}_\ell(\alpha) = 2M_L^{(\ell+1)}\int_0^\infty \dd\tau M_L^{(\ell+\alpha+3)}(t_r-\tau)\sum_{s=0}^{\alpha+1} \kappa^{(\ell,\alpha)}_s \ln^s\tau + \sum_{j=0}^{\alpha-1}\bigg( \int_0^\infty \dd\tau M_L^{(\ell+j+3)} \sum_{s_1=0}^{j+1}\kappa_{s_1}^{(\ell,j)}\ln^{s_1}\tau \bigg) \nonumber\\
&\quad \times \bigg( \int_0^\infty \dd\tau \dd\tau M_L^{(\ell+\alpha-j+2)} \sum_{s_2=0}^{\alpha-j}\kappa_{s_2}^{(\ell,\alpha-j-1)}\ln^{s_2}\tau \bigg) \\
&\mathcal{M}^{\Jflux}_\ell(0) = \epsilon_{zab} \bigg[ M_{a(L-1)}^{(\ell)}\int_0^\infty \dd\tau M_{b(L-1)}^{(\ell+3)}(t_r-\tau)\Big(\ln\tau + \kappa^{(\ell,0)}_0\Big) +  M_{b(L-1)}^{(\ell+1)}\int_0^\infty \dd\tau M_{a(L-1)}^{(\ell+2)}(t_r-\tau)\Big(\ln\tau + \kappa^{(\ell,0)}_0\Big) \bigg], \\
&\mathcal{M}^{\Jflux}_\ell(\alpha) = \epsilon_{zab}\Bigg[ M_{a(L-1)}^{(\ell)}\int_0^\infty \dd\tau M_{b(L-1)}^{(\ell +\alpha+3)} \sum_{s=0}^{\alpha+1} \kappa^{(\ell,\alpha)}_s \ln^s\tau + M_{b(L-1)}^{(\ell+1)}\int_0^\infty \dd\tau M_{a(L-1)}^{(\ell +\alpha+2)} \sum_{s=0}^{\alpha+1} \kappa^{(\ell,\alpha)}_s \ln^s\tau \nonumber\\
&\quad + \sum_{j=0}^{\alpha-1}\bigg( \int_0^\infty \dd\tau M_{a(L-1)}^{(\ell+j+2)} \sum_{s_1=0}^{j+1}\kappa_{s_1}^{(\ell,j)}\ln^{s_1}\tau \bigg)\bigg( \int_0^\infty \dd\tau M_{b(L-1)}^{(\ell+\alpha-j+2)} \sum_{s_2=0}^{\alpha-j}\kappa_{s_2}^{(\ell,\alpha-j-1)}\ln^{s_2}\tau \bigg) \Bigg]
\end{align}
where $\kappa_s^{(\ell,\alpha)}$ are constants. 
These integrals will appear in the tail-related calculations. 
The structure of the contribution part of moment $S_L$ is the same, just replace $M_L$ with $S_L$ and the associated constants $\kappa_s^{(\ell, n)}$.
In principle, when calculating flux involving high-order PNs, the derivatives of multipole moments should take into account the radiation reaction force.
We currently do not consider the effects of radiation reactions on these moment derivatives, keeping on adiabatic approximation and neglect possible dissipative pieces~\cite{Trestini_2025_Schott}. And nor more complicated moment interactions, such as tail-memory terms.
The purpose is to demonstrate the properties of the conserved part, through the new tools introduced in this work. Effects related to the radiation reaction force and the post-adiabatic approximation will be analyzed in future work.

Substituting Fourier expansion of $M_{L}$, and average over $\ell$, one would get the following summation
\begin{align}
    \sum_{p,p',m,m'}\ii^{a+b}(pn+mv^{3})^a(p'n+m'v^{3})^b\fmodeM{p}{m}{L}\fmodeM{p'}{m'}{L}\avg{\eexp{\ii[(p+p')l+(m+m')\lambda]}}\int_0^\infty \eexp{-\ii(p'+m'(k+1))n\tau}\Big(\sum_s \kappa^{(\ell,\alpha)}_s \ln^s\tau \Big)\dd{\tau}.\label{eq_tailint_general_form}
\end{align}
The average over $l$ reads
\begin{align}
    &\avg{\eexp{\ii\big(p+p'+m+m'+(m+m')k\big)l}}:=\frac{1}{2\pi}\int_0^{2\pi}\eexp{\ii\big(p+p'+m+m'+(m+m')k\big)l}\dd{l} \nonumber\\
    &=\delta_{(p+p'+m+m')0}+ 
    \left\{
    \begin{aligned}
    &\sum_{j=1}^\infty\frac{\big(\ii2\pi k (m+m') \big)^j}{(j+1)!}=\ii k(m+m')\pi+\order{k^2} & p+p'+m+m'=0 \\
    &\sum_{j=1}^{\infty}\sum_{j'=0}^{j-1}\frac{(-1)^{j'} \big(\ii2\pi k (m+m') \big)^j}{(j-j')! \big(\ii2\pi (p+p'+m+m') \big)^{j'+1}} = \frac{k(m+m')}{p+p'+m+m'}+\order{k^2} & p+p'+m+m'\neq 0
    \end{aligned}
    \right.\label{eq_avgell}
\end{align}
Note that averaging the $l$ here does not seem to completely limit the Fourier mode index $p+p'+m+m'$, but we find that in fact the higher PN order part still vanishes. 
This is because of the factor $(m+m')^j$ would cancel the summation,
\begin{align}
    &\sum_{m,m'} (m+m')^j m^a m'^b \fmodeM{p-m}{m}{L} \fmodeM{p'-m'}{m'}{L} = 0, \label{eq_balance_Mom_1}\\
    &\sum_{m,m'} (m+m')^j m^a m'^b \fmodeS{p-m}{m}{L} \fmodeS{p'-m'}{m'}{L} = 0, \label{eq_balance_Mom_2}\\
    \epsilon_{zab}&\sum_{m,m'} (m+m')^j m^a m'^b \fmodeM{p-m}{m}{a(L-1)} \fmodeM{p'-m'}{m'}{b(L-1)} = 0, \label{eq_balance_Mom_3}\\
	\epsilon_{zab}&\sum_{m,m'} (m+m')^j m^a m'^b \fmodeS{p-m}{m}{a(L-1)} \fmodeS{p'-m'}{m'}{b(L-1)} = 0. \label{eq_balance_Mom_4}
\end{align}
The 1PN property was shown in~\cite{Munna_2020}.
By inserting the Fourier modes of all the 3PN multipole moments we computed, it is easy to verify that they all hold for any positive $j$ and any non-negative $a,b$. 
This is not a coincidence; it is very likely true for higher PN orders as well. 
When $j$ is even, this can be proved using the symmetry of STF tensors; when $j$ is odd, it should involve properties of the Einstein field equations themselves.
Although we cannot prove it for all PN orders here, even restricting ourselves to the 3PN multipole moments $M_L,S_L$ greatly simplifies the final result.
We define the following auxiliary symbols for simplification,
\begin{align}
    \fmodeMp{p}{L} &: = \sum_{m=-\ell}^\ell \fmodeM{p-m}{m}{L}, \\
    \fmodemnMp{p}{L}{s} &: = \sum_{m=-\ell}^\ell m^s\fmodeM{p-m}{m}{L}.
\end{align}
Eq.~(\ref{eq_balance_Mom_1})-(\ref{eq_balance_Mom_4}) tells us
\begin{align}
    &\sum_{s=0}^j (-1)^{j-s+b}\binom{j}{s} \fmodemnMp{p}{L}{s+a} \fmodeconjmnMp{p'}{L}{j-s+b} = 0, \\
    &\epsilon_{zab}\sum_{s=0}^j (-1)^{j-s+b}\binom{j}{s} \fmodemnMp{p}{a(L-1)}{s+a} \fmodeconjmnMp{p'}{a(L-1)}{j-s+b} = 0.
\end{align}
Therefore one can expand Eq.~(\ref{eq_tailint_general_form}), and rewrite it by $\fmodeMp{p}{L}$ and $\fmodemnMp{p}{L}{s}$.
First, let's consider the simplest cases $\mathcal{M}^{\Eflux}_\ell(0),\mathcal{M}^{\Jflux}_\ell(0)$. Following the above steps and then sorting them out, we find they can be re-summed as very simple forms
\begin{align}
&\mathcal{M}^{\Eflux}_\ell(0) = \frac{\pi}{n}\sum_{p=1}^\infty \fmodexMp{p}{L} \fmodeconjMp{p}{L}, \\
&\mathcal{M}^{\Jflux}_\ell(0) = -\frac{\ii\pi}{n}\epsilon_{zab}\sum_{p=1}^\infty \fmodeyMp{p}{a(L-1)} \fmodeconjMp{p}{b(L-1)},
\end{align}
where
\begin{align}
&\fmodexMp{p}{L} := \sum_{m=-\ell}^\ell \frac{\big[(p-m)n+mv^{3}\big]^{2(\ell+2)}}{p+mk}\fmodeM{p-m}{m}{L}, \\
&\fmodeyMp{p}{L} := \sum_{m=-\ell}^\ell \frac{\big[(p-m)n+mv^{3}\big]^{2\ell+3}}{p+mk}\fmodeM{p-m}{m}{L}.
\end{align}
And we emphasize here $n=\dot{l}$.
Similarly, the higher order cases $\mathcal{M}^{\Eflux}_\ell(\alpha),\mathcal{M}^{\Jflux}_\ell(\alpha)$ are
\begin{align}
&\mathcal{M}^{\Eflux}_\ell(\alpha) = \frac{\pi}{n}\sum_{p=1}^\infty \fmodexaMp{p}{L}{\alpha} \fmodeconjMp{p}{L}, \\
&\mathcal{M}^{\Jflux}_\ell(\alpha) = -\frac{\ii\pi}{n}\epsilon_{zab}\sum_{p=1}^\infty \fmodeyaMp{p}{a(L-1)}{\alpha} \fmodeconjMp{p}{b(L-1)},
\end{align}
where
\begin{align}
    &\fmodexaMp{p}{L}{\alpha} := \sum_{m=-\ell}^\ell \frac{\big[(p-m)n+mv^{3}\big]^{2(\ell+2)+\alpha}}{p+mk}\Lambda^{(M)}_{\ell pm\alpha}\fmodeM{p-m}{m}{L}, \\
    &\fmodeyaMp{p}{L}{\alpha} := \sum_{m=-\ell}^\ell \frac{\big[(p-m)n+mv^{3}\big]^{2\ell+3+\alpha}}{p+mk}\Lambda^{(M)}_{\ell pm\alpha}\fmodeM{p-m}{m}{L}, \\
    &\Lambda^{(M)}_{\ell pm\alpha} = \frac{\ii^{\alpha+1}}{2\pi}\bigg[\sum_{s=0}^{\alpha+1}\kappa_s^{(l,\alpha)}\Big(\Lambda^{(+)}_{pms}-(-1)^\alpha\Lambda^{(-)}_{pms}\Big) - \frac{(p-m)n+mv^{3}}{2n(p+mk)}\sum_{j=0}^{\alpha-1}\sum_{s_1}^{j+1}\sum_{s_2=0}^{\alpha-j}(-1)^j \kappa_{s_1}^{(\ell,j)}\kappa_{s_2}^{(l,\alpha-j-1)} \nonumber\\
    &\qquad\qquad \times \Big( \Lambda^{(-)}_{pms_1}\Lambda^{(+)}_{pms_2} - (-1)^\alpha\Lambda^{(+)}_{pms_1}\Lambda^{(-)}_{pms_2} \Big)\bigg], \\
    &\Lambda^{(\pm)}_{pms}:=\sum_{s'=0}^s \binom{s}{s'} \cfgm{s-s'} \ln^{s'}\big(\pm\ii n(p+m k)\big).
\end{align}
We denote $\Lambda_{\ell pm0}=1$, then formally $\fmodexaMp{p}{L}{0}=\fmodexMp{p}{L}$.
The expression for the moment $S_L$ takes the same structure, one just need to replace the constant symbol $\kappa_s^{(\ell,j)}$ with $\pi_s^{(\ell,j)}$.
These results reproduce the work of~\cite{Arun_2008_tail,Arun_2009_tail}.
For example, when $\ell=2$,
\begin{align}
&\avg{\Eflux_{\tail,\mathrm{quad}}} = a_\ell \Big( 4\MADM \mathcal{M}^{\Eflux}_2(0) + \MADM^2 \mathcal{M}^{\Eflux}_2(1) \Big) = \frac{32v^{13}\nu^2}{5}\Bigg[ 4\pi \varphi(e) + \pi v^2\bigg( -\frac{428}{21}\alpha(e) + \frac{178}{21}\nu\theta(e) \bigg) \nonumber\\
&\qquad + v^3 \Bigg( \bigg(-\frac{515063}{11025}+\frac{16\pi^2}{3} - \frac{856}{35}\ln\bigg(\frac{v^2}{x_0'}\bigg)\bigg)F(e) - \frac{1712}{105}\chi(e)   \Bigg) + \order{c^{-4}} \Bigg] \\
&\avg{\Jflux^z_{\tail,\mathrm{quad}}} = \tilde{a}_\ell \Big( 4\MADM \mathcal{M}^{\Jflux}_2(0) + \MADM^2 \mathcal{M}^{\Jflux}_2(1) \Big) = \frac{32v^{10}\nu^2}{5}\Bigg[ 4\pi \tilde\varphi(e) + \pi v^2\bigg( -\frac{428}{21}\tilde\alpha(e) + \frac{178}{21}\nu \tilde\theta(e) \bigg) \nonumber\\
&\qquad + v^3 \Bigg( \bigg(-\frac{515063}{11025}+\frac{16\pi^2}{3} - \frac{856}{35}\ln\bigg(\frac{v^2}{x_0'}\bigg)\bigg)\tilde{F}(e) - \frac{1712}{105}\tilde\chi(e) \Bigg) + \order{c^{-4}} \Bigg].
\end{align}
The eccentricity enhancement function is defined by summing the product of the multipole moments,
\begin{align}
&\varphi(e) = \frac{v^8}{32\nu^2}\sum_{p=1}^\infty p^7 \mathrm{PN}_{0}\Big[\fmodeMp{p}{ij}\fmodeconjMp{p}{ij}\Big],\qquad 
\tilde\varphi(e) = -\ii\frac{v^8}{16\nu^2}\epsilon_{zab}\sum_{p=1}^\infty p^6 \mathrm{PN}_{0}\Big[\fmodeMp{p}{ak}\fmodeconjMp{p}{bk}\Big],\\
&F(e) = \frac{v^8}{64\nu^2}\sum_{p=1}^\infty p^8 \mathrm{PN}_{0}\Big[\fmodeMp{p}{ij}\fmodeconjMp{p}{ij}\Big],\qquad 
\tilde{F}(e) = -\ii\frac{v^8}{32\nu^2}\epsilon_{zab}\sum_{p=1}^\infty p^7 \mathrm{PN}_{0}\Big[\fmodeMp{p}{ak}\fmodeconjMp{p}{bk}\Big],\\
&\chi(e) = \frac{v^8}{64\nu^2}\sum_{p=1}^\infty p^8 \ln\bigg(\frac{p}{2}\bigg) \mathrm{PN}_{0}\Big[\fmodeMp{p}{ij}\fmodeconjMp{p}{ij}\Big],\qquad
\tilde\chi(e) = -\ii\frac{v^8}{32\nu^2}\epsilon_{zab}\sum_{p=1}^\infty p^7 \ln\bigg(\frac{p}{2}\bigg)\mathrm{PN}_{0}\Big[\fmodeMp{p}{ak}\fmodeconjMp{p}{bk}\Big].
\end{align}
$\mathrm{PN}_a$ denotes taking $\order{c^{-2a}}$, the $a$th PN part.
The explicit form of $\alpha(e),\tilde\alpha(e),\theta(e)$ and $\tilde\theta(e)$ are given by~\cite{Munna_2020}, here we reproduce them in terms of our convention
\begin{align}
&-\frac{428}{21}\alpha(e) + \frac{178}{21}\nu\theta(e) = 4\Big(\varphi_{(0,1)}(e) - \frac{1}{2}\nu \varphi(e) - \frac{21}{1-e^2}\big( \varphi(e) - \varphi_{(1,0)}(e) \big) \Big),\\
&-\frac{428}{21}\tilde\alpha(e) + \frac{178}{21}\nu\tilde\theta(e) = 4\Big(\tilde\varphi_{(0,1)}(e) - \frac{1}{2}\nu \tilde\varphi(e) - \frac{18}{1-e^2}\big( \tilde\varphi(e) - \tilde\varphi_{(1,0)}(e) \big) \Big), \label{eq_eccenhance}
\end{align}
where
\begin{align}
&\varphi_{(a,b)}(e) := \frac{v^{8-2b}}{32\nu^2} \sum_{p=1}^\infty p^{7-a}\mathrm{PN}_{b}\Big[\fmodemnMp{p}{ij}{a}\fmodeconjMp{p}{ij}\Big], \\
&\tilde\varphi_{(a,b)}(e) := -\ii\frac{v^{8-2b}}{16\nu^2} \epsilon_{zij}\sum_{p=1}^\infty p^{6-a}\mathrm{PN}_{b}\Big[\fmodemnMp{p}{ik}{a}\fmodeconjMp{p}{jk}\Big].
\end{align}
The reason for these symbols is that when $\fmodexMp{p}{L}$ is expanded, one get
\begin{align}
\fmodexMp{p}{L} = p^{2\ell+3}v^{6(\ell+2)} \bigg[ \fmodeMp{p}{L} + \frac{v^2}{1-e^2}\bigg(3(2\ell+3)\fmodemnMp{p}{L}{1} - 6(\ell+2)\fmodeMp{p}{L}\bigg) + \order{c^{-4}} \bigg].
\end{align}
One can check Eq.(\ref{eq_eccenhance}) is consistent with \cite{Arun_2008_tail,Arun_2009_tail} by substituting
\begin{align}
&\varphi_{(0,1)}(e) = -\frac{107}{21}\bigg(1 - \frac{119275e^2}{6848} - \frac{4666781e^4}{27392} - \frac{2938484437e^6}{3944448} \bigg) \nonumber\\
&\quad + \frac{55\nu}{21}\bigg( 1 + \frac{132247e^2}{10560} + \frac{531713e^4}{8448} + \frac{84766423e^6}{405504} \bigg) + \order{e^8} , \\
&\varphi_{(1,0)}(e) = 1 + \frac{209e^2}{32} + \frac{2415e^4}{128} + \frac{730751e^6}{18432} + \order{e^8}, \\
&\tilde\varphi_{(0,1)}(e) = -\frac{107}{21}\bigg(1 - \frac{33049e^2}{3424} - \frac{820263e^4}{13696} - \frac{366232601e^6}{1972224} \bigg) \nonumber\\
&\quad + \frac{55\nu}{21}\bigg( 1 + \frac{13133e^2}{1760} + \frac{36177e^4}{1408} + \frac{12848311e^6}{202752} \bigg) + \order{e^8},\\
&\varphi_{(1,0)}(e) = 1 + \frac{23e^2}{8} + \frac{19e^4}{4} + \frac{61105e^6}{9216} + \order{e^8},
\end{align}
into Eq.(\ref{eq_eccenhance}) then expand to reproduce the well-known results
\begin{align}
&\alpha(e) = 1 + \frac{2477e^2}{428} + \frac{76601e^4}{13696} - \frac{38169985e^6}{986112} + \order{e^8},  \\
&\theta(e) = 1 + \frac{215459e^2}{17088} + \frac{4415075e^4}{68352} + \frac{717366643e^6}{3280896} + \order{e^8}, \\
&\tilde\alpha(e) = 1 + \frac{11177e^2}{3424} + \frac{39687e^4}{13696} - \frac{12371303e^6}{1972224} + \order{e^8}. \\
&\tilde\theta(e) = 1 + \frac{21877e^2}{2848} + \frac{3495e^4}{128} + \frac{113137339e^6}{1640448} + \order{e^8}.
\end{align}
Similarly, for higher-order tail integrals, the corresponding expansions of $\fmodexaMp{p}{L}{\alpha}$ will also include similar terms. 
In our convention, these eccentricity enhancement functions can all be expressed as infinite sums of products of J-integrals. 
It is well known that some of these sums can be written in simple closed form,
\begin{align}
    &F(e) = \frac{1}{(1-e^2)^{13/2}}\bigg(1+\frac{85 e^2}{6}+\frac{5171 e^4}{192}+\frac{1751 e^6}{192}+\frac{297 e^8}{1024}\bigg), \\
    &\tilde{F}(e) = \frac{1}{(1-e^2)^5}\bigg(1+\frac{229 e^2}{32}+\frac{327 e^4}{64}+\frac{69 e^6}{256}\bigg).\label{eq_hgeneration_quadJ}
\end{align}
Consider the general form of summation that appears in eccentricity enhancement function expressions: $\sum_p p^n\fJint{b_1}{(p+m_1)pa_1}\fJint{b_2}{(p+m_2)pa_2}$, a closed-form expression exists only when the summation is carried out over $p$ from negative to positive infinity.
Using the generating function method, this sum can be calculated as follows,
\begin{align}
& \sum_{p=-\infty}^{\infty} p^n \fJint{b_1}{(p+m_1)pa_1} \fJint{b_2}{(p+m_2)pa_2} = \frac{1}{2\pi}\int_{-\pi}^{\pi}\ii^{b_1+b_2-n}\frac{\delta\chi(x)^{b_1}\eexp{\ii m_1 x}}{(1-e\cos{x})^{a_1}} \eval{\pdv[n]{z} \Big[\frac{\delta\chi(\sigma)^{b_2}\eexp{\ii m_2 \sigma}}{(1-e\cos{\sigma})^{a_2+1}}\Big]}_{z=0} \dd{x}
\end{align}
where $\sigma(z,x)$ is an implicit function determined by,
\begin{align}
    x+\sigma(z,x)+z=e(\sin{x}+\sin{\sigma(z,x)}).
\end{align}
In addition, due to the symmetry of the dynamic structure itself, after tedious arrangement, it can be found that some eccentricity enhancement functions can be written in closed form. For example, one can obtain the following results which will appear in the next PN order of $\mathcal{M}_2^{\Eflux}(1)$,
\begin{align}
&F_{(1,0)}(e) := \frac{v^8}{64\nu^2}\sum_{p=1}^\infty p^{7} \mathrm{PN}_{0}\Big[\fmodemnMp{p}{L}{1}\fmodeconjMp{p}{L}\Big] = \frac{1}{(1-e^2)^5}\bigg(1+\frac{229 e^2}{32}+\frac{327 e^4}{64}+\frac{69 e^6}{256}\bigg) = \tilde{F}(e), \\
&\tilde{F}_{(1,0)}(e) := -\ii\frac{v^8}{32\nu^2}\epsilon_{zab}\sum_{p=1}^\infty p^{6} \mathrm{PN}_{0}\Big[\fmodemnMp{p}{a(L-1)}{1}\fmodeconjMp{p}{b(L-1)}\Big] = \frac{1}{(1-e^2)^{7/2}}\bigg(1 + \frac{97}{32}e^2 + \frac{49}{128}e^4\bigg), \\
&F_{(0,1)}(e) := \frac{v^8}{64\nu^2}\sum_{p=1}^\infty p^{8} \mathrm{PN}_{1}\Big[\fmodeMp{p}{L}\fmodeconjMp{p}{L}\Big] = -\frac{107}{21(1-e^2)^8}\Bigg[\frac{507}{107}+\frac{10395e^2}{428} - \frac{33075e^4}{856} - \frac{45045e^6}{3424} + \frac{36855e^8}{1712} + \frac{4347e^{10}}{3424} \nonumber\\
&\quad - \sqrt{1-e^2}\bigg( \frac{397}{107} + \frac{216615 e^2}{3424} + +\frac{548117
   e^4}{3424} +\frac{616191 e^6}{6848} +\frac{69631 e^8}{6848} + \frac{43119 e^{10}}{438272}\bigg)\Bigg] \nonumber\\
&\quad + \frac{55\nu}{21(1-e^2)^{15/2}}\bigg(1+\frac{3031 e^2}{240}+\frac{60763 e^4}{3520}-\frac{267 e^6}{176} - \frac{67567e^8}{33792} - \frac{5427e^{10}}{112640}\bigg), \\
&\tilde{F}_{(0,1)}(e) := -\ii\frac{v^8}{32\nu^2}\epsilon_{zab}\sum_{p=1}^\infty p^{7} \mathrm{PN}_{1}\Big[\fmodeMp{p}{a(L-1)}\fmodeconjMp{p}{b(L-1)}\Big] = -\frac{107}{21(1-e^2)^8} \Bigg[ \sqrt{1-e^2} \bigg( \frac{441}{107}+\frac{28665 e^2}{3424}-\frac{149499 e^4}{13696} \nonumber\\
&\quad -\frac{21609 e^6}{13696} \bigg) -\bigg(\frac{334}{107}+\frac{101779
   e^2}{3424}+\frac{483379 e^4}{13696}+\frac{181765 e^6}{27392}+\frac{20241 e^8}{219136}\bigg) \Bigg] + \frac{55\nu}{21(1-e^2)^{6}} \bigg(1+\frac{11483 e^2}{1760}+\frac{2461 e^4}{880} \nonumber\\
&\quad-\frac{963 e^6}{1280}-\frac{1833 e^8}{56320}\bigg)
\end{align}
The above method is also applicable to the multipole moment $S_L$, which will not be elaborated here.
At the end of this section, we emphasize that although we have, using the new tools developed in this paper, derived a general re-summed expression for the tail part flux without including other complicated interaction terms, this expression is valid only when the conditions in Eq.~(\ref{eq_balance_Mom_1})-(\ref{eq_balance_Mom_4}) are satisfied. 
Substituting the 3PN multipole-moment results, we find that the condition indeed holds; therefore, at least for the tail contribution to the flux through 4.5PN order under the adiabatic approximation, our expression is reliable. 
Whether higher-PN multipole moments continue to satisfy this condition remains to be established by a rigorous proof.

\section{Summary}
With the rapid progress of GW observations and the planning of next‑generation detectors, the demand for more accurate waveform templates has intensified. 
In this paper, we systematically study several integrals that would appear in the Fourier expansion coefficients of the PN dynamics and gravitational radiation of the binaries.
We summarize the mathematical properties of these integrals and give a ready‑to‑use software package implementing the method.
This method is very promising for future calculations of high-precession and high-eccentricity GW waveforms.
We then apply the framework to the computation of the tail contribution to binary waveforms and derive a highly general re-summed expression.
This result streamlines the evaluation of tail terms in the fluxes at higher PN orders under the adiabatic approximation.

At present, our treatment is restricted to non-spinning binaries. 
The inclusion of spins renders the Fourier structure substantially more intricate; extending the method to generic spinning, possibly precessing, orbits is an important topic for future work. 
Moreover, the current method is highly compute-intensive. 
When the eccentricity becomes large, the convergence of the J-integrals slows markedly, and the overall cost of waveform generation grows almost exponentially. 
Developing more efficient algorithms to mitigate this scaling will thus be a key challenge for future work.
In addition, one proof step in our tail‑flux derivation remains open. 
We have not accounted for higher‑order hereditary effects arising from couplings among multipole moments or from tail–memory interactions, nor have we incorporated radiation‑reaction forces. 
Addressing these limitations will require further effort.

\section*{Acknowledgments}

This work was supported in part by the National Key Research and Development Program of China Grant No. 2021YFC2203001 and in part by the NSFC (No. 12475046, No. 12021003 and No. 12005016). 
X. Liu is supported by the Consolidaci\'on Investigadora 2022 grant CNS2022-135211, the I+D grant PID2023-149018NB-C42, and the Grant IFT Centro de Excelencia Severo Ochoa No CEX2020-001007-S, funded by MCIN/AEI/10.13039/501100011033. 
Z.Cao was supported by “the Fundamental Research Funds for the Central Universities”.

\begin{appendix}
\section{4PN contact transformations} \label{app_4PNct}
The conservative Lagrangian in SH coordinates depends not only on the positions and velocities of the component bodies, but also on their accelerations: $L({y}^i,{v}^i,{a}^i)$, where ${y}^i$ is the position and ${v}^i:=\dot{{y}}^i,{a}^i:=\dot{{v}}^i$. And the accelerations begin to appear from the 2PN order, and only the linear components of accelerations can be retained~\cite{Damour:1981ntn,Damour:1990jh}. Now the Lagrangian has already be derived in COM up to 4PN order~\cite{Bernard_2016,Bernard_2018}. And the Lagrangian in ADM coordinates can be found in~\cite{Damour_2014_nonlocal,Jaranowski_2013,Jaranowski_2015}. The accelerations and logarithmic terms can be canceled by applying a small contact transformation between SH and ADM~\cite{Andrade_2001},
\begin{align}
{y}^i_{\SH} \to {y}^i_{\ADM} + \delta{y}^i_{\SH\to\ADM}(y^i_{\ADM}).
\end{align}
The difference $\delta{y}$ starts at 2PN-order, we need to expand the variation of the Lagrangian up to $\order{\delta{y}^3}$,
\begin{align}
\delta{L}:= L({y}^i+\delta{y}^i)-L({y}^i) = \delta{L}^{(1)}+\frac{1}{2}\delta{L}^{(2)} + \order{\delta{y}^3},
\end{align}
where the $\order{\delta{y}}$ term reads
\begin{align}
&\delta{L}^{(1)} = \frac{\delta L}{\delta y^i}\delta{y}^i+\dv{Q^{(1)}}{t}, \\
&\frac{\delta L}{\delta y^i} = L_{y^i}-\dot{L}_{v^i}+\ddot{L}_{a^i} \\
&Q^{(1)}=(L_{v^i}-\dot{L}_{a^i})\delta{y}^i+L_{a^i}\delta{v}^i
\end{align}
We use the shorthands $L_Y:=\partial{L}/\partial{Y}$, and $\delta{v}^i:=\delta{\dot{y}}^i, \delta{a}^i:=\delta{\dot{v}}^i$.
The $\order{\delta{y}^2}$ term is
\begin{align}
&\delta{L}^{(2)} = \delta{y}^i\mathcal{J}^{ij}\delta{y}^j + \dv{Q^{(2)}}{t}, \\
&\mathcal{J}^{ij} = L_{y^iy^j}-\dv{t}L_{v^iv^j}\dv{t} + \dv[2]{t}L_{a^ia^j}\dv[2]{t} -2 \dv{t}L_{v^iy^j} + 2\dv[2]{t}L_{a^iy^j} + 2\dv[2]{t}L_{a^iv^j}\dv{t} \nonumber\\
&Q^{(2)} = L_{v^iv^j}\delta{y}^i\delta{v}^j + L_{a^ia^j}\delta{v}^i\delta{a}^j - \dv{t}\big(L_{a^ia^j}\delta{a}^j\big)\delta{y}^i + 2L_{v^iy^j}\delta{y}^i\delta{y}^j \nonumber\\
& + 2L_{a^iy^j}\delta{v}^i\delta{y}^j - 2\dv{t}\big(L_{a^iy^j}\delta{y}^j\big) \delta{y}^i + 2L_{a^iv^j}\delta{v}^i\delta{v}^j - 2\dv{t}\big(L_{a^iv^j}\delta{v}^j\big)\delta{y}^i.
\end{align}
The transformation of Lagrangian is composed of two part,
\begin{align}
L' = L + \dv{Q}{t} + \frac{\delta L}{\delta y^i}\delta y^i + \frac{1}{2}\delta{y}^i\mathcal{J}^{ij}\delta{y}^j + \order{\delta{y}^3},
\end{align}
We separate $L$ into three parts,
\begin{align}
    L = L' + \delta{L}_{\mathrm{na}} + L_{a^i}a^i,
\end{align}
where $\delta{L}_{\mathrm{na}}$ denotes the different terms that don't involve acceleration.
We can add any total-derivative term $\dd{F}/\dd{t}$ that don't affect the dynamics. One just need to make sure
\begin{align}
0=\delta{L}_{\mathrm{na}}+L_{a^i}a^i+F_{y^i}v^i+F_{v^i}a^i+F_{a^i}b^i+\frac{\delta L}{\delta y^i}\delta{y}^i + \frac{1}{2}\delta{y}^i\mathcal{J}^{ij}\delta{y}^j,
\end{align}
where $b^i:=\dot{a}^i$. At the 4PN level we must introduce acceleration into $F$, finally we get
\begin{align}
    &\delta{\pmb{y}}_{\SH\to\ADM} = \frac{1}{c^4} \bigg[\bigg(-\frac{\nu  \dot{r}^2}{8}+\bigg(3 \nu +\frac{1}{4}\bigg)\frac{1}{r}+\frac{5 \nu  v^2}{8}\bigg)\pnvec -\frac{9 \nu  \dot{r}}{4}\pvvec\bigg] + 
    \frac{1}{c^6}\bigg\{ \bigg[ \frac{1}{r^2}\bigg(\frac{22}{3}\ln\bigg(\frac{r}{r_0'}\bigg)-\frac{21 \pi ^2}{32}-\frac{2773}{280}\bigg) \nu \nonumber \\
    & + \bigg(\frac{\nu }{16}-\frac{5 \nu ^2}{16}\bigg) \dot{r}^4+\bigg(\frac{15 \nu ^2}{16}-\frac{5 \nu }{16}\bigg) \dot{r}^2 v^2 + \frac{1}{r}\bigg(\bigg(\frac{5 \nu ^2}{2}-\frac{161 \nu }{48}\bigg) \dot{r}^2+\bigg(\frac{3 \nu ^2}{8}+\frac{451 \nu }{48}\bigg) v^2\bigg)+\bigg(\frac{\nu }{2}-\frac{11 \nu ^2}{8}\bigg) v^4 \bigg]\pnvec \nonumber \\
    & + \bigg[ \bigg(\frac{5 \nu }{12}-\frac{29 \nu ^2}{24}\bigg) \dot{r}^3+\frac{1}{r}\bigg(-5 \nu ^2-\frac{43 \nu }{3}\bigg) \dot{r}+\bigg(\frac{21 \nu ^2}{4}-\frac{17 \nu }{8}\bigg) \dot{r} v^2 \bigg]\pvvec  \bigg\} + \frac{1}{c^8}\Bigg\{ \Bigg[ \frac{49}{4} \nu ^2 \dot{r} a_v+\bigg(-\frac{105 \nu ^3}{256}+\frac{35 \nu ^2}{128}-\frac{5 \nu }{128}\bigg) \dot{r}^6 \nonumber \\
    & + \bigg(-\frac{1143 \nu ^3}{256}+\frac{339 \nu ^2}{128}-\frac{53 \nu }{128}\bigg) \dot{r}^2 v^4+\bigg(\frac{555 \nu ^3}{256}-\frac{181 \nu ^2}{128}+\frac{27 \nu }{128}\bigg) \dot{r}^4 v^2+\bigg(\frac{1253 \nu ^3}{256}-\frac{361 \nu ^2}{128}+\frac{55 \nu }{128}\bigg) v^6 \nonumber\\
    & + \frac{1}{r} \bigg( \frac{9}{32} \nu ^2 r \dot{r}^2 a_r+v^2 \bigg[ \bigg(-\frac{535 \nu ^3}{64}+\frac{10247 \nu ^2}{384}-\frac{341 \nu }{40}\bigg) \dot{r}^2-\frac{171}{32} \nu ^2 r a_r\bigg]+\bigg(\frac{1337 \nu ^3}{384}-\frac{4765 \nu ^2}{256}+\frac{313 \nu }{480}\bigg) \dot{r}^4 \nonumber \\
    & + \bigg(-\frac{421 \nu ^3}{128}-\frac{18263 \nu ^2}{768}+\frac{2031 \nu }{160}\bigg) v^4 \bigg) + 
    \frac{1}{r^2}\bigg( \dot{r}^2 \bigg[\nu ^2 \bigg(-11 \ln\bigg(\frac{r}{r_0'}\bigg)+44 \ln\bigg(\frac{r}{r_0''}\bigg)-\frac{7449 \pi ^2}{16384}-\frac{1312259}{134400}\bigg) \nonumber \\
    & + \bigg(-11 \ln\bigg(\frac{r}{r_0''}\bigg)-\frac{10119 \pi ^2}{8192}-\frac{3629869}{44800}\bigg) \nu +\frac{71 \nu ^3}{8}\bigg]+v^2 \bigg[\bigg(22 \ln\bigg(\frac{r}{r_0'}\bigg)-\frac{6629 \pi ^2}{8192}+\frac{9427}{256}\bigg) \nu +\frac{\nu ^3}{8} \nonumber \\
    & + \bigg(-\frac{439447}{57600}-\frac{18107 \pi ^2}{16384}\bigg) \nu ^2\bigg] \bigg) + \frac{1}{r^3}\bigg( \nu ^2 \bigg[-120 \ln\bigg(\frac{r}{r_0'}\bigg)+240 \ln\bigg(\frac{r}{r_0''}\bigg)-\frac{26389 \pi ^2}{3072}+\frac{3439}{144}\bigg] \nonumber\\
    & + \nu  \bigg(\frac{104}{3} \ln\bigg(\frac{r}{r_0'}\bigg)-124 \ln\bigg(\frac{r}{r_0''}\bigg)+\frac{8925 \pi ^2}{1024}-\frac{27169}{1440}\bigg)-16 \ln\bigg(\frac{r}{r_0'}\bigg)+16 \ln\bigg(\frac{r}{r_0''}\bigg) \bigg) \Bigg]\pnvec + \Bigg[ -\frac{115 \nu ^2 a_v}{32} \nonumber \\
    & + \bigg(-\frac{121 \nu ^3}{128}+\frac{63 \nu ^2}{64}-\frac{13 \nu }{64}\bigg) \dot{r}^5+\bigg(-\frac{2449 \nu ^3}{128}+\frac{789 \nu ^2}{64}-\frac{135 \nu }{64}\bigg) \dot{r} v^4+\bigg(\frac{1193 \nu ^3}{192}-\frac{431 \nu ^2}{96}+\frac{77 \nu }{96}\bigg) \dot{r}^3 v^2 \nonumber \\
    & + \frac{1}{r}\bigg( \bigg(-\frac{85 \nu ^3}{32}-\frac{615 \nu ^2}{64}+\frac{6737 \nu }{480}\bigg) \dot{r}^3+\bigg(\frac{803 \nu ^3}{32}+\frac{4199 \nu ^2}{192}-\frac{4851 \nu }{160}\bigg) \dot{r} v^2 \bigg) 
    + \frac{\dot{r}}{r^2} \bigg(\nu ^2 \bigg(\frac{44}{3} \ln\bigg(\frac{r}{r_0'}\bigg)-\frac{88}{3} \ln\bigg(\frac{r}{r_0''}\bigg) \nonumber\\
    & - \frac{28603 \pi ^2}{8192}+\frac{4066291}{201600}\bigg)+\nu  \bigg(-44 \ln\bigg(\frac{r}{r_0'}\bigg)+\frac{22}{3} \ln\bigg(\frac{r}{r_0''}\bigg)-\frac{1253 \pi ^2}{4096}+\frac{22800709}{201600}\bigg)-\frac{21 \nu ^3}{2}\bigg)\Bigg]\pvvec \Bigg\} + \order{c^{-10}}.
\end{align}
It is equivalent to~\cite{blanchet_2025_ISCO}. 
The Lagrangian in SH coordinate used in this work is given by~\cite{Bernard_2018}.
The $r'_0$ and $r''_0$ are gauge constants of Eq.(2.4) of~\cite{Bernard_2018}, which are set equal in~\cite{blanchet_2025_ISCO}.
One can check that after applying the transformation, the new Lagrangian is consistent with the Lagrangian given by Eq.(A3) of~\cite{blanchet_2025_ISCO}.
They only differ by a total derivative.

The transformation between SH and MH is
\begin{align}
    &\delta{\pmb{y}}_{\SH\to\MH} = \frac{22\nu}{3r^2c^6}\ln\bigg(\frac{r}{r_0'}\bigg)\pnvec + \frac{1}{r^2c^8}\Bigg\{ \Bigg[\dot{r}^2 \bigg(\nu ^2 (44 \ln\bigg(\frac{r}{r_0''}\bigg)-11 \ln\bigg(\frac{r}{r_0'}\bigg))-11 \ln\bigg(\frac{r}{r_0''}\bigg) \nu \bigg)+22 \ln\bigg(\frac{r}{r_0'}\bigg) \nu  v^2 \nonumber\\
    & + \frac{1}{r}\bigg( \nu ^2 \bigg(240 \ln\bigg(\frac{r}{r_0''}\bigg)-120 \ln\bigg(\frac{r}{r_0'}\bigg)\bigg)+\nu  \bigg(\frac{104}{3} \ln\bigg(\frac{r}{r_0'}\bigg)-124 \ln\bigg(\frac{r}{r_0''}\bigg)\bigg)-16 \ln\bigg(\frac{r}{r_0'}\bigg)+16 \ln\bigg(\frac{r}{r_0''}\bigg) \bigg)\Bigg]\pnvec \nonumber\\
    & + \frac{\dot{r}}{r^2} \bigg(\nu ^2 \bigg(\frac{44}{3} \ln\bigg(\frac{r}{r_0'}\bigg)-\frac{88}{3} \ln\bigg(\frac{r}{r_0''}\bigg)\bigg)+\nu  \bigg(\frac{22}{3} \ln\bigg(\frac{r}{r_0''}\bigg)-44 \ln\bigg(\frac{r}{r_0'}\bigg)\bigg)\bigg)\pvvec\Bigg\} + \order{c^{-10}}.
\end{align}


\section{Quask-Keplerian parameterization in MH coordinates} \label{app_QKMH}
Using the contact transformation in Appendix~\ref{app_4PNct} and the instantaneous Lagrangian in 4PN SH coordinates~\cite{Bernard_2018}, we can obtain the instantaneous Lagrangian in 4PN MH coordinates. 
The conserved energy $E$ and angular momentum $J$ of the system can be obtained by
\begin{align}
& E = \pmb{v} \cdot \bigg( \pdv{L}{\pmb{v}} - \dv{t}\pdv{L}{\pmb{a}} \bigg) + \pmb{a} \cdot \pdv{L}{\pmb{a}} - L ,\\
& J = \pdv{L}{\Omega}.
\end{align}
The 3PN results can be found in Eq.(23) of~\cite{Memmesheimer_2004_QK}, the 4PN terms are
\begin{align}
& E_{\fPN{4}} = \frac{1}{r^5}\bigg(11 \pi ^2 \nu ^2-\frac{55111 \nu ^2}{720}-\frac{105 \pi ^2 \nu }{32}-\frac{1697177 \nu }{25200}-\frac{3}{8}\bigg) + \frac{\rdot^2}{r^4}\bigg( \frac{15 \nu ^4}{2}+\frac{213 \nu ^3}{8}-\frac{1367 \pi ^2 \nu ^2}{32}+\frac{50009 \nu ^2}{2520} \nonumber\\
&\quad +\frac{2645 \pi ^2 \nu }{96}-\frac{2470837 \nu }{12600}+\frac{9}{4} \bigg) + \frac{v^2}{r^4}\bigg( \frac{\nu ^4}{2}-\frac{29 \nu ^3}{8}+\frac{311 \pi ^2 \nu ^2}{32}+\frac{22963 \nu ^2}{5040}-\frac{149 \pi ^2 \nu }{32}+\frac{1859363 \nu }{16800}+\frac{15}{16} \bigg) \nonumber\\
&\quad + \frac{\rdot^4}{r^3}\bigg( \frac{87 \nu ^4}{4}+\frac{69 \nu ^3}{32}+\frac{615 \pi ^2 \nu ^2}{128}-\frac{95813 \nu ^2}{224}-\frac{6465 \pi ^2 \nu }{1024}-\frac{2673127 \nu }{6720} \bigg) + \frac{\rdot^2v^2}{r^3}\bigg( -\frac{141 \nu ^4}{2}-\frac{2437 \nu ^3}{16} \nonumber\\
&\quad +\frac{123 \pi ^2 \nu ^2}{64}+\frac{537959 \nu ^2}{1680}+\frac{333 \pi ^2 \nu }{512}+\frac{1080763 \nu }{1680}+\frac{21}{4} \bigg) + \frac{v^4}{r^3}\bigg( -\frac{45 \nu ^4}{4}+\frac{2373 \nu ^3}{32}-\frac{205 \pi ^2 \nu ^2}{128}+\frac{521063 \nu ^2}{10080} \nonumber\\
&\quad +\frac{1071 \pi ^2 \nu }{1024}-\frac{22649399 \nu }{100800}+\frac{273}{16} \bigg) + \frac{\rdot^6}{r^2}\bigg( \frac{15 \nu ^4}{2}-\frac{17 \nu ^3}{2}-\frac{461 \nu ^2}{8}-\frac{4771 \nu }{640} \bigg) + \frac{\rdot^4v^2}{r^2}\bigg( -\frac{135 \nu ^4}{2}-\frac{439 \nu ^3}{8} \nonumber\\
&\quad+\frac{19465 \nu ^2}{96}+\frac{5347 \nu }{384} \bigg) + \frac{\rdot^2v^4}{r^2}\bigg( \frac{2845 \nu ^4}{16}+\frac{18511 \nu ^3}{64}-\frac{12995 \nu ^2}{64}-\frac{5893 \nu }{128}+\frac{15}{16} \bigg) + \frac{v^6}{r^2}\bigg( \frac{975 \nu ^4}{16}-\frac{8289 \nu ^3}{64} \nonumber\\
&\quad +\frac{5129 \nu ^2}{64}-\frac{4489 \nu }{128}+\frac{575}{32} \bigg) + \frac{\rdot^8}{r} \bigg( \frac{245 \nu ^4}{128}-\frac{245 \nu ^3}{64}+\frac{245 \nu ^2}{128}-\frac{35 \nu }{128} \bigg) + \frac{\rdot^6v^2}{r}\bigg( -\frac{595 \nu ^4}{32}+\frac{185 \nu ^3}{8}-\frac{125 \nu ^2}{16}+\frac{25 \nu }{32} \bigg) \nonumber\\
&\quad + \frac{\rdot^4v^4}{r}\bigg( \frac{4851 \nu ^4}{64}-\frac{1683 \nu ^3}{32}+\frac{243 \nu ^2}{32}+\frac{27 \nu }{64} \bigg) + \frac{\rdot^2v^6}{r}\bigg( -\frac{4655 \nu ^4}{32}+\frac{423 \nu ^3}{8}+\frac{369 \nu ^2}{32}-\frac{147 \nu }{32} \bigg) \nonumber\\
&\quad + \frac{v^8}{r}\bigg( -\frac{15827 \nu ^4}{128}-\frac{357 \nu ^3}{64}+\frac{9507 \nu ^2}{128}-\frac{4011 \nu }{128}+\frac{525}{128} \bigg) + v^{10}\bigg( \frac{21735 \nu ^4}{256}-\frac{10143 \nu ^3}{128}+\frac{7065 \nu ^2}{256}-\frac{1089 \nu }{256}+\frac{63}{256} \bigg), \\
&\frac{1}{|\pmb{v}\times\pmb{v}|}J_{\fPN{4}} =\frac{1}{r^4}\bigg(\nu ^4-\frac{15 \nu ^3}{4}+\frac{663 \pi ^2 \nu ^2}{32}-\frac{20131 \nu ^2}{420}-\frac{85 \pi ^2 \nu }{8}+\frac{3809041 \nu }{25200}+\frac{15}{8}\bigg) + \frac{\rdot^2}{r^3}\bigg(-47 \nu ^4-\frac{1025 \nu ^3}{8} \nonumber\\
&\quad+\frac{302747 \nu ^2}{1680}+\frac{447 \pi ^2 \nu }{256}+\frac{7283177 \nu }{16800}+\frac{7}{2}\bigg) + \frac{v^2}{r^3}\bigg(-15 \nu ^4+\frac{637 \nu ^3}{8}-\frac{41 \pi ^2 \nu ^2}{16}+\frac{276433 \nu ^2}{5040}+\frac{469 \pi ^2 \nu }{256}-\frac{13576009 \nu }{50400}+\frac{91}{4}\bigg) \nonumber\\
&\quad + \frac{\rdot^4}{r^2}\bigg(-27 \nu ^4-\frac{155 \nu ^3}{4}+\frac{3235 \nu ^2}{48}+\frac{14773 \nu }{320}\bigg) + \frac{\rdot^2 v^2}{r^2}\bigg(\frac{569 \nu ^4}{4}+\frac{4459 \nu ^3}{16}-\frac{256 \nu ^2}{3}-\frac{5551 \nu }{60}+\frac{3}{4}\bigg) + \frac{v^4}{r^2}\bigg(\frac{585 \nu ^4}{8} \nonumber\\
&\quad -\frac{3845 \nu ^3}{32}+\frac{12427 \nu ^2}{96}-\frac{65491 \nu }{960}+\frac{345}{16}\bigg) + \frac{\rdot^6}{r}\bigg(-\frac{85 \nu ^4}{16}+\frac{45 \nu ^3}{16}+\frac{15 \nu ^2}{8}-\frac{5 \nu }{8}\bigg) + \frac{\rdot^4 v^2}{r}\bigg(\frac{693 \nu ^4}{16}-\frac{135 \nu ^3}{16}-\frac{45 \nu ^2}{4}+3 \nu\bigg) \nonumber\\
&\quad + \frac{\rdot^2 v^4}{r}\bigg(-\frac{1995 \nu ^4}{16}+\frac{299 \nu ^3}{16}+\frac{423 \nu ^2}{16}-\frac{53 \nu }{8}\bigg) + \frac{v^6}{r}\bigg(-\frac{2261 \nu ^4}{16}-\frac{425 \nu ^3}{16}+\frac{1553 \nu ^2}{16}-\frac{151 \nu }{4}+\frac{75}{16}\bigg) \nonumber\\
&\quad + v^8\bigg(\frac{12075 \nu ^4}{128}-\frac{5635 \nu ^3}{64}+\frac{3925 \nu ^2}{128}-\frac{605 \nu }{128}+\frac{35}{128}\bigg).
\end{align}
The 4PN part of eccentricities $e_r,e_\phi,e_t$, averaged orbital frequency $n$, velocity $v$ and periastron advance $k$ in terms of $E,J$ are
\begin{align}
&e^2_{r,\fPN{4}} = (-2E)^4\bigg[ (-2EJ^2)\bigg(\frac{45 \nu ^3}{64}-\frac{95 \nu ^2}{8}+\frac{93 \nu }{2}-\frac{201}{4}\bigg)+\frac{1}{(-2EJ^2)^3}\bigg(-44 \nu ^3+\frac{499 \pi ^2 \nu ^2}{4}+\frac{418904 \nu ^2}{315} \nonumber\\
&\quad +\frac{229 \pi ^2 \nu }{16}-\frac{6896084 \nu }{1575}+1280\bigg)+\frac{1}{(-2EJ^2)^2}\bigg(69 \nu ^3-\frac{3199 \pi ^2 \nu ^2}{16}-\frac{88072 \nu ^2}{105}+\frac{2315 \pi ^2 \nu }{32}+\frac{2082211 \nu }{1050}-352\bigg) \nonumber\\
&\quad +\frac{1}{(-2EJ^2)}\bigg(-17 \nu ^3+\frac{1203 \pi ^2 \nu ^2}{16}-\frac{363389 \nu ^2}{1260}-\frac{16195 \pi ^2 \nu }{256}+\frac{977288 \nu }{1575}-76\bigg)+\bigg(\frac{5 \nu ^3}{4}-\frac{123 \pi ^2 \nu ^2}{128} \nonumber\\
&\quad -\frac{25283 \nu ^2}{1120}+\frac{2967 \pi ^2 \nu }{512}+\frac{10957 \nu }{200}-\frac{615}{16}\bigg) \bigg],\\
&e_{\phi, \fPN{4}} = (-2E)^4\bigg[ (-2EJ^2)\bigg(-\frac{307 \nu ^4}{4096}-\frac{317 \nu ^3}{2048}-\frac{16147 \nu ^2}{4096}+\frac{98049 \nu }{4096}-\frac{201}{4}\bigg)+\frac{1}{(-2EJ^2)^3}\bigg(-\frac{639 \nu ^4}{4096} \nonumber\\
&\quad -\frac{16689 \nu ^3}{2048}-\frac{114371 \pi ^2 \nu ^2}{512}+\frac{416339139 \nu ^2}{143360}+\frac{3149843 \pi ^2 \nu }{12288}-\frac{35793568793 \nu }{6451200}+2298\bigg)+\frac{1}{(-2EJ^2)^2}\bigg(-\frac{87 \nu ^4}{256} \nonumber\\
&\quad -\frac{363 \nu ^3}{256}+\frac{1129 \pi ^2 \nu ^2}{4}-\frac{17631107 \nu ^2}{4480}-\frac{3543143 \pi ^2 \nu }{12288}+\frac{5317939433 \nu }{1612800}-646\bigg)+\frac{1}{(-2EJ^2)}\bigg(-\frac{323 \nu ^4}{2048} \nonumber\\
&\quad +\frac{13679 \nu ^3}{1024}-\frac{30305 \pi ^2 \nu ^2}{512}+\frac{44423585 \nu ^2}{43008}+\frac{234271 \pi ^2 \nu }{4096}+\frac{64295457 \nu }{71680}-313\bigg)+\bigg(\frac{231 \nu ^4}{512}-\frac{263 \nu ^3}{64}-\frac{1605199 \nu ^2}{53760} \nonumber\\
&\quad -\frac{7227 \pi ^2 \nu }{4096}-\frac{8044559 \nu }{179200}-\frac{2737}{16}\bigg) \bigg],\\
&e_{t, \fPN{4}} = (-2E)^4\bigg[ (-2EJ^2)\bigg(-\frac{25 \nu ^4}{16}+\frac{21 \nu ^3}{4}-\frac{237 \nu ^2}{16}+\frac{563 \nu }{16}-\frac{457}{4}\bigg)+\frac{1}{(-2EJ^2)^{5/2}}\bigg(\frac{105 \nu ^3}{4}+\frac{123 \pi ^2 \nu ^2}{8} \nonumber\\
&\quad -\frac{7013 \nu ^2}{8}-\frac{51439 \pi ^2 \nu }{1024}+\frac{293413 \nu }{120}-\frac{9009}{8}\bigg)+\frac{1}{(-2EJ^2)^{3/2}}\bigg(-\frac{705 \nu ^3}{8}-\frac{5453 \pi ^2 \nu ^2}{256}+\frac{4490 \nu ^2}{3}+\frac{102569 \pi ^2 \nu }{1536} \nonumber\\
&\quad -\frac{636827 \nu }{180}+\frac{12879}{8}\bigg)+\frac{1}{\sqrt{(-2EJ^2)}}\bigg(\frac{5913 \nu ^3}{64}+\frac{1435 \pi ^2 \nu ^2}{256}-\frac{290843 \nu ^2}{384}-\frac{59677 \pi ^2 \nu }{3072}+\frac{120641 \nu }{72}-\frac{129645}{128}\bigg) \nonumber\\
&\quad +\sqrt{(-2EJ^2)}\bigg(-\frac{1893 \nu ^3}{64}+\frac{18909 \nu ^2}{128}-\frac{2877 \nu }{8}+\frac{56385}{128}\bigg)+\frac{1}{(-2EJ^2)^3}\bigg(-22 \nu ^3+\frac{499 \pi ^2 \nu ^2}{8}+\frac{209452 \nu ^2}{315} \nonumber\\
&\quad +\frac{229 \pi ^2 \nu }{32}-\frac{3448042 \nu }{1575}+640\bigg)+\frac{1}{(-2EJ^2)^2}\bigg(\frac{175 \nu ^3}{2}-\frac{2167 \pi ^2 \nu ^2}{32}-\frac{1577777 \nu ^2}{1260}-\frac{1373 \pi ^2 \nu }{128}+\frac{17709101 \nu }{6300}-672\bigg) \nonumber\\
&\quad +\frac{1}{(-2EJ^2)}\bigg(-119 \nu ^3+\frac{465 \pi ^2 \nu ^2}{64}+\frac{26429 \nu ^2}{36}+\frac{7097 \pi ^2 \nu }{512}-\frac{9057163 \nu }{6300}+\frac{1795}{4}\bigg)+\bigg(\frac{35 \nu ^4}{16}+\frac{1369 \nu ^3}{32}-\frac{5713 \nu ^2}{32} \nonumber\\
&\quad +\frac{4007 \nu }{10}-\frac{3067}{16}\bigg) \bigg],\\
&n_{\fPN{4}} = (-2E)^{11/2}\bigg[ \frac{1}{(-2EJ^2)^{5/2}}\bigg(\frac{105 \nu ^3}{8}+\frac{123 \pi ^2 \nu ^2}{16}-\frac{7013 \nu ^2}{16}-\frac{51439 \pi ^2 \nu }{2048}+\frac{293413 \nu }{240}-\frac{9009}{16}\bigg) \nonumber\\
&\quad +\frac{1}{(-2EJ^2)^{3/2}}\bigg(-\frac{135 \nu ^3}{8}-\frac{451 \pi ^2 \nu ^2}{256}+\frac{1289 \nu ^2}{6}+\frac{50329 \pi ^2 \nu }{6144}-\frac{5111 \nu }{9}+\frac{4725}{16}\bigg)+\frac{1}{\sqrt{(-2EJ^2)}}\bigg(\frac{57 \nu ^3}{16}-\frac{525 \nu ^2}{32} \nonumber\\
&\quad +\frac{825 \nu }{16}-\frac{3375}{32}\bigg)+\frac{1}{(-2EJ^2)}\bigg(9 \nu ^2-45 \nu +\frac{225}{4}\bigg)+1\bigg(\frac{723 \nu ^4}{32768}-\frac{345 \nu ^3}{8192}+\frac{6105 \nu ^2}{16384}+\frac{48975 \nu }{8192}+\frac{698643}{32768}\bigg) \bigg],\\
&k_{\fPN{4}} = (-2E)^{4}\bigg[ \frac{1}{(-2EJ^2)^4}\bigg(-\frac{315 \nu ^3}{16}-\frac{7175 \pi ^2 \nu ^2}{256}+\frac{132475 \nu ^2}{96}+\frac{2975735 \pi ^2 \nu }{24576}-\frac{1736399 \nu }{288}+\frac{225225}{64}\bigg) \nonumber\\
&\quad +\frac{1}{(-2EJ^2)^3}\bigg(\frac{525 \nu ^3}{16}+\frac{615 \pi ^2 \nu ^2}{32}-\frac{35065 \nu ^2}{32}-\frac{257195 \pi ^2 \nu }{4096}+\frac{293413 \nu }{96}-\frac{45045}{32}\bigg)+\frac{1}{(-2EJ^2)^2}\bigg(-\frac{45 \nu ^3}{4} \nonumber\\
&\quad -\frac{615 \pi ^2 \nu ^2}{512}+\frac{4045 \nu ^2}{32}+\frac{35569 \pi ^2 \nu }{8192}-\frac{20323 \nu }{96}+\frac{4725}{64}\bigg)+\frac{1}{(-2EJ^2)}\bigg(\frac{3 \nu ^3}{8}-\frac{15 \nu ^2}{32}\bigg) \bigg], \\
&v^2_{\fPN{4}} = (-2E)^{5}\bigg[ \frac{1}{(-2EJ^2)^{5/2}}\bigg(\frac{35 \nu ^3}{4}+\frac{41 \pi ^2 \nu ^2}{8}-\frac{7493 \nu ^2}{24}-\frac{18021 \pi ^2 \nu }{1024}+\frac{38797 \nu }{40}-\frac{4223}{8}\bigg) \nonumber\\
&\quad +\frac{1}{(-2EJ^2)^{3/2}}\bigg(-\frac{265 \nu ^3}{24}-\frac{2665 \pi ^2 \nu ^2}{2304}+\frac{31301 \nu ^2}{216}+\frac{47869 \pi ^2 \nu }{9216}-\frac{81311 \nu }{216}+\frac{895}{4}\bigg)+\frac{1}{\sqrt{(-2EJ^2)}}\bigg(\frac{1279 \nu ^3}{576} \nonumber\\
&\quad -\frac{10535 \nu ^2}{1152}+\frac{625 \nu }{24}-\frac{6875}{128}\bigg)+\frac{1}{(-2EJ^2)^4}\bigg(-\frac{105 \nu ^3}{8}-\frac{7175 \pi ^2 \nu ^2}{384}+\frac{130315 \nu ^2}{144}+\frac{2857655 \pi ^2 \nu }{36864}-\frac{1631819 \nu }{432} \nonumber\\
&\quad +\frac{202531}{96}\bigg)+\frac{1}{(-2EJ^2)^3}\bigg(\frac{1085 \nu ^3}{48}+\frac{10045 \pi ^2 \nu ^2}{768}-\frac{53885 \nu ^2}{72}-\frac{277859 \pi ^2 \nu }{6144}+\frac{319793 \nu }{144}-\frac{52295}{48}\bigg) \nonumber\\
&\quad +\frac{1}{(-2EJ^2)^2}\bigg(-\frac{605 \nu ^3}{72}-\frac{41 \pi ^2 \nu ^2}{48}+\frac{14219 \nu ^2}{144}+\frac{45409 \pi ^2 \nu }{12288}-\frac{17429 \nu }{72}+\frac{4825}{32}\bigg)+\frac{1}{(-2EJ^2)}\bigg(\frac{65 \nu ^3}{162}+\frac{575 \nu ^2}{144}\nonumber\\
&\quad -\frac{1165 \nu }{48}+\frac{1385}{96}\bigg)+1\bigg(\frac{25 \nu ^4}{1944}-\frac{85 \nu ^3}{5184}+\frac{55 \nu ^2}{288}+\frac{235 \nu }{72}+\frac{241}{24}\bigg) \bigg].
\end{align}
The new terms in the 4PN Kepler equation are
\begin{align}
&g_{8t} = (-2E)^4\bigg[ \frac{1}{(-2EJ^2)^{5/2}}\bigg(-\frac{105 \nu ^3}{8}-\frac{123 \pi ^2 \nu ^2}{16}+\frac{7013 \nu ^2}{16}+\frac{51439 \pi ^2 \nu }{2048}-\frac{293413 \nu }{240}+\frac{9009}{16}\bigg) \nonumber\\
&\quad + \frac{1}{(-2EJ^2)^{3/2}}\bigg(\frac{255 \nu ^3}{16}+\frac{861 \pi ^2 \nu ^2}{512}-\frac{3067 \nu ^2}{16}-\frac{42949 \pi ^2 \nu }{6144}+\frac{61211 \nu }{144}-\frac{1575}{8}\bigg)+\frac{1}{\sqrt{(-2EJ^2)}}\bigg(-\frac{369 \nu ^3}{128} \nonumber\\
&\quad +\frac{2265 \nu ^2}{256}-\frac{75 \nu }{4}+\frac{10125}{256}\bigg)+\frac{1}{(-2EJ^2)}\bigg(-9 \nu ^2+45 \nu -\frac{225}{4}\bigg) \bigg],\\
&f_{8t} = \frac{(-2E)^4}{(1+2EJ^2)^{3/2}}\bigg[ (-2EJ^2)^{3/2}\bigg(-\frac{451 \nu ^4}{1024}+\frac{1355 \nu ^3}{512}-\frac{5071 \nu ^2}{1024}+\frac{10493 \nu }{1024}\bigg)+(-2EJ^2)\bigg(-\frac{3 \nu ^3}{8} +\frac{105 \nu ^2}{16} \nonumber\\
&\quad -\frac{225 \nu }{16}\bigg)+\frac{1}{(-2EJ^2)^{5/2}}\bigg(\frac{135 \nu ^4}{1024}+\frac{4573 \nu ^3}{512}-\frac{15051 \pi ^2 \nu ^2}{512}+\frac{8669641 \nu ^2}{64512}+\frac{321233 \pi ^2 \nu }{12288}-\frac{29451283 \nu }{179200}+\frac{505}{4}\bigg) \nonumber\\
&\quad +\frac{1}{(-2EJ^2)^{3/2}}\bigg(-\frac{5 \nu ^4}{64}-\frac{9521 \nu ^3}{256}+\frac{7341 \pi ^2 \nu ^2}{128}-\frac{4931281 \nu ^2}{23040}-\frac{628627 \pi ^2 \nu }{12288}+\frac{25861391 \nu }{100800}-\frac{1065}{4}\bigg) \nonumber\\
&\quad +\frac{1}{\sqrt{(-2EJ^2)}}\bigg(-\frac{723 \nu ^4}{1024}+\frac{50567 \nu ^3}{1024}-\frac{851 \pi ^2 \nu ^2}{32}+\frac{3862021 \nu ^2}{322560}+\frac{284699 \pi ^2 \nu }{12288}+\frac{18840839 \nu }{1612800}+165\bigg) \nonumber\\
&\quad +\sqrt{(-2EJ^2)}\bigg(\frac{35 \nu ^4}{32}-\frac{1521 \nu ^3}{64}-\frac{697 \pi ^2 \nu ^2}{512}+\frac{1668967 \nu ^2}{23040}+\frac{7565 \pi ^2 \nu }{4096}-\frac{1660729 \nu }{14400}-25\bigg) \nonumber\\
&\quad +\frac{1}{(-2EJ^2)}\bigg(-\frac{3 \nu ^3}{8}+\frac{105 \nu ^2}{16}-\frac{225 \nu }{16}\bigg)+1\bigg(\frac{3 \nu ^3}{4}-\frac{105 \nu ^2}{8}+\frac{225 \nu }{8}\bigg) \bigg],\\
&i_{8t} = (-2E)^4\bigg[ \frac{1}{(-2EJ^2)^{5/2}}\bigg(\frac{35 \nu ^4}{256}+\frac{443 \nu ^3}{64}-\frac{135 \pi ^2 \nu ^2}{32}+\frac{1535 \nu ^2}{768}+\frac{60715 \pi ^2 \nu }{12288}+\frac{1286507 \nu }{50400}+\frac{95}{32}\bigg) \nonumber\\
&\quad +\frac{1}{(-2EJ^2)^{3/2}}\bigg(\frac{49 \nu ^4}{256}-\frac{1593 \nu ^3}{128}+\frac{135 \pi ^2 \nu ^2}{32}+\frac{6815 \nu ^2}{768}-\frac{60715 \pi ^2 \nu }{12288}-\frac{813091 \nu }{25200}-\frac{95}{32}\bigg) \nonumber\\
&\quad +\frac{1}{\sqrt{(-2EJ^2)}}\bigg(-\frac{9 \nu ^4}{32}+\frac{663 \nu ^3}{128}-\frac{6125 \nu ^2}{384}+\frac{2213 \nu }{96}\bigg) \bigg],\\
&h_{8t} = (-2E)^4\nu\sqrt{1+2EJ^2}\bigg[ \frac{1}{(-2EJ^2)^{5/2}}\bigg(\frac{125 \nu ^3}{2048}+\frac{1767 \nu ^2}{1024}+\frac{41 \pi ^2 \nu }{512}+\frac{60355 \nu }{129024}+\frac{903 \pi ^2}{4096}-\frac{5719117}{3225600}\bigg) \nonumber\\
&\quad +\frac{1}{(-2EJ^2)^{3/2}}\bigg(\frac{95 \nu ^3}{1024}-\frac{3503 \nu ^2}{1536}-\frac{41 \pi ^2 \nu }{512}-\frac{46925 \nu }{64512}-\frac{903 \pi ^2}{4096}+\frac{3080321}{1612800}\bigg)+\frac{1}{\sqrt{(-2EJ^2)}}\bigg(-\frac{263 \nu ^3}{2048} \nonumber\\
&\quad +\frac{1969 \nu ^2}{3072}-\frac{6013 \nu }{6144}+\frac{1643}{6144}\bigg) \bigg],\\
&j_{8t} = (-2E)^4\nu \sqrt{1+2EJ^2}\bigg[ \frac{1}{(-2EJ^2)^{5/2}}\bigg(-\frac{33 \nu ^3}{2048}+\frac{93 \nu ^2}{1024}-\frac{161 \nu }{2048}+\frac{31}{2048}\bigg)+\frac{1}{(-2EJ^2)^{3/2}}\bigg(\frac{33 \nu ^3}{1024}-\frac{93 \nu ^2}{512} \nonumber\\
&\quad +\frac{161 \nu }{1024}-\frac{31}{1024}\bigg)+\frac{1}{\sqrt{(-2EJ^2)}}\bigg(-\frac{33 \nu ^3}{2048}+\frac{93 \nu ^2}{1024}-\frac{161 \nu }{2048}+\frac{31}{2048}\bigg) \bigg]
\end{align}
Other parameters, including the semi-major axis $a_r$, the relationship between several angles $\chi,\psi,\zeta,l$ and the conserved quantities $E,J$ in SH, MH, ADM, and EOB coordinates, and the Lagrangian and Hamiltonian used are listed in the Supplementary Material `supp\_hlmFourierCoefficientsResults.m'.

\end{appendix}

\bibliography{refs}

@article{advLIGO,
doi = {10.1088/0264-9381/32/7/074001},
url = {https://dx.doi.org/10.1088/0264-9381/32/7/074001},
year = {2015},
month = {mar},
publisher = {IOP Publishing},
volume = {32},
number = {7},
pages = {074001},
author = {The LIGO Scientific Collaboration and J Aasi and B P Abbott and R Abbott and T Abbott and M R Abernathy and K Ackley and C Adams and T Adams and P Addesso and R X Adhikari and V Adya and C Affeldt and N Aggarwal and O D Aguiar and A Ain and P Ajith and A Alemic and B Allen and D Amariutei and S B Anderson and W G Anderson and K Arai and M C Araya and C Arceneaux and J S Areeda and G Ashton and S Ast and S M Aston and P Aufmuth and C Aulbert and B E Aylott and S Babak and P T Baker and S W Ballmer and J C Barayoga and M Barbet and S Barclay and B C Barish and D Barker and B Barr and L Barsotti and J Bartlett and M A Barton and I Bartos and R Bassiri and J C Batch and C Baune and B Behnke and A S Bell and C Bell and M Benacquista and J Bergman and G Bergmann and C P L Berry and J Betzwieser and S Bhagwat and R Bhandare and I A Bilenko and G Billingsley and J Birch and S Biscans and C Biwer and J K Blackburn and L Blackburn and C D Blair and D Blair and O Bock and T P Bodiya and P Bojtos and C Bond and R Bork and M Born and Sukanta Bose and P R Brady and V B Braginsky and J E Brau and D O Bridges and M Brinkmann and A F Brooks and D A Brown and D D Brown and N M Brown and S Buchman and A Buikema and A Buonanno and L Cadonati and J Calderón Bustillo and J B Camp and K C Cannon and J Cao and C D Capano and S Caride and S Caudill and M Cavaglià and C Cepeda and R Chakraborty and T Chalermsongsak and S J Chamberlin and S Chao and P Charlton and Y Chen and H S Cho and M Cho and J H Chow and N Christensen and Q Chu and S Chung and G Ciani and F Clara and J A Clark and C Collette and L Cominsky and M Constancio and D Cook and T R Corbitt and N Cornish and A Corsi and C A Costa and M W Coughlin and S Countryman and P Couvares and D M Coward and M J Cowart and D C Coyne and R Coyne and K Craig and J D E Creighton and T D Creighton and J Cripe and S G Crowder and A Cumming and L Cunningham and C Cutler and K Dahl and T Dal Canton and M Damjanic and S L Danilishin and K Danzmann and L Dartez and I Dave and H Daveloza and G S Davies and E J Daw and D DeBra and W Del Pozzo and T Denker and T Dent and V Dergachev and R T DeRosa and R DeSalvo and S Dhurandhar and M D´ıaz and I Di Palma and G Dojcinoski and E Dominguez and F Donovan and K L Dooley and S Doravari and R Douglas and T P Downes and J C Driggers and Z Du and S Dwyer and T Eberle and T Edo and M Edwards and M Edwards and A Effler and H.-B Eggenstein and P Ehrens and J Eichholz and S S Eikenberry and R Essick and T Etzel and M Evans and T Evans and M Factourovich and S Fairhurst and X Fan and Q Fang and B Farr and W M Farr and M Favata and M Fays and H Fehrmann and M M Fejer and D Feldbaum and E C Ferreira and R P Fisher and Z Frei and A Freise and R Frey and T T Fricke and P Fritschel and V V Frolov and S Fuentes-Tapia and P Fulda and M Fyffe and J R Gair and S Gaonkar and N Gehrels and L Á Gergely´ and J A Giaime and K D Giardina and J Gleason and E Goetz and R Goetz and L Gondan and G González and N Gordon and M L Gorodetsky and S Gossan and S Goßler and C Gräf and P B Graff and A Grant and S Gras and C Gray and R J S Greenhalgh and A M Gretarsson and H Grote and S Grunewald and C J Guido and X Guo and K Gushwa and E K Gustafson and R Gustafson and J Hacker and E D Hall and G Hammond and M Hanke and J Hanks and C Hanna and M D Hannam and J Hanson and T Hardwick and G M Harry and I W Harry and M Hart and M T Hartman and C-J Haster and K Haughian and S Hee and M Heintze and G Heinzel and M Hendry and I S Heng and A W Heptonstall and M Heurs and M Hewitson and S Hild and D Hoak and K A Hodge and S E Hollitt and K Holt and P Hopkins and D J Hosken and J Hough and E Houston and E J Howell and Y M Hu and E Huerta and B Hughey and S Husa and S H Huttner and M Huynh and T Huynh-Dinh and A Idrisy and N Indik and D R Ingram and R Inta and G Islas and J C Isler and T Isogai and B R Iyer and K Izumi and M Jacobson and H Jang and S Jawahar and Y Ji and F Jiménez-Forteza and W W Johnson and D I Jones and R Jones and L Ju and K Haris and V Kalogera and S Kandhasamy and G Kang and J B Kanner and E Katsavounidis and W Katzman and H Kaufer and S Kaufer and T Kaur and K Kawabe and F Kawazoe and G M Keiser and D Keitel and D B Kelley and W Kells and D G Keppel and J S Key and A Khalaidovski and F Y Khalili and E A Khazanov and C Kim and K Kim and N G Kim and N Kim and Y.-M Kim and E J King and P J King and D L Kinzel and J S Kissel and S Klimenko and J Kline and S Koehlenbeck and K Kokeyama and V Kondrashov and M Korobko and W Z Korth and D B Kozak and V Kringel and B Krishnan and C Krueger and G Kuehn and A Kumar and P Kumar and L Kuo and M Landry and B Lantz and S Larson and P D Lasky and A Lazzarini and C Lazzaro and J Le and P Leaci and S Leavey and E O Lebigot and C H Lee and H K Lee and H M Lee and J R Leong and Y Levin and B Levine and J Lewis and T G F Li and K Libbrecht and A Libson and A C Lin and T B Littenberg and N A Lockerbie and V Lockett and J Logue and A L Lombardi and M Lormand and J Lough and M J Lubinski and H Lück and A P Lundgren and R Lynch and Y Ma and J Macarthur and T MacDonald and B Machenschalk and M MacInnis and D M Macleod and F Magaña-Sandoval and R Magee and M Mageswaran and C Maglione and K Mailand and I Mandel and V Mandic and V Mangano and G L Mansell and S Márka and Z Márka and A Markosyan and E Maros and I W Martin and R M Martin and D Martynov and J N Marx and K Mason and T J Massinger and F Matichard and L Matone and N Mavalvala and N Mazumder and G Mazzolo and R McCarthy and D E McClelland and S McCormick and S C McGuire and G McIntyre and J McIver and K McLin and S McWilliams and G D Meadors and M Meinders and A Melatos and G Mendell and R A Mercer and S Meshkov and C Messenger and P M Meyers and H Miao and H Middleton and E E Mikhailov and A Miller and J Miller and M Millhouse and J Ming and S Mirshekari and C Mishra and S Mitra and V P Mitrofanov and G Mitselmakher and R Mittleman and B Moe and S D Mohanty and S R P Mohapatra and B Moore and D Moraru and G Moreno and S R Morriss and K Mossavi and C M Mow-Lowry and C L Mueller and G Mueller and S Mukherjee and A Mullavey and J Munch and D Murphy and P G Murray and A Mytidis and T Nash and R K Nayak and V Necula and K Nedkova and G Newton and T Nguyen and A B Nielsen and S Nissanke and A H Nitz and D Nolting and M E N Normandin and L K Nuttall and E Ochsner and J O’Dell and E Oelker and G H Ogin and J J Oh and S H Oh and F Ohme and P Oppermann and R Oram and B O’Reilly and W Ortega and R O’Shaughnessy and C Osthelder and C D Ott and D J Ottaway and R S Ottens and H Overmier and B J Owen and C Padilla and A Pai and S Pai and O Palashov and A Pal-Singh and H Pan and C Pankow and F Pannarale and B C Pant and M A Papa and H Paris and Z Patrick and M Pedraza and L Pekowsky and A Pele and S Penn and A Perreca and M Phelps and V Pierro and I M Pinto and M Pitkin and J Poeld and A Post and A Poteomkin and J Powell and J Prasad and V Predoi and S Premachandra and T Prestegard and L R Price and M Principe and S Privitera and R Prix and L Prokhorov and O Puncken and M Pürrer and J Qin and V Quetschke and E Quintero and G Quiroga and R Quitzow-James and F J Raab and D S Rabeling and H Radkins and P Raffai and S Raja and G Rajalakshmi and M Rakhmanov and K Ramirez and V Raymond and C M Reed and S Reid and D H Reitze and O Reula and K Riles and N A Robertson and R Robie and J G Rollins and V Roma and J D Romano and G Romanov and J H Romie and S Rowan and A Rüdiger and K Ryan and S Sachdev and T Sadecki and L Sadeghian and M Saleem and F Salemi and L Sammut and V Sandberg and J R Sanders and V Sannibale and I Santiago-Prieto and B S Sathyaprakash and P R Saulson and R Savage and A Sawadsky and J Scheuer and R Schilling and P Schmidt and R Schnabel and R M S Schofield and E Schreiber and D Schuette and B F Schutz and J Scott and S M Scott and D Sellers and A S Sengupta and A Sergeev and G Serna and A Sevigny and D A Shaddock and M S Shahriar and M Shaltev and Z Shao and B Shapiro and P Shawhan and D H Shoemaker and T L Sidery and X Siemens and D Sigg and A D Silva and D Simakov and A Singer and L Singer and R Singh and A M Sintes and B J J Slagmolen and J R Smith and M R Smith and R J E Smith and N D Smith-Lefebvre and E J Son and B Sorazu and T Souradeep and A Staley and J Stebbins and M Steinke and J Steinlechner and S Steinlechner and D Steinmeyer and B C Stephens and S Steplewski and S Stevenson and R Stone and K A Strain and S Strigin and R Sturani and A L Stuver and T Z Summerscales and P J Sutton and M Szczepanczyk and G Szeifert and D Talukder and D B Tanner and M Tápai and S P Tarabrin and A Taracchini and R Taylor and G Tellez and T Theeg and M P Thirugnanasambandam and M Thomas and P Thomas and K A Thorne and K S Thorne and E Thrane and V Tiwari and C Tomlinson and C V Torres and C I Torrie and G Traylor and M Tse and D Tshilumba and D Ugolini and C S Unnikrishnan and A L Urban and S A Usman and H Vahlbruch and G Vajente and G Valdes and M Vallisneri and A A van Veggel and S Vass and R Vaulin and A Vecchio and J Veitch and P J Veitch and K Venkateswara and R Vincent-Finley and S Vitale and T Vo and C Vorvick and W D Vousden and S P Vyatchanin and A R Wade and L Wade and M Wade and M Walker and L Wallace and S Walsh and H Wang and M Wang and X Wang and R L Ward and J Warner and M Was and B Weaver and M Weinert and A J Weinstein and R Weiss and T Welborn and L Wen and P Wessels and T Westphal and K Wette and J T Whelan and S E Whitcomb and D J White and B F Whiting and C Wilkinson and L Williams and R Williams and A R Williamson and J L Willis and B Willke and M Wimmer and W Winkler and C C Wipf and H Wittel and G Woan and J Worden and S Xie and J Yablon and I Yakushin and W Yam and H Yamamoto and C C Yancey and Q Yang and M Zanolin and Fan Zhang and L Zhang and M Zhang and Y Zhang and C Zhao and M Zhou and X J Zhu and M E Zucker and S Zuraw and J Zweizig},
title = {Advanced LIGO},
journal = {Classical and Quantum Gravity}
}

@article{AdvVirgo,
doi = {10.1088/0264-9381/32/2/024001},
url = {https://dx.doi.org/10.1088/0264-9381/32/2/024001},
year = {2014},
month = {dec},
publisher = {IOP Publishing},
volume = {32},
number = {2},
pages = {024001},
author = {F Acernese and M Agathos and K Agatsuma and D Aisa and N Allemandou and A Allocca and J Amarni and P Astone and G Balestri and G Ballardin and F Barone and J-P Baronick and M Barsuglia and A Basti and F Basti and Th S Bauer and V Bavigadda and M Bejger and M G Beker and C Belczynski and D Bersanetti and A Bertolini and M Bitossi and M A Bizouard and S Bloemen and M Blom and M Boer and G Bogaert and D Bondi and F Bondu and L Bonelli and R Bonnand and V Boschi and L Bosi and T Bouedo and C Bradaschia and M Branchesi and T Briant and A Brillet and V Brisson and T Bulik and H J Bulten and D Buskulic and C Buy and G Cagnoli and E Calloni and C Campeggi and B Canuel and F Carbognani and F Cavalier and R Cavalieri and G Cella and E Cesarini and E Chassande- Mottin and A Chincarini and A Chiummo and S Chua and F Cleva and E Coccia and P-F Cohadon and A Colla and M Colombini and A Conte and J-P Coulon and E Cuoco and A Dalmaz and S D’Antonio and V Dattilo and M Davier and R Day and G Debreczeni and J Degallaix and S Deléglise and W Del Pozzo and H Dereli and R De Rosa and L Di Fiore and A Di Lieto and A Di Virgilio and M Doets and V Dolique and M Drago and M Ducrot and G Endrőczi and V Fafone and S Farinon and I Ferrante and F Ferrini and F Fidecaro and I Fiori and R Flaminio and J-D Fournier and S Franco and S Frasca and F Frasconi and L Gammaitoni and F Garufi and M Gaspard and A Gatto and G Gemme and B Gendre and E Genin and A Gennai and S Ghosh and L Giacobone and A Giazotto and R Gouaty and M Granata and G Greco and P Groot and G M Guidi and J Harms and A Heidmann and H Heitmann and P Hello and G Hemming and E Hennes and D Hofman and P Jaranowski and R J G Jonker and M Kasprzack and F Kéfélian and I Kowalska and M Kraan and A Królak and A Kutynia and C Lazzaro and M Leonardi and N Leroy and N Letendre and T G F Li and B Lieunard and M Lorenzini and V Loriette and G Losurdo and C Magazzù and E Majorana and I Maksimovic and V Malvezzi and N Man and V Mangano and M Mantovani and F Marchesoni and F Marion and J Marque and F Martelli and L Martellini and A Masserot and D Meacher and J Meidam and F Mezzani and C Michel and L Milano and Y Minenkov and A Moggi and M Mohan and M Montani and N Morgado and B Mours and F Mul and M F Nagy and I Nardecchia and L Naticchioni and G Nelemans and I Neri and M Neri and F Nocera and E Pacaud and C Palomba and F Paoletti and A Paoli and A Pasqualetti and R Passaquieti and D Passuello and M Perciballi and S Petit and M Pichot and F Piergiovanni and G Pillant and A Piluso and L Pinard and R Poggiani and M Prijatelj and G A Prodi and M Punturo and P Puppo and D S Rabeling and I Rácz and P Rapagnani and M Razzano and V Re and T Regimbau and F Ricci and F Robinet and A Rocchi and L Rolland and R Romano and D Rosińska and P Ruggi and E Saracco and B Sassolas and F Schimmel and D Sentenac and V Sequino and S Shah and K Siellez and N Straniero and B Swinkels and M Tacca and M Tonelli and F Travasso and M Turconi and G Vajente and N van Bakel and M van Beuzekom and J F J van den Brand and C Van Den Broeck and M V van der Sluys and J van Heijningen and M Vasúth and G Vedovato and J Veitch and D Verkindt and F Vetrano and A Viceré and J-Y Vinet and G Visser and H Vocca and R Ward and M Was and L-W Wei and M Yvert and A Zadro żny and J-P Zendri},
title = {Advanced Virgo: a second-generation interferometric gravitational wave detector},
journal = {Classical and Quantum Gravity}
}

@article{KAGRA,
    author = {Akutsu, T and Ando, M and Arai, K and Arai, Y and Araki, S and Araya, A and Aritomi, N and Aso, Y and Bae, S and Bae, Y and Baiotti, L and Bajpai, R and Barton, M A and Cannon, K and Capocasa, E and Chan, M and Chen, C and Chen, K and Chen, Y and Chu, H and Chu, Y -K and Eguchi, S and Enomoto, Y and Flaminio, R and Fujii, Y and Fukunaga, M and Fukushima, M and Ge, G and Hagiwara, A and Haino, S and Hasegawa, K and Hayakawa, H and Hayama, K and Himemoto, Y and Hiranuma, Y and Hirata, N and Hirose, E and Hong, Z and Hsieh, B H and Huang, C -Z and Huang, P and Huang, Y and Ikenoue, B and Imam, S and Inayoshi, K and Inoue, Y and Ioka, K and Itoh, Y and Izumi, K and Jung, K and Jung, P and Kajita, T and Kamiizumi, M and Kanda, N and Kang, G and Kawaguchi, K and Kawai, N and Kawasaki, T and Kim, C and Kim, J C and Kim, W S and Kim, Y -M and Kimura, N and Kita, N and Kitazawa, H and Kojima, Y and Kokeyama, K and Komori, K and Kong, A K H and Kotake, K and Kozakai, C and Kozu, R and Kumar, R and Kume, J and Kuo, C and Kuo, H -S and Kuroyanagi, S and Kusayanagi, K and Kwak, K and Lee, H K and Lee, H W and Lee, R and Leonardi, M and Lin, L C -C and Lin, C -Y and Lin, F -L and Liu, G C and Luo, L -W and Marchio, M and Michimura, Y and Mio, N and Miyakawa, O and Miyamoto, A and Miyazaki, Y and Miyo, K and Miyoki, S and Morisaki, S and Moriwaki, Y and Nagano, K and Nagano, S and Nakamura, K and Nakano, H and Nakano, M and Nakashima, R and Narikawa, T and Negishi, R and Ni, W -T and Nishizawa, A and Obuchi, Y and Ogaki, W and Oh, J J and Oh, S H and Ohashi, M and Ohishi, N and Ohkawa, M and Okutomi, K and Oohara, K and Ooi, C P and Oshino, S and Pan, K and Pang, H and Park, J and Arellano, F E Peña and Pinto, I and Sago, N and Saito, S and Saito, Y and Sakai, K and Sakai, Y and Sakuno, Y and Sato, S and Sato, T and Sawada, T and Sekiguchi, T and Sekiguchi, Y and Shibagaki, S and Shimizu, R and Shimoda, T and Shimode, K and Shinkai, H and Shishido, T and Shoda, A and Somiya, K and Son, E J and Sotani, H and Sugimoto, R and Suzuki, T and Suzuki, T and Tagoshi, H and Takahashi, H and Takahashi, R and Takamori, A and Takano, S and Takeda, H and Takeda, M and Tanaka, H and Tanaka, K and Tanaka, K and Tanaka, T and Tanaka, T and Tanioka, S and Tapia San Martin, E N and Telada, S and Tomaru, T and Tomigami, Y and Tomura, T and Travasso, F and Trozzo, L and Tsang, T and Tsubono, K and Tsuchida, S and Tsuzuki, T and Tuyenbayev, D and Uchikata, N and Uchiyama, T and Ueda, A and Uehara, T and Ueno, K and Ueshima, G and Uraguchi, F and Ushiba, T and van Putten, M H P M and Vocca, H and Wang, J and Wu, C and Wu, H and Wu, S and Xu, W- R and Yamada, T and Yamamoto, K and Yamamoto, K and Yamamoto, T and Yokogawa, K and Yokoyama, J and Yokozawa, T and Yoshioka, T and Yuzurihara, H and Zeidler, S and Zhao, Y and Zhu, Z -H},
    title = "{Overview of KAGRA: Detector design and construction history}",
    journal = {Progress of Theoretical and Experimental Physics},
    volume = {2021},
    number = {5},
    pages = {05A101},
    year = {2020},
    month = {08},
    issn = {2050-3911},
    doi = {10.1093/ptep/ptaa125},
    url = {https://doi.org/10.1093/ptep/ptaa125},
    eprint = {https://academic.oup.com/ptep/article-pdf/2021/5/05A101/37974994/ptaa125.pdf},
}

@article{GW150914,
   title={Observation of Gravitational Waves from a Binary Black Hole Merger},
   volume={116},
   ISSN={1079-7114},
   url={http://dx.doi.org/10.1103/PhysRevLett.116.061102},
   DOI={10.1103/physrevlett.116.061102},
   number={6},
   journal={Physical Review Letters},
   publisher={American Physical Society (APS)},
   author={Abbott, B. P. and Abbott, R. and Abbott, T. D. and Abernathy, M. R. and Acernese, F. and Ackley, K. and Adams, C. and Adams, T. and Addesso, P. and Adhikari, R. X. and Adya, V. B. and Affeldt, C. and Agathos, M. and Agatsuma, K. and Aggarwal, N. and Aguiar, O. D. and Aiello, L. and Ain, A. and Ajith, P. and Allen, B. and Allocca, A. and Altin, P. A. and Anderson, S. B. and Anderson, W. G. and Arai, K. and Arain, M. A. and Araya, M. C. and Arceneaux, C. C. and Areeda, J. S. and Arnaud, N. and Arun, K. G. and Ascenzi, S. and Ashton, G. and Ast, M. and Aston, S. M. and Astone, P. and Aufmuth, P. and Aulbert, C. and Babak, S. and Bacon, P. and Bader, M. K. M. and Baker, P. T. and Baldaccini, F. and Ballardin, G. and Ballmer, S. W. and Barayoga, J. C. and Barclay, S. E. and Barish, B. C. and Barker, D. and Barone, F. and Barr, B. and Barsotti, L. and Barsuglia, M. and Barta, D. and Bartlett, J. and Barton, M. A. and Bartos, I. and Bassiri, R. and Basti, A. and Batch, J. C. and Baune, C. and Bavigadda, V. and Bazzan, M. and Behnke, B. and Bejger, M. and Belczynski, C. and Bell, A. S. and Bell, C. J. and Berger, B. K. and Bergman, J. and Bergmann, G. and Berry, C. P. L. and Bersanetti, D. and Bertolini, A. and Betzwieser, J. and Bhagwat, S. and Bhandare, R. and Bilenko, I. A. and Billingsley, G. and Birch, J. and Birney, I. A. and Birnholtz, O. and Biscans, S. and Bisht, A. and Bitossi, M. and Biwer, C. and Bizouard, M. A. and Blackburn, J. K. and Blair, C. D. and Blair, D. G. and Blair, R. M. and Bloemen, S. and Bock, O. and Bodiya, T. P. and Boer, M. and Bogaert, G. and Bogan, C. and Bohe, A. and Bojtos, P. and Bond, C. and Bondu, F. and Bonnand, R. and Boom, B. A. and Bork, R. and Boschi, V. and Bose, S. and Bouffanais, Y. and Bozzi, A. and Bradaschia, C. and Brady, P. R. and Braginsky, V. B. and Branchesi, M. and Brau, J. E. and Briant, T. and Brillet, A. and Brinkmann, M. and Brisson, V. and Brockill, P. and Brooks, A. F. and Brown, D. A. and Brown, D. D. and Brown, N. M. and Buchanan, C. C. and Buikema, A. and Bulik, T. and Bulten, H. J. and Buonanno, A. and Buskulic, D. and Buy, C. and Byer, R. L. and Cabero, M. and Cadonati, L. and Cagnoli, G. and Cahillane, C. and Bustillo, J. Calderón and Callister, T. and Calloni, E. and Camp, J. B. and Cannon, K. C. and Cao, J. and Capano, C. D. and Capocasa, E. and Carbognani, F. and Caride, S. and Diaz, J. Casanueva and Casentini, C. and Caudill, S. and Cavaglià, M. and Cavalier, F. and Cavalieri, R. and Cella, G. and Cepeda, C. B. and Baiardi, L. Cerboni and Cerretani, G. and Cesarini, E. and Chakraborty, R. and Chalermsongsak, T. and Chamberlin, S. J. and Chan, M. and Chao, S. and Charlton, P. and Chassande-Mottin, E. and Chen, H. Y. and Chen, Y. and Cheng, C. and Chincarini, A. and Chiummo, A. and Cho, H. S. and Cho, M. and Chow, J. H. and Christensen, N. and Chu, Q. and Chua, S. and Chung, S. and Ciani, G. and Clara, F. and Clark, J. A. and Cleva, F. and Coccia, E. and Cohadon, P.-F. and Colla, A. and Collette, C. G. and Cominsky, L. and Constancio, M. and Conte, A. and Conti, L. and Cook, D. and Corbitt, T. R. and Cornish, N. and Corsi, A. and Cortese, S. and Costa, C. A. and Coughlin, M. W. and Coughlin, S. B. and Coulon, J.-P. and Countryman, S. T. and Couvares, P. and Cowan, E. E. and Coward, D. M. and Cowart, M. J. and Coyne, D. C. and Coyne, R. and Craig, K. and Creighton, J. D. E. and Creighton, T. D. and Cripe, J. and Crowder, S. G. and Cruise, A. M. and Cumming, A. and Cunningham, L. and Cuoco, E. and Canton, T. Dal and Danilishin, S. L. and D’Antonio, S. and Danzmann, K. and Darman, N. S. and Da Silva Costa, C. F. and Dattilo, V. and Dave, I. and Daveloza, H. P. and Davier, M. and Davies, G. S. and Daw, E. J. and Day, R. and De, S. and DeBra, D. and Debreczeni, G. and Degallaix, J. and De Laurentis, M. and Deléglise, S. and Del Pozzo, W. and Denker, T. and Dent, T. and Dereli, H. and Dergachev, V. and DeRosa, R. T. and De Rosa, R. and DeSalvo, R. and Dhurandhar, S. and Díaz, M. C. and Di Fiore, L. and Di Giovanni, M. and Di Lieto, A. and Di Pace, S. and Di Palma, I. and Di Virgilio, A. and Dojcinoski, G. and Dolique, V. and Donovan, F. and Dooley, K. L. and Doravari, S. and Douglas, R. and Downes, T. P. and Drago, M. and Drever, R. W. P. and Driggers, J. C. and Du, Z. and Ducrot, M. and Dwyer, S. E. and Edo, T. B. and Edwards, M. C. and Effler, A. and Eggenstein, H.-B. and Ehrens, P. and Eichholz, J. and Eikenberry, S. S. and Engels, W. and Essick, R. C. and Etzel, T. and Evans, M. and Evans, T. M. and Everett, R. and Factourovich, M. and Fafone, V. and Fair, H. and Fairhurst, S. and Fan, X. and Fang, Q. and Farinon, S. and Farr, B. and Farr, W. M. and Favata, M. and Fays, M. and Fehrmann, H. and Fejer, M. M. and Feldbaum, D. and Ferrante, I. and Ferreira, E. C. and Ferrini, F. and Fidecaro, F. and Finn, L. S. and Fiori, I. and Fiorucci, D. and Fisher, R. P. and Flaminio, R. and Fletcher, M. and Fong, H. and Fournier, J.-D. and Franco, S. and Frasca, S. and Frasconi, F. and Frede, M. and Frei, Z. and Freise, A. and Frey, R. and Frey, V. and Fricke, T. T. and Fritschel, P. and Frolov, V. V. and Fulda, P. and Fyffe, M. and Gabbard, H. A. G. and Gair, J. R. and Gammaitoni, L. and Gaonkar, S. G. and Garufi, F. and Gatto, A. and Gaur, G. and Gehrels, N. and Gemme, G. and Gendre, B. and Genin, E. and Gennai, A. and George, J. and Gergely, L. and Germain, V. and Ghosh, Abhirup and Ghosh, Archisman and Ghosh, S. and Giaime, J. A. and Giardina, K. D. and Giazotto, A. and Gill, K. and Glaefke, A. and Gleason, J. R. and Goetz, E. and Goetz, R. and Gondan, L. and González, G. and Castro, J. M. Gonzalez and Gopakumar, A. and Gordon, N. A. and Gorodetsky, M. L. and Gossan, S. E. and Gosselin, M. and Gouaty, R. and Graef, C. and Graff, P. B. and Granata, M. and Grant, A. and Gras, S. and Gray, C. and Greco, G. and Green, A. C. and Greenhalgh, R. J. S. and Groot, P. and Grote, H. and Grunewald, S. and Guidi, G. M. and Guo, X. and Gupta, A. and Gupta, M. K. and Gushwa, K. E. and Gustafson, E. K. and Gustafson, R. and Hacker, J. J. and Hall, B. R. and Hall, E. D. and Hammond, G. and Haney, M. and Hanke, M. M. and Hanks, J. and Hanna, C. and Hannam, M. D. and Hanson, J. and Hardwick, T. and Harms, J. and Harry, G. M. and Harry, I. W. and Hart, M. J. and Hartman, M. T. and Haster, C.-J. and Haughian, K. and Healy, J. and Heefner, J. and Heidmann, A. and Heintze, M. C. and Heinzel, G. and Heitmann, H. and Hello, P. and Hemming, G. and Hendry, M. and Heng, I. S. and Hennig, J. and Heptonstall, A. W. and Heurs, M. and Hild, S. and Hoak, D. and Hodge, K. A. and Hofman, D. and Hollitt, S. E. and Holt, K. and Holz, D. E. and Hopkins, P. and Hosken, D. J. and Hough, J. and Houston, E. A. and Howell, E. J. and Hu, Y. M. and Huang, S. and Huerta, E. A. and Huet, D. and Hughey, B. and Husa, S. and Huttner, S. H. and Huynh-Dinh, T. and Idrisy, A. and Indik, N. and Ingram, D. R. and Inta, R. and Isa, H. N. and Isac, J.-M. and Isi, M. and Islas, G. and Isogai, T. and Iyer, B. R. and Izumi, K. and Jacobson, M. B. and Jacqmin, T. and Jang, H. and Jani, K. and Jaranowski, P. and Jawahar, S. and Jiménez-Forteza, F. and Johnson, W. W. and Johnson-McDaniel, N. K. and Jones, D. I. and Jones, R. and Jonker, R. J. G. and Ju, L. and Haris, K. and Kalaghatgi, C. V. and Kalogera, V. and Kandhasamy, S. and Kang, G. and Kanner, J. B. and Karki, S. and Kasprzack, M. and Katsavounidis, E. and Katzman, W. and Kaufer, S. and Kaur, T. and Kawabe, K. and Kawazoe, F. and Kéfélian, F. and Kehl, M. S. and Keitel, D. and Kelley, D. B. and Kells, W. and Kennedy, R. and Keppel, D. G. and Key, J. S. and Khalaidovski, A. and Khalili, F. Y. and Khan, I. and Khan, S. and Khan, Z. and Khazanov, E. A. and Kijbunchoo, N. and Kim, C. and Kim, J. and Kim, K. and Kim, Nam-Gyu and Kim, Namjun and Kim, Y.-M. and King, E. J. and King, P. J. and Kinzel, D. L. and Kissel, J. S. and Kleybolte, L. and Klimenko, S. and Koehlenbeck, S. M. and Kokeyama, K. and Koley, S. and Kondrashov, V. and Kontos, A. and Koranda, S. and Korobko, M. and Korth, W. Z. and Kowalska, I. and Kozak, D. B. and Kringel, V. and Krishnan, B. and Królak, A. and Krueger, C. and Kuehn, G. and Kumar, P. and Kumar, R. and Kuo, L. and Kutynia, A. and Kwee, P. and Lackey, B. D. and Landry, M. and Lange, J. and Lantz, B. and Lasky, P. D. and Lazzarini, A. and Lazzaro, C. and Leaci, P. and Leavey, S. and Lebigot, E. O. and Lee, C. H. and Lee, H. K. and Lee, H. M. and Lee, K. and Lenon, A. and Leonardi, M. and Leong, J. R. and Leroy, N. and Letendre, N. and Levin, Y. and Levine, B. M. and Li, T. G. F. and Libson, A. and Littenberg, T. B. and Lockerbie, N. A. and Logue, J. and Lombardi, A. L. and London, L. T. and Lord, J. E. and Lorenzini, M. and Loriette, V. and Lormand, M. and Losurdo, G. and Lough, J. D. and Lousto, C. O. and Lovelace, G. and Lück, H. and Lundgren, A. P. and Luo, J. and Lynch, R. and Ma, Y. and MacDonald, T. and Machenschalk, B. and MacInnis, M. and Macleod, D. M. and Magaña-Sandoval, F. and Magee, R. M. and Mageswaran, M. and Majorana, E. and Maksimovic, I. and Malvezzi, V. and Man, N. and Mandel, I. and Mandic, V. and Mangano, V. and Mansell, G. L. and Manske, M. and Mantovani, M. and Marchesoni, F. and Marion, F. and Márka, S. and Márka, Z. and Markosyan, A. S. and Maros, E. and Martelli, F. and Martellini, L. and Martin, I. W. and Martin, R. M. and Martynov, D. V. and Marx, J. N. and Mason, K. and Masserot, A. and Massinger, T. J. and Masso-Reid, M. and Matichard, F. and Matone, L. and Mavalvala, N. and Mazumder, N. and Mazzolo, G. and McCarthy, R. and McClelland, D. E. and McCormick, S. and McGuire, S. C. and McIntyre, G. and McIver, J. and McManus, D. J. and McWilliams, S. T. and Meacher, D. and Meadors, G. D. and Meidam, J. and Melatos, A. and Mendell, G. and Mendoza-Gandara, D. and Mercer, R. A. and Merilh, E. and Merzougui, M. and Meshkov, S. and Messenger, C. and Messick, C. and Meyers, P. M. and Mezzani, F. and Miao, H. and Michel, C. and Middleton, H. and Mikhailov, E. E. and Milano, L. and Miller, J. and Millhouse, M. and Minenkov, Y. and Ming, J. and Mirshekari, S. and Mishra, C. and Mitra, S. and Mitrofanov, V. P. and Mitselmakher, G. and Mittleman, R. and Moggi, A. and Mohan, M. and Mohapatra, S. R. P. and Montani, M. and Moore, B. C. and Moore, C. J. and Moraru, D. and Moreno, G. and Morriss, S. R. and Mossavi, K. and Mours, B. and Mow-Lowry, C. M. and Mueller, C. L. and Mueller, G. and Muir, A. W. and Mukherjee, Arunava and Mukherjee, D. and Mukherjee, S. and Mukund, N. and Mullavey, A. and Munch, J. and Murphy, D. J. and Murray, P. G. and Mytidis, A. and Nardecchia, I. and Naticchioni, L. and Nayak, R. K. and Necula, V. and Nedkova, K. and Nelemans, G. and Neri, M. and Neunzert, A. and Newton, G. and Nguyen, T. T. and Nielsen, A. B. and Nissanke, S. and Nitz, A. and Nocera, F. and Nolting, D. and Normandin, M. E. N. and Nuttall, L. K. and Oberling, J. and Ochsner, E. and O’Dell, J. and Oelker, E. and Ogin, G. H. and Oh, J. J. and Oh, S. H. and Ohme, F. and Oliver, M. and Oppermann, P. and Oram, Richard J. and O’Reilly, B. and O’Shaughnessy, R. and Ott, C. D. and Ottaway, D. J. and Ottens, R. S. and Overmier, H. and Owen, B. J. and Pai, A. and Pai, S. A. and Palamos, J. R. and Palashov, O. and Palomba, C. and Pal-Singh, A. and Pan, H. and Pan, Y. and Pankow, C. and Pannarale, F. and Pant, B. C. and Paoletti, F. and Paoli, A. and Papa, M. A. and Paris, H. R. and Parker, W. and Pascucci, D. and Pasqualetti, A. and Passaquieti, R. and Passuello, D. and Patricelli, B. and Patrick, Z. and Pearlstone, B. L. and Pedraza, M. and Pedurand, R. and Pekowsky, L. and Pele, A. and Penn, S. and Perreca, A. and Pfeiffer, H. P. and Phelps, M. and Piccinni, O. and Pichot, M. and Pickenpack, M. and Piergiovanni, F. and Pierro, V. and Pillant, G. and Pinard, L. and Pinto, I. M. and Pitkin, M. and Poeld, J. H. and Poggiani, R. and Popolizio, P. and Post, A. and Powell, J. and Prasad, J. and Predoi, V. and Premachandra, S. S. and Prestegard, T. and Price, L. R. and Prijatelj, M. and Principe, M. and Privitera, S. and Prix, R. and Prodi, G. A. and Prokhorov, L. and Puncken, O. and Punturo, M. and Puppo, P. and Pürrer, M. and Qi, H. and Qin, J. and Quetschke, V. and Quintero, E. A. and Quitzow-James, R. and Raab, F. J. and Rabeling, D. S. and Radkins, H. and Raffai, P. and Raja, S. and Rakhmanov, M. and Ramet, C. R. and Rapagnani, P. and Raymond, V. and Razzano, M. and Re, V. and Read, J. and Reed, C. M. and Regimbau, T. and Rei, L. and Reid, S. and Reitze, D. H. and Rew, H. and Reyes, S. D. and Ricci, F. and Riles, K. and Robertson, N. A. and Robie, R. and Robinet, F. and Rocchi, A. and Rolland, L. and Rollins, J. G. and Roma, V. J. and Romano, J. D. and Romano, R. and Romanov, G. and Romie, J. H. and Rosińska, D. and Rowan, S. and Rüdiger, A. and Ruggi, P. and Ryan, K. and Sachdev, S. and Sadecki, T. and Sadeghian, L. and Salconi, L. and Saleem, M. and Salemi, F. and Samajdar, A. and Sammut, L. and Sampson, L. M. and Sanchez, E. J. and Sandberg, V. and Sandeen, B. and Sanders, G. H. and Sanders, J. R. and Sassolas, B. and Sathyaprakash, B. S. and Saulson, P. R. and Sauter, O. and Savage, R. L. and Sawadsky, A. and Schale, P. and Schilling, R. and Schmidt, J. and Schmidt, P. and Schnabel, R. and Schofield, R. M. S. and Schönbeck, A. and Schreiber, E. and Schuette, D. and Schutz, B. F. and Scott, J. and Scott, S. M. and Sellers, D. and Sengupta, A. S. and Sentenac, D. and Sequino, V. and Sergeev, A. and Serna, G. and Setyawati, Y. and Sevigny, A. and Shaddock, D. A. and Shaffer, T. and Shah, S. and Shahriar, M. S. and Shaltev, M. and Shao, Z. and Shapiro, B. and Shawhan, P. and Sheperd, A. and Shoemaker, D. H. and Shoemaker, D. M. and Siellez, K. and Siemens, X. and Sigg, D. and Silva, A. D. and Simakov, D. and Singer, A. and Singer, L. P. and Singh, A. and Singh, R. and Singhal, A. and Sintes, A. M. and Slagmolen, B. J. J. and Smith, J. R. and Smith, M. R. and Smith, N. D. and Smith, R. J. E. and Son, E. J. and Sorazu, B. and Sorrentino, F. and Souradeep, T. and Srivastava, A. K. and Staley, A. and Steinke, M. and Steinlechner, J. and Steinlechner, S. and Steinmeyer, D. and Stephens, B. C. and Stevenson, S. P. and Stone, R. and Strain, K. A. and Straniero, N. and Stratta, G. and Strauss, N. A. and Strigin, S. and Sturani, R. and Stuver, A. L. and Summerscales, T. Z. and Sun, L. and Sutton, P. J. and Swinkels, B. L. and Szczepańczyk, M. J. and Tacca, M. and Talukder, D. and Tanner, D. B. and Tápai, M. and Tarabrin, S. P. and Taracchini, A. and Taylor, R. and Theeg, T. and Thirugnanasambandam, M. P. and Thomas, E. G. and Thomas, M. and Thomas, P. and Thorne, K. A. and Thorne, K. S. and Thrane, E. and Tiwari, S. and Tiwari, V. and Tokmakov, K. V. and Tomlinson, C. and Tonelli, M. and Torres, C. V. and Torrie, C. I. and Töyrä, D. and Travasso, F. and Traylor, G. and Trifirò, D. and Tringali, M. C. and Trozzo, L. and Tse, M. and Turconi, M. and Tuyenbayev, D. and Ugolini, D. and Unnikrishnan, C. S. and Urban, A. L. and Usman, S. A. and Vahlbruch, H. and Vajente, G. and Valdes, G. and Vallisneri, M. and van Bakel, N. and van Beuzekom, M. and van den Brand, J. F. J. and Van Den Broeck, C. and Vander-Hyde, D. C. and van der Schaaf, L. and van Heijningen, J. V. and van Veggel, A. A. and Vardaro, M. and Vass, S. and Vasúth, M. and Vaulin, R. and Vecchio, A. and Vedovato, G. and Veitch, J. and Veitch, P. J. and Venkateswara, K. and Verkindt, D. and Vetrano, F. and Viceré, A. and Vinciguerra, S. and Vine, D. J. and Vinet, J.-Y. and Vitale, S. and Vo, T. and Vocca, H. and Vorvick, C. and Voss, D. and Vousden, W. D. and Vyatchanin, S. P. and Wade, A. R. and Wade, L. E. and Wade, M. and Waldman, S. J. and Walker, M. and Wallace, L. and Walsh, S. and Wang, G. and Wang, H. and Wang, M. and Wang, X. and Wang, Y. and Ward, H. and Ward, R. L. and Warner, J. and Was, M. and Weaver, B. and Wei, L.-W. and Weinert, M. and Weinstein, A. J. and Weiss, R. and Welborn, T. and Wen, L. and Weßels, P. and Westphal, T. and Wette, K. and Whelan, J. T. and Whitcomb, S. E. and White, D. J. and Whiting, B. F. and Wiesner, K. and Wilkinson, C. and Willems, P. A. and Williams, L. and Williams, R. D. and Williamson, A. R. and Willis, J. L. and Willke, B. and Wimmer, M. H. and Winkelmann, L. and Winkler, W. and Wipf, C. C. and Wiseman, A. G. and Wittel, H. and Woan, G. and Worden, J. and Wright, J. L. and Wu, G. and Yablon, J. and Yakushin, I. and Yam, W. and Yamamoto, H. and Yancey, C. C. and Yap, M. J. and Yu, H. and Yvert, M. and Zadrożny, A. and Zangrando, L. and Zanolin, M. and Zendri, J.-P. and Zevin, M. and Zhang, F. and Zhang, L. and Zhang, M. and Zhang, Y. and Zhao, C. and Zhou, M. and Zhou, Z. and Zhu, X. J. and Zucker, M. E. and Zuraw, S. E. and Zweizig, J.},
   year={2016},
   month=feb }

@Misc{GWTC_2,
  author = {LIGO},
  title  = {GWTC-2 Data Release Documentation},
  doi    = {10.7935/99gf-ax93},
  url    = {https://www.gw-openscience.org/GWTC-2/},
}

@Article{ET_2010,
  author    = {M Punturo and M Abernathy and F Acernese and B Allen and N Andersson and K Arun and F Barone and B Barr and M Barsuglia and M Beker and N Beveridge and S Birindelli and S Bose and L Bosi and S Braccini and C Bradaschia and T Bulik and E Calloni and G Cella and E Chassande Mottin and S Chelkowski and A Chincarini and J Clark and E Coccia and C Colacino and J Colas and A Cumming and L Cunningham and E Cuoco and S Danilishin and K Danzmann and G De Luca and R De Salvo and T Dent and R De Rosa and L Di Fiore and A Di Virgilio and M Doets and V Fafone and P Falferi and R Flaminio and J Franc and F Frasconi and A Freise and P Fulda and J Gair and G Gemme and A Gennai and A Giazotto and K Glampedakis and M Granata and H Grote and G Guidi and G Hammond and M Hannam and J Harms and D Heinert and M Hendry and I Heng and E Hennes and S Hild and J Hough and S Husa and S Huttner and G Jones and F Khalili and K Kokeyama and K Kokkotas and B Krishnan and M Lorenzini and H Lück and E Majorana and I Mandel and V Mandic and I Martin and C Michel and Y Minenkov and N Morgado and S Mosca and B Mours and H Müller{\textendash}Ebhardt and P Murray and R Nawrodt and J Nelson and R Oshaughnessy and C D Ott and C Palomba and A Paoli and G Parguez and A Pasqualetti and R Passaquieti and D Passuello and L Pinard and R Poggiani and P Popolizio and M Prato and P Puppo and D Rabeling and P Rapagnani and J Read and T Regimbau and H Rehbein and S Reid and L Rezzolla and F Ricci and F Richard and A Rocchi and S Rowan and A Rüdiger and B Sassolas and B Sathyaprakash and R Schnabel and C Schwarz and P Seidel and A Sintes and K Somiya and F Speirits and K Strain and S Strigin and P Sutton and S Tarabrin and A Thüring and J van den Brand and C van Leewen and M van Veggel and C van den Broeck and A Vecchio and J Veitch and F Vetrano and A Vicere and S Vyatchanin and B Willke and G Woan and P Wolfango and K Yamamoto},
  journal   = {Classical and Quantum Gravity},
  title     = {The Einstein Telescope: a third-generation gravitational wave observatory},
  year      = {2010},
  month     = {sep},
  number    = {19},
  pages     = {194002},
  volume    = {27},
  doi       = {10.1088/0264-9381/27/19/194002},
  publisher = {{IOP} Publishing},
  url       = {https://iopscience.iop.org/article/10.1088/0264-9381/27/19/194002},
}

@Article{LISA_1997,
  author    = {K Danzmann for the LISA Study Team},
  journal   = {Classical and Quantum Gravity},
  title     = {{LISA} - an {ESA} cornerstone mission for a gravitational wave observatory},
  year      = {1997},
  month     = {jun},
  number    = {6},
  pages     = {1399--1404},
  volume    = {14},
  doi       = {10.1088/0264-9381/14/6/002},
  publisher = {{IOP} Publishing},
  url       = {https://iopscience.iop.org/article/10.1088/0264-9381/14/6/002},
}

@misc{CosmicExplorer,
      title={A Horizon Study for Cosmic Explorer: Science, Observatories, and Community}, 
      author={Matthew Evans and Rana X Adhikari and Chaitanya Afle and Stefan W. Ballmer and Sylvia Biscoveanu and Ssohrab Borhanian and Duncan A. Brown and Yanbei Chen and Robert Eisenstein and Alexandra Gruson and Anuradha Gupta and Evan D. Hall and Rachael Huxford and Brittany Kamai and Rahul Kashyap and Jeff S. Kissel and Kevin Kuns and Philippe Landry and Amber Lenon and Geoffrey Lovelace and Lee McCuller and Ken K. Y. Ng and Alexander H. Nitz and Jocelyn Read and B. S. Sathyaprakash and David H. Shoemaker and Bram J. J. Slagmolen and Joshua R. Smith and Varun Srivastava and Ling Sun and Salvatore Vitale and Rainer Weiss},
      year={2021},
      eprint={2109.09882},
      archivePrefix={arXiv},
      primaryClass={astro-ph.IM}
}

@article{Tianqin,
   title={TianQin: a space-borne gravitational wave detector},
   volume={33},
   ISSN={1361-6382},
   url={http://dx.doi.org/10.1088/0264-9381/33/3/035010},
   DOI={10.1088/0264-9381/33/3/035010},
   number={3},
   journal={Classical and Quantum Gravity},
   publisher={IOP Publishing},
   author={Luo, Jun and Chen, Li-Sheng and Duan, Hui-Zong and Gong, Yun-Gui and Hu, Shoucun and Ji, Jianghui and Liu, Qi and Mei, Jianwei and Milyukov, Vadim and Sazhin, Mikhail and Shao, Cheng-Gang and Toth, Viktor T and Tu, Hai-Bo and Wang, Yamin and Wang, Yan and Yeh, Hsien-Chi and Zhan, Ming-Sheng and Zhang, Yonghe and Zharov, Vladimir and Zhou, Ze-Bing},
   year={2016},
   month=jan, pages={035010} }

@article{Taiji,
   title={Taiji program: Gravitational-wave sources},
   volume={35},
   ISSN={1793-656X},
   url={http://dx.doi.org/10.1142/S0217751X2050075X},
   DOI={10.1142/s0217751x2050075x},
   number={17},
   journal={International Journal of Modern Physics A},
   publisher={World Scientific Pub Co Pte Lt},
   author={Ruan, Wen-Hong and Guo, Zong-Kuan and Cai, Rong-Gen and Zhang, Yuan-Zhong},
   year={2020},
   month=jun, pages={2050075} }

@article{DECIGO,
    author = "Kawamura, S. and others",
    editor = "Mio, N.",
    title = "{The Japanese space gravitational wave antenna DECIGO}",
    doi = "10.1088/0264-9381/23/8/S17",
    journal = "Class. Quant. Grav.",
    volume = "23",
    pages = "S125--S132",
    year = "2006"
}

@article{GWTC3,
   title={GWTC-3: Compact Binary Coalescences Observed by LIGO and Virgo during the Second Part of the Third Observing Run},
   volume={13},
   ISSN={2160-3308},
   url={http://dx.doi.org/10.1103/PhysRevX.13.041039},
   DOI={10.1103/physrevx.13.041039},
   number={4},
   journal={Physical Review X},
   publisher={American Physical Society (APS)},
   author={Abbott, R. and Abbott, T. D. and Acernese, F. and Ackley, K. and Adams, C. and Adhikari, N. and Adhikari, R. X. and Adya, V. B. and Affeldt, C. and Agarwal, D. and Agathos, M. and Agatsuma, K. and Aggarwal, N. and Aguiar, O. D. and Aiello, L. and Ain, A. and Ajith, P. and Akcay, S. and Akutsu, T. and Albanesi, S. and Allocca, A. and Altin, P. A. and Amato, A. and Anand, C. and Anand, S. and Ananyeva, A. and Anderson, S. B. and Anderson, W. G. and Ando, M. and Andrade, T. and Andres, N. and Andrić, T. and Angelova, S. V. and Ansoldi, S. and Antelis, J. M. and Antier, S. and Appert, S. and Arai, Koji and Arai, Koya and Arai, Y. and Araki, S. and Araya, A. and Araya, M. C. and Areeda, J. S. and Arène, M. and Aritomi, N. and Arnaud, N. and Arogeti, M. and Aronson, S. M. and Arun, K. G. and Asada, H. and Asali, Y. and Ashton, G. and Aso, Y. and Assiduo, M. and Aston, S. M. and Astone, P. and Aubin, F. and Austin, C. and Babak, S. and Badaracco, F. and Bader, M. K. M. and Badger, C. and Bae, S. and Bae, Y. and Baer, A. M. and Bagnasco, S. and Bai, Y. and Baiotti, L. and Baird, J. and Bajpai, R. and Ball, M. and Ballardin, G. and Ballmer, S. W. and Balsamo, A. and Baltus, G. and Banagiri, S. and Bankar, D. and Barayoga, J. C. and Barbieri, C. and Barish, B. C. and Barker, D. and Barneo, P. and Barone, F. and Barr, B. and Barsotti, L. and Barsuglia, M. and Barta, D. and Bartlett, J. and Barton, M. A. and Bartos, I. and Bassiri, R. and Basti, A. and Bawaj, M. and Bayley, J. C. and Baylor, A. C. and Bazzan, M. and Bécsy, B. and Bedakihale, V. M. and Bejger, M. and Belahcene, I. and Benedetto, V. and Beniwal, D. and Bennett, T. F. and Bentley, J. D. and BenYaala, M. and Bergamin, F. and Berger, B. K. and Bernuzzi, S. and Berry, C. P. L. and Bersanetti, D. and Bertolini, A. and Betzwieser, J. and Beveridge, D. and Bhandare, R. and Bhardwaj, U. and Bhattacharjee, D. and Bhaumik, S. and Bilenko, I. A. and Billingsley, G. and Bini, S. and Birney, R. and Birnholtz, O. and Biscans, S. and Bischi, M. and Biscoveanu, S. and Bisht, A. and Biswas, B. and Bitossi, M. and Bizouard, M.-A. and Blackburn, J. K. and Blair, C. D. and Blair, D. G. and Blair, R. M. and Bobba, F. and Bode, N. and Boer, M. and Bogaert, G. and Boldrini, M. and Bonavena, L. D. and Bondu, F. and Bonilla, E. and Bonnand, R. and Booker, P. and Boom, B. A. and Bork, R. and Boschi, V. and Bose, N. and Bose, S. and Bossilkov, V. and Boudart, V. and Bouffanais, Y. and Bozzi, A. and Bradaschia, C. and Brady, P. R. and Bramley, A. and Branch, A. and Branchesi, M. and Brandt, J. and Brau, J. E. and Breschi, M. and Briant, T. and Briggs, J. H. and Brillet, A. and Brinkmann, M. and Brockill, P. and Brooks, A. F. and Brooks, J. and Brown, D. D. and Brunett, S. and Bruno, G. and Bruntz, R. and Bryant, J. and Bulik, T. and Bulten, H. J. and Buonanno, A. and Buscicchio, R. and Buskulic, D. and Buy, C. and Byer, R. L. and Davies, G. S. Cabourn and Cadonati, L. and Cagnoli, G. and Cahillane, C. and Bustillo, J. Calderón and Callaghan, J. D. and Callister, T. A. and Calloni, E. and Cameron, J. and Camp, J. B. and Canepa, M. and Canevarolo, S. and Cannavacciuolo, M. and Cannon, K. C. and Cao, H. and Cao, Z. and Capocasa, E. and Capote, E. and Carapella, G. and Carbognani, F. and Carlin, J. B. and Carney, M. F. and Carpinelli, M. and Carrillo, G. and Carullo, G. and Carver, T. L. and Diaz, J. Casanueva and Casentini, C. and Castaldi, G. and Caudill, S. and Cavaglià, M. and Cavalier, F. and Cavalieri, R. and Ceasar, M. and Cella, G. and Cerdá-Durán, P. and Cesarini, E. and Chaibi, W. and Chakravarti, K. and Subrahmanya, S. Chalathadka and Champion, E. and Chan, C.-H. and Chan, C. and Chan, C. L. and Chan, K. and Chan, M. and Chandra, K. and Chanial, P. and Chao, S. and Chapman-Bird, C. E. A. and Charlton, P. and Chase, E. A. and Chassande-Mottin, E. and Chatterjee, C. and Chatterjee, Debarati and Chatterjee, Deep and Chaturvedi, M. and Chaty, S. and Chatziioannou, K. and Chen, C. and Chen, H. Y. and Chen, J. and Chen, K. and Chen, X. and Chen, Y.-B. and Chen, Y.-R. and Chen, Z. and Cheng, H. and Cheong, C. K. and Cheung, H. Y. and Chia, H. Y. and Chiadini, F. and Chiang, C-Y. and Chiarini, G. and Chierici, R. and Chincarini, A. and Chiofalo, M. L. and Chiummo, A. and Cho, G. and Cho, H. S. and Choudhary, R. K. and Choudhary, S. and Christensen, N. and Chu, H. and Chu, Q. and Chu, Y-K. and Chua, S. and Chung, K. W. and Ciani, G. and Ciecielag, P. and Cieślar, M. and Cifaldi, M. and Ciobanu, A. A. and Ciolfi, R. and Cipriano, F. and Cirone, A. and Clara, F. and Clark, E. N. and Clark, J. A. and Clarke, L. and Clearwater, P. and Clesse, S. and Cleva, F. and Coccia, E. and Codazzo, E. and Cohadon, P.-F. and Cohen, D. E. and Cohen, L. and Colleoni, M. and Collette, C. G. and Colombo, A. and Colpi, M. and Compton, C. M. and Constancio, M. and Conti, L. and Cooper, S. J. and Corban, P. and Corbitt, T. R. and Cordero-Carrión, I. and Corezzi, S. and Corley, K. R. and Cornish, N. and Corre, D. and Corsi, A. and Cortese, S. and Costa, C. A. and Cotesta, R. and Coughlin, M. W. and Coulon, J.-P. and Countryman, S. T. and Cousins, B. and Couvares, P. and Coward, D. M. and Cowart, M. J. and Coyne, D. C. and Coyne, R. and Creighton, J. D. E. and Creighton, T. D. and Criswell, A. W. and Croquette, M. and Crowder, S. G. and Cudell, J. R. and Cullen, T. J. and Cumming, A. and Cummings, R. and Cunningham, L. and Cuoco, E. and Curyło, M. and Dabadie, P. and Canton, T. Dal and Dall’Osso, S. and Dálya, G. and Dana, A. and DaneshgaranBajastani, L. M. and D’Angelo, B. and Danila, B. and Danilishin, S. and D’Antonio, S. and Danzmann, K. and Darsow-Fromm, C. and Dasgupta, A. and Datrier, L. E. H. and Dattilo, V. and Dave, I. and Davier, M. and Davis, D. and Davis, M. C. and Daw, E. J. and de Alarcón, P. F. and Dean, R. and DeBra, D. and Deenadayalan, M. and Degallaix, J. and De Laurentis, M. and Deléglise, S. and Del Favero, V. and De Lillo, F. and De Lillo, N. and Del Pozzo, W. and DeMarchi, L. M. and De Matteis, F. and D’Emilio, V. and Demos, N. and Dent, T. and Depasse, A. and De Pietri, R. and De Rosa, R. and De Rossi, C. and DeSalvo, R. and De Simone, R. and Dhurandhar, S. and Díaz, M. C. and Diaz-Ortiz, M. and Didio, N. A. and Dietrich, T. and Di Fiore, L. and Di Fronzo, C. and Di Giorgio, C. and Di Giovanni, F. and Di Giovanni, M. and Di Girolamo, T. and Di Lieto, A. and Ding, B. and Di Pace, S. and Di Palma, I. and Di Renzo, F. and Divakarla, A. K. and Dmitriev, A. and Doctor, Z. and D’Onofrio, L. and Donovan, F. and Dooley, K. L. and Doravari, S. and Dorrington, I. and Drago, M. and Driggers, J. C. and Drori, Y. and Ducoin, J.-G. and Dupej, P. and Durante, O. and D’Urso, D. and Duverne, P.-A. and Dwyer, S. E. and Eassa, C. and Easter, P. J. and Ebersold, M. and Eckhardt, T. and Eddolls, G. and Edelman, B. and Edo, T. B. and Edy, O. and Effler, A. and Eguchi, S. and Eichholz, J. and Eikenberry, S. S. and Eisenmann, M. and Eisenstein, R. A. and Ejlli, A. and Engelby, E. and Enomoto, Y. and Errico, L. and Essick, R. C. and Estellés, H. and Estevez, D. and Etienne, Z. and Etzel, T. and Evans, M. and Evans, T. M. and Ewing, B. E. and Fafone, V. and Fair, H. and Fairhurst, S. and Farah, A. M. and Farinon, S. and Farr, B. and Farr, W. M. and Farrow, N. W. and Fauchon-Jones, E. J. and Favaro, G. and Favata, M. and Fays, M. and Fazio, M. and Feicht, J. and Fejer, M. M. and Fenyvesi, E. and Ferguson, D. L. and Fernandez-Galiana, A. and Ferrante, I. and Ferreira, T. A. and Fidecaro, F. and Figura, P. and Fiori, I. and Fishbach, M. and Fisher, R. P. and Fittipaldi, R. and Fiumara, V. and Flaminio, R. and Floden, E. and Fong, H. and Font, J. A. and Fornal, B. and Forsyth, P. W. F. and Franke, A. and Frasca, S. and Frasconi, F. and Frederick, C. and Freed, J. P. and Frei, Z. and Freise, A. and Frey, R. and Fritschel, P. and Frolov, V. V. and Fronzé, G. G. and Fujii, Y. and Fujikawa, Y. and Fukunaga, M. and Fukushima, M. and Fulda, P. and Fyffe, M. and Gabbard, H. A. and Gabella, W. E. and Gadre, B. U. and Gair, J. R. and Gais, J. and Galaudage, S. and Gamba, R. and Ganapathy, D. and Ganguly, A. and Gao, D. and Gaonkar, S. G. and Garaventa, B. and García, F. and García-Núñez, C. and García-Quirós, C. and Garufi, F. and Gateley, B. and Gaudio, S. and Gayathri, V. and Ge, G.-G. and Gemme, G. and Gennai, A. and George, J. and George, R. N. and Gerberding, O. and Gergely, L. and Gewecke, P. and Ghonge, S. and Ghosh, Abhirup and Ghosh, Archisman and Ghosh, Shaon and Ghosh, Shrobana and Giacomazzo, B. and Giacoppo, L. and Giaime, J. A. and Giardina, K. D. and Gibson, D. R. and Gier, C. and Giesler, M. and Giri, P. and Gissi, F. and Glanzer, J. and Gleckl, A. E. and Godwin, P. and Goetz, E. and Goetz, R. and Gohlke, N. and Golomb, J. and Goncharov, B. and González, G. and Gopakumar, A. and Gosselin, M. and Gouaty, R. and Gould, D. W. and Grace, B. and Grado, A. and Granata, M. and Granata, V. and Grant, A. and Gras, S. and Grassia, P. and Gray, C. and Gray, R. and Greco, G. and Green, A. C. and Green, R. and Gretarsson, A. M. and Gretarsson, E. M. and Griffith, D. and Griffiths, W. and Griggs, H. L. and Grignani, G. and Grimaldi, A. and Grimm, S. J. and Grote, H. and Grunewald, S. and Gruning, P. and Guerra, D. and Guidi, G. M. and Guimaraes, A. R. and Guixé, G. and Gulati, H. K. and Guo, H.-K. and Guo, Y. and Gupta, Anchal and Gupta, Anuradha and Gupta, P. and Gustafson, E. K. and Gustafson, R. and Guzman, F. and Ha, S. and Haegel, L. and Hagiwara, A. and Haino, S. and Halim, O. and Hall, E. D. and Hamilton, E. Z. and Hammond, G. and Han, W.-B. and Haney, M. and Hanks, J. and Hanna, C. and Hannam, M. D. and Hannuksela, O. and Hansen, H. and Hansen, T. J. and Hanson, J. and Harder, T. and Hardwick, T. and Haris, K. and Harms, J. and Harry, G. M. and Harry, I. W. and Hartwig, D. and Hasegawa, K. and Haskell, B. and Hasskew, R. K. and Haster, C.-J. and Hattori, K. and Haughian, K. and Hayakawa, H. and Hayama, K. and Hayes, F. J. and Healy, J. and Heidmann, A. and Heidt, A. and Heintze, M. C. and Heinze, J. and Heinzel, J. and Heitmann, H. and Hellman, F. and Hello, P. and Helmling-Cornell, A. F. and Hemming, G. and Hendry, M. and Heng, I. S. and Hennes, E. and Hennig, J. and Hennig, M. H. and Hernandez, A. G. and Hernandez Vivanco, F. and Heurs, M. and Hild, S. and Hill, P. and Himemoto, Y. and Hines, A. S. and Hiranuma, Y. and Hirata, N. and Hirose, E. and Hochheim, S. and Hofman, D. and Hohmann, J. N. and Holcomb, D. G. and Holland, N. A. and Holley-Bockelmann, K. and Hollows, I. J. and Holmes, Z. J. and Holt, K. and Holz, D. E. and Hong, Z. and Hopkins, P. and Hough, J. and Hourihane, S. and Howell, E. J. and Hoy, C. G. and Hoyland, D. and Hreibi, A. and Hsieh, B-H. and Hsu, Y. and Huang, G-Z. and Huang, H-Y. and Huang, P. and Huang, Y-C. and Huang, Y.-J. and Huang, Y. and Hübner, M. T. and Huddart, A. D. and Hughey, B. and Hui, D. C. Y. and Hui, V. and Husa, S. and Huttner, S. H. and Huxford, R. and Huynh-Dinh, T. and Ide, S. and Idzkowski, B. and Iess, A. and Ikenoue, B. and Imam, S. and Inayoshi, K. and Ingram, C. and Inoue, Y. and Ioka, K. and Isi, M. and Isleif, K. and Ito, K. and Itoh, Y. and Iyer, B. R. and Izumi, K. and JaberianHamedan, V. and Jacqmin, T. and Jadhav, S. J. and Jadhav, S. P. and James, A. L. and Jan, A. Z. and Jani, K. and Janquart, J. and Janssens, K. and Janthalur, N. N. and Jaranowski, P. and Jariwala, D. and Jaume, R. and Jenkins, A. C. and Jenner, K. and Jeon, C. and Jeunon, M. and Jia, W. and Jin, H.-B. and Johns, G. R. and Johnson-McDaniel, N. K. and Jones, A. W. and Jones, D. I. and Jones, J. D. and Jones, P. and Jones, R. and Jonker, R. J. G. and Ju, L. and Jung, P. and Jung, K. and Junker, J. and Juste, V. and Kaihotsu, K. and Kajita, T. and Kakizaki, M. and Kalaghatgi, C. V. and Kalogera, V. and Kamai, B. and Kamiizumi, M. and Kanda, N. and Kandhasamy, S. and Kang, G. and Kanner, J. B. and Kao, Y. and Kapadia, S. J. and Kapasi, D. P. and Karat, S. and Karathanasis, C. and Karki, S. and Kashyap, R. and Kasprzack, M. and Kastaun, W. and Katsanevas, S. and Katsavounidis, E. and Katzman, W. and Kaur, T. and Kawabe, K. and Kawaguchi, K. and Kawai, N. and Kawasaki, T. and Kéfélian, F. and Keitel, D. and Key, J. S. and Khadka, S. and Khalili, F. Y. and Khan, S. and Khazanov, E. A. and Khetan, N. and Khursheed, M. and Kijbunchoo, N. and Kim, C. and Kim, J. C. and Kim, J. and Kim, K. and Kim, W. S. and Kim, Y.-M. and Kimball, C. and Kimura, N. and Kinley-Hanlon, M. and Kirchhoff, R. and Kissel, J. S. and Kita, N. and Kitazawa, H. and Kleybolte, L. and Klimenko, S. and Knee, A. M. and Knowles, T. D. and Knyazev, E. and Koch, P. and Koekoek, G. and Kojima, Y. and Kokeyama, K. and Koley, S. and Kolitsidou, P. and Kolstein, M. and Komori, K. and Kondrashov, V. and Kong, A. K. H. and Kontos, A. and Koper, N. and Korobko, M. and Kotake, K. and Kovalam, M. and Kozak, D. B. and Kozakai, C. and Kozu, R. and Kringel, V. and Krishnendu, N. V. and Królak, A. and Kuehn, G. and Kuei, F. and Kuijer, P. and Kulkarni, S. and Kumar, A. and Kumar, P. and Kumar, Rahul and Kumar, Rakesh and Kume, J. and Kuns, K. and Kuo, C. and Kuo, H-S. and Kuromiya, Y. and Kuroyanagi, S. and Kusayanagi, K. and Kuwahara, S. and Kwak, K. and Lagabbe, P. and Laghi, D. and Lalande, E. and Lam, T. L. and Lamberts, A. and Landry, M. and Lane, B. B. and Lang, R. N. and Lange, J. and Lantz, B. and La Rosa, I. and Lartaux-Vollard, A. and Lasky, P. D. and Laxen, M. and Lazzarini, A. and Lazzaro, C. and Leaci, P. and Leavey, S. and Lecoeuche, Y. K. and Lee, H. K. and Lee, H. M. and Lee, H. W. and Lee, J. and Lee, K. and Lee, R. and Lehmann, J. and Lemaître, A. and Leonardi, M. and Leroy, N. and Letendre, N. and Levesque, C. and Levin, Y. and Leviton, J. N. and Leyde, K. and Li, A. K. Y. and Li, B. and Li, J. and Li, K. L. and Li, T. G. F. and Li, X. and Lin, C-Y. and Lin, F-K. and Lin, F-L. and Lin, H. L. and Lin, L. C.-C. and Linde, F. and Linker, S. D. and Linley, J. N. and Littenberg, T. B. and Liu, G. C. and Liu, J. and Liu, K. and Liu, X. and Llamas, F. and Llorens-Monteagudo, M. and Lo, R. K. L. and Lockwood, A. and Loh, M. and London, L. T. and Longo, A. and Lopez, D. and Portilla, M. Lopez and Lorenzini, M. and Loriette, V. and Lormand, M. and Losurdo, G. and Lott, T. P. and Lough, J. D. and Lousto, C. O. and Lovelace, G. and Lucaccioni, J. F. and Lück, H. and Lumaca, D. and Lundgren, A. P. and Luo, L.-W. and Lynam, J. E. and Macas, R. and MacInnis, M. and Macleod, D. M. and MacMillan, I. A. O. and Macquet, A. and Hernandez, I. Magaña and Magazzù, C. and Magee, R. M. and Maggiore, R. and Magnozzi, M. and Mahesh, S. and Majorana, E. and Makarem, C. and Maksimovic, I. and Maliakal, S. and Malik, A. and Man, N. and Mandic, V. and Mangano, V. and Mango, J. L. and Mansell, G. L. and Manske, M. and Mantovani, M. and Mapelli, M. and Marchesoni, F. and Marchio, M. and Marion, F. and Mark, Z. and Márka, S. and Márka, Z. and Markakis, C. and Markosyan, A. S. and Markowitz, A. and Maros, E. and Marquina, A. and Marsat, S. and Martelli, F. and Martin, I. W. and Martin, R. M. and Martinez, M. and Martinez, V. A. and Martinez, V. and Martinovic, K. and Martynov, D. V. and Marx, E. J. and Masalehdan, H. and Mason, K. and Massera, E. and Masserot, A. and Massinger, T. J. and Masso-Reid, M. and Mastrogiovanni, S. and Matas, A. and Mateu-Lucena, M. and Matichard, F. and Matiushechkina, M. and Mavalvala, N. and McCann, J. J. and McCarthy, R. and McClelland, D. E. and McClincy, P. K. and McCormick, S. and McCuller, L. and McGhee, G. I. and McGuire, S. C. and McIsaac, C. and McIver, J. and McRae, T. and McWilliams, S. T. and Meacher, D. and Mehmet, M. and Mehta, A. K. and Meijer, Q. and Melatos, A. and Melchor, D. A. and Mendell, G. and Menendez-Vazquez, A. and Menoni, C. S. and Mercer, R. A. and Mereni, L. and Merfeld, K. and Merilh, E. L. and Merritt, J. D. and Merzougui, M. and Meshkov, S. and Messenger, C. and Messick, C. and Meyers, P. M. and Meylahn, F. and Mhaske, A. and Miani, A. and Miao, H. and Michaloliakos, I. and Michel, C. and Michimura, Y. and Middleton, H. and Milano, L. and Miller, A. L. and Miller, A. and Miller, B. and Millhouse, M. and Mills, J. C. and Milotti, E. and Minazzoli, O. and Minenkov, Y. and Mio, N. and Mir, Ll. M. and Miravet-Tenés, M. and Mishra, C. and Mishra, T. and Mistry, T. and Mitra, S. and Mitrofanov, V. P. and Mitselmakher, G. and Mittleman, R. and Miyakawa, O. and Miyamoto, A. and Miyazaki, Y. and Miyo, K. and Miyoki, S. and Mo, Geoffrey and Modafferi, L. M. and Moguel, E. and Mogushi, K. and Mohapatra, S. R. P. and Mohite, S. R. and Molina, I. and Molina-Ruiz, M. and Mondin, M. and Montani, M. and Moore, C. J. and Moraru, D. and Morawski, F. and More, A. and Moreno, C. and Moreno, G. and Mori, Y. and Morisaki, S. and Moriwaki, Y. and Morrás, G. and Mours, B. and Mow-Lowry, C. M. and Mozzon, S. and Muciaccia, F. and Mukherjee, Arunava and Mukherjee, D. and Mukherjee, Soma and Mukherjee, Subroto and Mukherjee, Suvodip and Mukund, N. and Mullavey, A. and Munch, J. and Muñiz, E. A. and Murray, P. G. and Musenich, R. and Muusse, S. and Nadji, S. L. and Nagano, K. and Nagano, S. and Nagar, A. and Nakamura, K. and Nakano, H. and Nakano, M. and Nakashima, R. and Nakayama, Y. and Napolano, V. and Nardecchia, I. and Narikawa, T. and Naticchioni, L. and Nayak, B. and Nayak, R. K. and Negishi, R. and Neil, B. F. and Neilson, J. and Nelemans, G. and Nelson, T. J. N. and Nery, M. and Neubauer, P. and Neunzert, A. and Ng, K. Y. and Ng, S. W. S. and Nguyen, C. and Nguyen, P. and Nguyen, T. and Quynh, L. Nguyen and Ni, W.-T. and Nichols, S. A. and Nishizawa, A. and Nissanke, S. and Nitoglia, E. and Nocera, F. and Norman, M. and North, C. and Nozaki, S. and Siles, J. F. Nuño and Nuttall, L. K. and Oberling, J. and O’Brien, B. D. and Obuchi, Y. and O’Dell, J. and Oelker, E. and Ogaki, W. and Oganesyan, G. and Oh, J. J. and Oh, K. and Oh, S. H. and Ohashi, M. and Ohishi, N. and Ohkawa, M. and Ohme, F. and Ohta, H. and Okada, M. A. and Okutani, Y. and Okutomi, K. and Olivetto, C. and Oohara, K. and Ooi, C. and Oram, R. and O’Reilly, B. and Ormiston, R. G. and Ormsby, N. D. and Ortega, L. F. and O’Shaughnessy, R. and O’Shea, E. and Oshino, S. and Ossokine, S. and Osthelder, C. and Otabe, S. and Ottaway, D. J. and Overmier, H. and Pace, A. E. and Pagano, G. and Page, M. A. and Pagliaroli, G. and Pai, A. and Pai, S. A. and Palamos, J. R. and Palashov, O. and Palomba, C. and Pan, H. and Pan, K. and Panda, P. K. and Pang, H. and Pang, P. T. H. and Pankow, C. and Pannarale, F. and Pant, B. C. and Panther, F. H. and Paoletti, F. and Paoli, A. and Paolone, A. and Parisi, A. and Park, H. and Park, J. and Parker, W. and Pascucci, D. and Pasqualetti, A. and Passaquieti, R. and Passuello, D. and Patel, M. and Pathak, M. and Patricelli, B. and Patron, A. S. and Paul, S. and Payne, E. and Pedraza, M. and Pegoraro, M. and Pele, A. and Arellano, F. E. Peña and Penn, S. and Perego, A. and Pereira, A. and Pereira, T. and Perez, C. J. and Périgois, C. and Perkins, C. C. and Perreca, A. and Perriès, S. and Petermann, J. and Petterson, D. and Pfeiffer, H. P. and Pham, K. A. and Phukon, K. S. and Piccinni, O. J. and Pichot, M. and Piendibene, M. and Piergiovanni, F. and Pierini, L. and Pierro, V. and Pillant, G. and Pillas, M. and Pilo, F. and Pinard, L. and Pinto, I. M. and Pinto, M. and Piotrzkowski, B. and Piotrzkowski, K. and Pirello, M. and Pitkin, M. D. and Placidi, E. and Planas, L. and Plastino, W. and Pluchar, C. and Poggiani, R. and Polini, E. and Pong, D. Y. T. and Ponrathnam, S. and Popolizio, P. and Porter, E. K. and Poulton, R. and Powell, J. and Pracchia, M. and Pradier, T. and Prajapati, A. K. and Prasai, K. and Prasanna, R. and Pratten, G. and Principe, M. and Prodi, G. A. and Prokhorov, L. and Prosposito, P. and Prudenzi, L. and Puecher, A. and Punturo, M. and Puosi, F. and Puppo, P. and Pürrer, M. and Qi, H. and Quetschke, V. and Quitzow-James, R. and Qutob, N. and Raab, F. J. and Raaijmakers, G. and Radkins, H. and Radulesco, N. and Raffai, P. and Rail, S. X. and Raja, S. and Rajan, C. and Ramirez, K. E. and Ramirez, T. D. and Ramos-Buades, A. and Rana, J. and Rapagnani, P. and Rapol, U. D. and Ray, A. and Raymond, V. and Raza, N. and Razzano, M. and Read, J. and Rees, L. A. and Regimbau, T. and Rei, L. and Reid, S. and Reid, S. W. and Reitze, D. H. and Relton, P. and Renzini, A. and Rettegno, P. and Reza, A. and Rezac, M. and Ricci, F. and Richards, D. and Richardson, J. W. and Richardson, L. and Riemenschneider, G. and Riles, K. and Rinaldi, S. and Rink, K. and Rizzo, M. and Robertson, N. A. and Robie, R. and Robinet, F. and Rocchi, A. and Rodriguez, S. and Rolland, L. and Rollins, J. G. and Romanelli, M. and Romano, R. and Romel, C. L. and Romero-Rodríguez, A. and Romero-Shaw, I. M. and Romie, J. H. and Ronchini, S. and Rosa, L. and Rose, C. A. and Rosińska, D. and Ross, M. P. and Rowan, S. and Rowlinson, S. J. and Roy, S. and Roy, Santosh and Roy, Soumen and Rozza, D. and Ruggi, P. and Ruiz-Rocha, K. and Ryan, K. and Sachdev, S. and Sadecki, T. and Sadiq, J. and Sago, N. and Saito, S. and Saito, Y. and Sakai, K. and Sakai, Y. and Sakellariadou, M. and Sakuno, Y. and Salafia, O. S. and Salconi, L. and Saleem, M. and Salemi, F. and Samajdar, A. and Sanchez, E. J. and Sanchez, J. H. and Sanchez, L. E. and Sanchis-Gual, N. and Sanders, J. R. and Sanuy, A. and Saravanan, T. R. and Sarin, N. and Sassolas, B. and Satari, H. and Sathyaprakash, B. S. and Sato, S. and Sato, T. and Sauter, O. and Savage, R. L. and Sawada, T. and Sawant, D. and Sawant, H. L. and Sayah, S. and Schaetzl, D. and Scheel, M. and Scheuer, J. and Schiworski, M. and Schmidt, P. and Schmidt, S. and Schnabel, R. and Schneewind, M. and Schofield, R. M. S. and Schönbeck, A. and Schulte, B. W. and Schutz, B. F. and Schwartz, E. and Scott, J. and Scott, S. M. and Seglar-Arroyo, M. and Sekiguchi, T. and Sekiguchi, Y. and Sellers, D. and Sengupta, A. S. and Sentenac, D. and Seo, E. G. and Sequino, V. and Sergeev, A. and Setyawati, Y. and Shaffer, T. and Shahriar, M. S. and Shams, B. and Shao, L. and Sharma, A. and Sharma, P. and Shawhan, P. and Shcheblanov, N. S. and Shibagaki, S. and Shikauchi, M. and Shimizu, R. and Shimoda, T. and Shimode, K. and Shinkai, H. and Shishido, T. and Shoda, A. and Shoemaker, D. H. and Shoemaker, D. M. and ShyamSundar, S. and Sieniawska, M. and Sigg, D. and Singer, L. P. and Singh, D. and Singh, N. and Singha, A. and Sintes, A. M. and Sipala, V. and Skliris, V. and Slagmolen, B. J. J. and Slaven-Blair, T. J. and Smetana, J. and Smith, J. R. and Smith, R. J. E. and Soldateschi, J. and Somala, S. N. and Somiya, K. and Son, E. J. and Soni, K. and Soni, S. and Sordini, V. and Sorrentino, F. and Sorrentino, N. and Sotani, H. and Soulard, R. and Souradeep, T. and Sowell, E. and Spagnuolo, V. and Spencer, A. P. and Spera, M. and Srinivasan, R. and Srivastava, A. K. and Srivastava, V. and Staats, K. and Stachie, C. and Steer, D. A. and Steinhoff, J. and Steinlechner, J. and Steinlechner, S. and Stevenson, S. P. and Stops, D. J. and Stover, M. and Strain, K. A. and Strang, L. C. and Stratta, G. and Strunk, A. and Sturani, R. and Stuver, A. L. and Sudhagar, S. and Sudhir, V. and Sugimoto, R. and Suh, H. G. and Sullivan, A. G. and Sullivan, J. M. and Summerscales, T. Z. and Sun, H. and Sun, L. and Sunil, S. and Sur, A. and Suresh, J. and Sutton, P. J. and Suzuki, Takamasa and Suzuki, Toshikazu and Swinkels, B. L. and Szczepańczyk, M. J. and Szewczyk, P. and Tacca, M. and Tagoshi, H. and Tait, S. C. and Takahashi, H. and Takahashi, R. and Takamori, A. and Takano, S. and Takeda, H. and Takeda, M. and Talbot, C. J. and Talbot, C. and Tanaka, H. and Tanaka, Kazuyuki and Tanaka, Kenta and Tanaka, Taiki and Tanaka, Takahiro and Tanasijczuk, A. J. and Tanioka, S. and Tanner, D. B. and Tao, D. and Tao, L. and Martín, E. N. Tapia San and Taranto, C. and Tasson, J. D. and Telada, S. and Tenorio, R. and Terhune, J. E. and Terkowski, L. and Thirugnanasambandam, M. P. and Thomas, L. and Thomas, M. and Thomas, P. and Thompson, J. E. and Thondapu, S. R. and Thorne, K. A. and Thrane, E. and Tiwari, Shubhanshu and Tiwari, Srishti and Tiwari, V. and Toivonen, A. M. and Toland, K. and Tolley, A. E. and Tomaru, T. and Tomigami, Y. and Tomura, T. and Tonelli, M. and Torres-Forné, A. and Torrie, C. I. and e Melo, I. Tosta and Töyrä, D. and Trapananti, A. and Travasso, F. and Traylor, G. and Trevor, M. and Tringali, M. C. and Tripathee, A. and Troiano, L. and Trovato, A. and Trozzo, L. and Trudeau, R. J. and Tsai, D. S. and Tsai, D. and Tsang, K. W. and Tsang, T. and Tsao, J-S. and Tse, M. and Tso, R. and Tsubono, K. and Tsuchida, S. and Tsukada, L. and Tsuna, D. and Tsutsui, T. and Tsuzuki, T. and Turbang, K. and Turconi, M. and Tuyenbayev, D. and Ubhi, A. S. and Uchikata, N. and Uchiyama, T. and Udall, R. P. and Ueda, A. and Uehara, T. and Ueno, K. and Ueshima, G. and Unnikrishnan, C. S. and Uraguchi, F. and Urban, A. L. and Ushiba, T. and Utina, A. and Vahlbruch, H. and Vajente, G. and Vajpeyi, A. and Valdes, G. and Valentini, M. and Valsan, V. and van Bakel, N. and van Beuzekom, M. and van den Brand, J. F. J. and Van Den Broeck, C. and Vander-Hyde, D. C. and van der Schaaf, L. and van Heijningen, J. V. and Vanosky, J. and van Putten, M. H. P. M. and van Remortel, N. and Vardaro, M. and Vargas, A. F. and Varma, V. and Vasúth, M. and Vecchio, A. and Vedovato, G. and Veitch, J. and Veitch, P. J. and Venneberg, J. and Venugopalan, G. and Verkindt, D. and Verma, P. and Verma, Y. and Veske, D. and Vetrano, F. and Viceré, A. and Vidyant, S. and Viets, A. D. and Vijaykumar, A. and Villa-Ortega, V. and Vinet, J.-Y. and Virtuoso, A. and Vitale, S. and Vo, T. and Vocca, H. and von Reis, E. R. G. and von Wrangel, J. S. A. and Vorvick, C. and Vyatchanin, S. P. and Wade, L. E. and Wade, M. and Wagner, K. J. and Walet, R. C. and Walker, M. and Wallace, G. S. and Wallace, L. and Walsh, S. and Wang, J. and Wang, J. Z. and Wang, W. H. and Ward, R. L. and Warner, J. and Was, M. and Washimi, T. and Washington, N. Y. and Watchi, J. and Weaver, B. and Webster, S. A. and Weinert, M. and Weinstein, A. J. and Weiss, R. and Weller, C. M. and Weller, R. A. and Wellmann, F. and Wen, L. and Weßels, P. and Wette, K. and Whelan, J. T. and White, D. D. and Whiting, B. F. and Whittle, C. and Wilken, D. and Williams, D. and Williams, M. J. and Williams, N. and Williamson, A. R. and Willis, J. L. and Willke, B. and Wilson, D. J. and Winkler, W. and Wipf, C. C. and Wlodarczyk, T. and Woan, G. and Woehler, J. and Wofford, J. K. and Wong, I. C. F. and Wu, C. and Wu, D. S. and Wu, H. and Wu, S. and Wysocki, D. M. and Xiao, L. and Xu, W-R. and Yamada, T. and Yamamoto, H. and Yamamoto, Kazuhiro and Yamamoto, Kohei and Yamamoto, T. and Yamashita, K. and Yamazaki, R. and Yang, F. W. and Yang, L. and Yang, Y. and Yang, Yang and Yang, Z. and Yap, M. J. and Yeeles, D. W. and Yelikar, A. B. and Ying, M. and Yokogawa, K. and Yokoyama, J. and Yokozawa, T. and Yoo, J. and Yoshioka, T. and Yu, Hang and Yu, Haocun and Yuzurihara, H. and Zadrożny, A and Zanolin, M. and Zeidler, S. and Zelenova, T. and Zendri, J.-P. and Zevin, M. and Zhan, M. and Zhang, H. and Zhang, J. and Zhang, L. and Zhang, T. and Zhang, Y. and Zhao, C. and Zhao, G. and Zhao, Y. and Zhao, Yue and Zheng, Y. and Zhou, R. and Zhou, Z. and Zhu, X. J. and Zhu, Z.-H. and Zimmerman, A. B. and Zlochower, Y. and Zucker, M. E. and Zweizig, J.},
   year={2023},
   month=dec }

@article{Blanchet_2008,
   title={The third post-Newtonian gravitational wave polarizations and associated spherical harmonic modes for inspiralling compact binaries in quasi-circular orbits},
   volume={25},
   ISSN={1361-6382},
   url={http://dx.doi.org/10.1088/0264-9381/25/16/165003},
   DOI={10.1088/0264-9381/25/16/165003},
   number={16},
   journal={Classical and Quantum Gravity},
   publisher={IOP Publishing},
   author={Blanchet, Luc and Faye, Guillaume and Iyer, Bala R and Sinha, Siddhartha},
   year={2008},
   month=jul, pages={165003} }

@article{seobnrv4,
   title={Improved effective-one-body model of spinning, nonprecessing binary black holes for the era of gravitational-wave astrophysics with advanced detectors},
   volume={95},
   ISSN={2470-0029},
   url={http://dx.doi.org/10.1103/PhysRevD.95.044028},
   DOI={10.1103/physrevd.95.044028},
   number={4},
   journal={Physical Review D},
   publisher={American Physical Society (APS)},
   author={Bohé, Alejandro and Shao, Lijing and Taracchini, Andrea and Buonanno, Alessandra and Babak, Stanislav and Harry, Ian W. and Hinder, Ian and Ossokine, Serguei and Pürrer, Michael and Raymond, Vivien and Chu, Tony and Fong, Heather and Kumar, Prayush and Pfeiffer, Harald P. and Boyle, Michael and Hemberger, Daniel A. and Kidder, Lawrence E. and Lovelace, Geoffrey and Scheel, Mark A. and Szilágyi, Béla},
   year={2017},
   month=feb }

@article{Buonanno_EOB_1999,
   title={Effective one-body approach to general relativistic two-body dynamics},
   volume={59},
   ISSN={1089-4918},
   url={http://dx.doi.org/10.1103/PhysRevD.59.084006},
   DOI={10.1103/physrevd.59.084006},
   number={8},
   journal={Physical Review D},
   publisher={American Physical Society (APS)},
   author={Buonanno, A. and Damour, T.},
   year={1999},
   month=mar }

@article{Pan_SEOBNRv1_2011,
   title={Inspiral-merger-ringdown multipolar waveforms of nonspinning black-hole binaries using the effective-one-body formalism},
   volume={84},
   ISSN={1550-2368},
   url={http://dx.doi.org/10.1103/PhysRevD.84.124052},
   DOI={10.1103/physrevd.84.124052},
   number={12},
   journal={Physical Review D},
   publisher={American Physical Society (APS)},
   author={Pan, Yi and Buonanno, Alessandra and Boyle, Michael and Buchman, Luisa T. and Kidder, Lawrence E. and Pfeiffer, Harald P. and Scheel, Mark A.},
   year={2011},
   month=dec }

@article{RevModPhys.52.299,
  title = {Multipole expansions of gravitational radiation},
  author = {Thorne, Kip S.},
  journal = {Rev. Mod. Phys.},
  volume = {52},
  issue = {2},
  pages = {299--339},
  numpages = {0},
  year = {1980},
  month = {Apr},
  publisher = {American Physical Society},
  doi = {10.1103/RevModPhys.52.299},
  url = {https://link.aps.org/doi/10.1103/RevModPhys.52.299}
}

@article{Boetzel_2017_PNKeplerEQ,
  title = {Solving post-Newtonian accurate Kepler equation},
  author = {Boetzel, Yannick and Susobhanan, Abhimanyu and Gopakumar, Achamveedu and Klein, Antoine and Jetzer, Philippe},
  journal = {Phys. Rev. D},
  volume = {96},
  issue = {4},
  pages = {044011},
  numpages = {22},
  year = {2017},
  month = {Aug},
  publisher = {American Physical Society},
  doi = {10.1103/PhysRevD.96.044011},
  url = {https://link.aps.org/doi/10.1103/PhysRevD.96.044011}
}

@article{Mishra_2015,
   title={Third post-Newtonian gravitational waveforms for compact binary systems in general orbits: Instantaneous terms},
   volume={91},
   ISSN={1550-2368},
   url={http://dx.doi.org/10.1103/PhysRevD.91.084040},
   DOI={10.1103/physrevd.91.084040},
   number={8},
   journal={Physical Review D},
   publisher={American Physical Society (APS)},
   author={Mishra, Chandra Kant and Arun, K. G. and Iyer, Bala R.},
   year={2015},
   month=apr }

@article{Memmesheimer_2004_QK,
  title = {Third post-Newtonian accurate generalized quasi-Keplerian parametrization for compact binaries in eccentric orbits},
  author = {Memmesheimer, Raoul-Martin and Gopakumar, Achamveedu and Sch\"afer, Gerhard},
  journal = {Phys. Rev. D},
  volume = {70},
  issue = {10},
  pages = {104011},
  numpages = {17},
  year = {2004},
  month = {Nov},
  publisher = {American Physical Society},
  doi = {10.1103/PhysRevD.70.104011},
  url = {https://link.aps.org/doi/10.1103/PhysRevD.70.104011}
}

@article{Damour:1981ntn,
    author = "Damour, Thibault and Deruelle, Nathalie",
    title = "{Generalized lagrangian of two point masses in the post-post newtonian approximation of general relativity}",
    journal = "Comptes Rendus des Seances de l'Academie des Sciences. Serie 2",
    volume = "293",
    number = "8",
    pages = "537--540",
    year = "1981"
}

@article{Bernard_2016,
  title = {Fokker action of nonspinning compact binaries at the fourth post-Newtonian approximation},
  author = {Bernard, Laura and Blanchet, Luc and Boh\'e, Alejandro and Faye, Guillaume and Marsat, Sylvain},
  journal = {Phys. Rev. D},
  volume = {93},
  issue = {8},
  pages = {084037},
  numpages = {33},
  year = {2016},
  month = {Apr},
  publisher = {American Physical Society},
  doi = {10.1103/PhysRevD.93.084037},
  url = {https://link.aps.org/doi/10.1103/PhysRevD.93.084037}
}

@article{Andrade_2001,
   title={Third post-Newtonian dynamics of compact binaries: Noetherian conserved quantities and equivalence between the harmonic-coordinate and ADM-Hamiltonian formalisms},
   volume={18},
   ISSN={1361-6382},
   url={http://dx.doi.org/10.1088/0264-9381/18/5/301},
   DOI={10.1088/0264-9381/18/5/301},
   number={5},
   journal={Classical and Quantum Gravity},
   publisher={IOP Publishing},
   author={Andrade, Vanessa C de and Blanchet, Luc and Faye, Guillaume},
   year={2001},
   month=feb, pages={753–778} }

@article{Bernard_2018,
  title = {Center-of-mass equations of motion and conserved integrals of compact binary systems at the fourth post-Newtonian order},
  author = {Bernard, Laura and Blanchet, Luc and Faye, Guillaume and Marchand, Tanguy},
  journal = {Phys. Rev. D},
  volume = {97},
  issue = {4},
  pages = {044037},
  numpages = {30},
  year = {2018},
  month = {Feb},
  publisher = {American Physical Society},
  doi = {10.1103/PhysRevD.97.044037},
  url = {https://link.aps.org/doi/10.1103/PhysRevD.97.044037}
}

@article{Damour:1990jh,
    author = "Damour, Thibault and Schaefer, Gerhard",
    title = "{Redefinition of position variables and the reduction of higher order Lagrangians}",
    reportNumber = "IHES/P/90/32",
    doi = "10.1063/1.529135",
    journal = "J. Math. Phys.",
    volume = "32",
    pages = "127--134",
    year = "1991"
}

@article{Damour_2014_nonlocal,
  title = {Nonlocal-in-time action for the fourth post-Newtonian conservative dynamics of two-body systems},
  author = {Damour, Thibault and Jaranowski, Piotr and Sch\"afer, Gerhard},
  journal = {Phys. Rev. D},
  volume = {89},
  issue = {6},
  pages = {064058},
  numpages = {17},
  year = {2014},
  month = {Mar},
  publisher = {American Physical Society},
  doi = {10.1103/PhysRevD.89.064058},
  url = {https://link.aps.org/doi/10.1103/PhysRevD.89.064058}
}

@article{Damour_2015_4PNEOB,
   title={Fourth post-Newtonian effective one-body dynamics},
   volume={91},
   ISSN={1550-2368},
   url={http://dx.doi.org/10.1103/PhysRevD.91.084024},
   DOI={10.1103/physrevd.91.084024},
   number={8},
   journal={Physical Review D},
   publisher={American Physical Society (APS)},
   author={Damour, Thibault and Jaranowski, Piotr and Schäfer, Gerhard},
   year={2015},
   month=apr }

@article{Jaranowski_2013,
   title={Dimensional regularization of local singularities in the fourth post-Newtonian two-point-mass Hamiltonian},
   volume={87},
   ISSN={1550-2368},
   url={http://dx.doi.org/10.1103/PhysRevD.87.081503},
   DOI={10.1103/physrevd.87.081503},
   number={8},
   journal={Physical Review D},
   publisher={American Physical Society (APS)},
   author={Jaranowski, Piotr and Schäfer, Gerhard},
   year={2013},
   month=apr }

@article{Jaranowski_2015,
   title={Derivation of local-in-time fourth post-Newtonian ADM Hamiltonian for spinless compact binaries},
   volume={92},
   ISSN={1550-2368},
   url={http://dx.doi.org/10.1103/PhysRevD.92.124043},
   DOI={10.1103/physrevd.92.124043},
   number={12},
   journal={Physical Review D},
   publisher={American Physical Society (APS)},
   author={Jaranowski, Piotr and Schäfer, Gerhard},
   year={2015},
   month=dec }

@article{Bini_2020_5PNEOB,
   title={Binary dynamics at the fifth and fifth-and-a-half post-Newtonian orders},
   volume={102},
   ISSN={2470-0029},
   url={http://dx.doi.org/10.1103/PhysRevD.102.024062},
   DOI={10.1103/physrevd.102.024062},
   number={2},
   journal={Physical Review D},
   publisher={American Physical Society (APS)},
   author={Bini, Donato and Damour, Thibault and Geralico, Andrea},
   year={2020},
   month=jul }

@misc{khalil2023_5PNEOB,
      title={Theoretical groundwork supporting the precessing-spin two-body dynamics of the effective-one-body waveform models SEOBNRv5}, 
      author={Mohammed Khalil and Alessandra Buonanno and Héctor Estellés and Deyan P. Mihaylov and Serguei Ossokine and Lorenzo Pompili and Antoni Ramos-Buades},
      year={2023},
      eprint={2303.18143},
      archivePrefix={arXiv},
      primaryClass={gr-qc},
      url={https://arxiv.org/abs/2303.18143}, 
}

@book{hypergeometric_function,
   title =     {Theory of Hypergeometric Functions},
   author =    {Kazuhiko Aomoto, Michitake Kita},
   publisher = {Springer Tokyo},
   year =      {2011},
   edition =   {1},
   url =       {https://doi.org/10.1007/978-4-431-53938-4}
}

@article{Blanchet_1998,
   title={On the multipole expansion of the gravitational field},
   volume={15},
   ISSN={1361-6382},
   url={http://dx.doi.org/10.1088/0264-9381/15/7/013},
   DOI={10.1088/0264-9381/15/7/013},
   number={7},
   journal={Classical and Quantum Gravity},
   publisher={IOP Publishing},
   author={Blanchet, Luc},
   year={1998},
   month=jul, pages={1971–1999} }

@article{Arun_2008,
   title={Inspiralling compact binaries in quasi-elliptical orbits: The complete third post-Newtonian energy flux},
   volume={77},
   ISSN={1550-2368},
   url={http://dx.doi.org/10.1103/PhysRevD.77.064035},
   DOI={10.1103/physrevd.77.064035},
   number={6},
   journal={Physical Review D},
   publisher={American Physical Society (APS)},
   author={Arun, K. G. and Blanchet, Luc and Iyer, Bala R. and Qusailah, Moh’d S. S.},
   year={2008},
   month=mar }

@article{Arun_2008_tail,
   title={Tail effects in the third post-Newtonian gravitational wave energy flux of compact binaries in quasi-elliptical orbits},
   volume={77},
   ISSN={1550-2368},
   url={http://dx.doi.org/10.1103/PhysRevD.77.064034},
   DOI={10.1103/physrevd.77.064034},
   number={6},
   journal={Physical Review D},
   publisher={American Physical Society (APS)},
   author={Arun, K. G. and Blanchet, Luc and Iyer, Bala R. and Qusailah, Moh’d S. S.},
   year={2008},
   month=mar }

@article{Marchand_2020,
   title={The mass quadrupole moment of compact binary systems at the fourth post-Newtonian order},
   volume={37},
   ISSN={1361-6382},
   url={http://dx.doi.org/10.1088/1361-6382/ab9ce1},
   DOI={10.1088/1361-6382/ab9ce1},
   number={21},
   journal={Classical and Quantum Gravity},
   publisher={IOP Publishing},
   author={Marchand, Tanguy and Henry, Quentin and Larrouturou, François and Marsat, Sylvain and Faye, Guillaume and Blanchet, Luc},
   year={2020},
   month=oct, pages={215006} }

@article{Arun_2009_tail,
   title={Third post-Newtonian angular momentum flux and the secular evolution of orbital elements for inspiralling compact binaries in quasi-elliptical orbits},
   volume={80},
   ISSN={1550-2368},
   url={http://dx.doi.org/10.1103/PhysRevD.80.124018},
   DOI={10.1103/physrevd.80.124018},
   number={12},
   journal={Physical Review D},
   publisher={American Physical Society (APS)},
   author={Arun, K. G. and Blanchet, Luc and Iyer, Bala R. and Sinha, Siddhartha},
   year={2009},
   month=dec }

@article{PhysRev.136.B1224,
  title = {Gravitational Radiation and the Motion of Two Point Masses},
  author = {Peters, P. C.},
  journal = {Phys. Rev.},
  volume = {136},
  issue = {4B},
  pages = {B1224--B1232},
  numpages = {0},
  year = {1964},
  month = {Nov},
  publisher = {American Physical Society},
  doi = {10.1103/PhysRev.136.B1224},
  url = {https://link.aps.org/doi/10.1103/PhysRev.136.B1224}
}

@article{Boetzel_2019,
   title={Gravitational-wave amplitudes for compact binaries in eccentric orbits at the third post-Newtonian order: Tail contributions and postadiabatic corrections},
   volume={100},
   ISSN={2470-0029},
   url={http://dx.doi.org/10.1103/PhysRevD.100.044018},
   DOI={10.1103/physrevd.100.044018},
   number={4},
   journal={Physical Review D},
   publisher={American Physical Society (APS)},
   author={Boetzel, Yannick and Mishra, Chandra Kant and Faye, Guillaume and Gopakumar, Achamveedu and Iyer, Bala R.},
   year={2019},
   month=aug }

@book{book:celestMechanic,
   title =     {Methods of celestial mechanics Volume 1},
   author =    {Gerhard Beutler, Leos Mervart, Andreas Verdun},
   publisher = {Springer},
   isbn =      {3540407499; 9783540407492; 3540407502; 9783540407508},
   year =      {2005},
   series =    {Astronomy and astrophysics library},
   edition =   {1}
}

@book{book:BesselFunction,
   title =     {A Treatise on the Theory of Bessel Functions},
   author =    {G. N. Watson},
   publisher = {Cambridge},
   isbn =      {9780521483919},
   year =      {1995},
   series =    {Cambridge Mathematical Library},
   edition =   {2}
}

@article{Blanchet_2000,
   title={Hadamard regularization},
   volume={41},
   ISSN={1089-7658},
   url={http://dx.doi.org/10.1063/1.1308506},
   DOI={10.1063/1.1308506},
   number={11},
   journal={Journal of Mathematical Physics},
   publisher={AIP Publishing},
   author={Blanchet, Luc and Faye, Guillaume},
   year={2000},
   month=nov, pages={7675–7714} }

@article{Blanchet_2023,
   title={Gravitational-wave flux and quadrupole modes from quasicircular nonspinning compact binaries to the fourth post-Newtonian order},
   volume={108},
   ISSN={2470-0029},
   url={http://dx.doi.org/10.1103/PhysRevD.108.064041},
   DOI={10.1103/physrevd.108.064041},
   number={6},
   journal={Physical Review D},
   publisher={American Physical Society (APS)},
   author={Blanchet, Luc and Faye, Guillaume and Henry, Quentin and Larrouturou, François and Trestini, David},
   year={2023},
   month=sep }

@article{Marchand_2016,
   title={Gravitational-wave tail effects to quartic non-linear order},
   volume={33},
   ISSN={1361-6382},
   url={http://dx.doi.org/10.1088/0264-9381/33/24/244003},
   DOI={10.1088/0264-9381/33/24/244003},
   number={24},
   journal={Classical and Quantum Gravity},
   publisher={IOP Publishing},
   author={Marchand, Tanguy and Blanchet, Luc and Faye, Guillaume},
   year={2016},
   month=nov, pages={244003} }

@article{Mik_czi_2015,
   title={First order post-Newtonian gravitational waveforms of binaries on eccentric orbits with Hansen coefficients},
   volume={92},
   ISSN={1550-2368},
   url={http://dx.doi.org/10.1103/PhysRevD.92.044038},
   DOI={10.1103/physrevd.92.044038},
   number={4},
   journal={Physical Review D},
   publisher={American Physical Society (APS)},
   author={Mikóczi, Balázs and Forgács, Péter and Vasúth, Mátyás},
   year={2015},
   month=aug }

@article{Munna_2020,
   title={Eccentric-orbit extreme-mass-ratio-inspiral radiation. II. 1PN correction to leading-logarithm and subleading-logarithm flux sequences and the entire perturbative 4PN flux},
   volume={102},
   ISSN={2470-0029},
   url={http://dx.doi.org/10.1103/PhysRevD.102.104006},
   DOI={10.1103/physrevd.102.104006},
   number={10},
   journal={Physical Review D},
   publisher={American Physical Society (APS)},
   author={Munna, Christopher and Evans, Charles R.},
   year={2020},
   month=nov }

@article{Munna_2022,
   title={High-order post-Newtonian expansion of the redshift invariant for eccentric-orbit nonspinning extreme-mass-ratio inspirals},
   volume={106},
   ISSN={2470-0029},
   url={http://dx.doi.org/10.1103/PhysRevD.106.044004},
   DOI={10.1103/physrevd.106.044004},
   number={4},
   journal={Physical Review D},
   publisher={American Physical Society (APS)},
   author={Munna, Christopher and Evans, Charles R.},
   year={2022},
   month=aug }

@article{Ledesma_2020,
   title={Spherical-harmonic tensors},
   volume={2},
   ISSN={2643-1564},
   url={http://dx.doi.org/10.1103/PhysRevResearch.2.043061},
   DOI={10.1103/physrevresearch.2.043061},
   number={4},
   journal={Physical Review Research},
   publisher={American Physical Society (APS)},
   author={Ledesma, Francisco Gonzalez and Mewes, Matthew},
   year={2020},
   month=oct }

@misc{blanchet_PNReview,
      title={Post-Newtonian Theory for Gravitational Waves}, 
      author={Luc Blanchet},
      year={2024},
      eprint={1310.1528},
      archivePrefix={arXiv},
      primaryClass={gr-qc},
      url={https://arxiv.org/abs/1310.1528}, 
}

@article{Damour_1985,
   title={Lagrangians forn point masses at the second post-Newtonian approximation of general relativity},
   volume={17},
   ISSN={1572-9532},
   url={https://doi.org/10.1007/BF00773685},
   DOI={10.1007/BF00773685},
   number={9},
   journal={General Relativity and Gravitation},
   publisher={American Physical Society (APS)},
   author={Damour Thibault and Schäfer Gerhard},
   year={1085},
   month=sep }

@article{Blanchet_1988_tail,
  title = {Tail-transported temporal correlations in the dynamics of a gravitating system},
  author = {Blanchet, Luc and Damour, Thibault},
  journal = {Phys. Rev. D},
  volume = {37},
  issue = {6},
  pages = {1410--1435},
  numpages = {0},
  year = {1988},
  month = {Mar},
  publisher = {American Physical Society},
  doi = {10.1103/PhysRevD.37.1410},
  url = {https://link.aps.org/doi/10.1103/PhysRevD.37.1410}
}

@article{Blanchet_2002_3PNflux,
  title = {Gravitational waves from inspiraling compact binaries: Energy flux to third post-Newtonian order},
  author = {Blanchet, Luc and Iyer, Bala R. and Joguet, Benoit},
  journal = {Phys. Rev. D},
  volume = {65},
  issue = {6},
  pages = {064005},
  numpages = {41},
  year = {2002},
  month = {Feb},
  publisher = {American Physical Society},
  doi = {10.1103/PhysRevD.65.064005},
  url = {https://link.aps.org/doi/10.1103/PhysRevD.65.064005}
}

@article{Damour_1985_1PN,
     author = {Damour, T. and Deruelle, N.},
     title = {General relativistic celestial mechanics of binary systems. {I.} {The} post-newtonian motion},
     journal = {Annales de l'I.H.P. Physique th\'eorique},
     pages = {107--132},
     publisher = {Gauthier-Villars},
     volume = {43},
     number = {1},
     year = {1985},
     mrnumber = {813140},
     zbl = {0585.70010},
     language = {en},
     url = {https://www.numdam.org/item/AIHPA_1985__43_1_107_0/}
}

@article{Damour_1988_RPA,
  title={Higher-order relativistic periastron advances and binary pulsars},
  author={Damour, Thibault and Sch{\"a}eer, G},
  journal={Il Nuovo Cimento B (1971-1996)},
  volume={101},
  number={2},
  pages={127--176},
  year={1988},
  publisher={Springer}
}

@article{Schafer_1993_2PN,
  title={Second post-Newtonian motion of compact binaries},
  author={Sch{\"a}fer, Gerhard and Wex, Norbert},
  journal={Physics Letters A},
  volume={174},
  number={3},
  pages={196--205},
  year={1993},
  publisher={Elsevier}
}

@article{Hinder_2010,
   title={Comparisons of eccentric binary black hole simulations with post-Newtonian models},
   volume={82},
   ISSN={1550-2368},
   url={http://dx.doi.org/10.1103/PhysRevD.82.024033},
   DOI={10.1103/physrevd.82.024033},
   number={2},
   journal={Physical Review D},
   publisher={American Physical Society (APS)},
   author={Hinder, Ian and Herrmann, Frank and Laguna, Pablo and Shoemaker, Deirdre},
   year={2010},
   month=jul }

@article{Moore_taylorf2ecc_2016,
   title={Gravitational-wave phasing for low-eccentricity inspiralling compact binaries to 3PN order},
   volume={93},
   ISSN={2470-0029},
   url={http://dx.doi.org/10.1103/PhysRevD.93.124061},
   DOI={10.1103/physrevd.93.124061},
   number={12},
   journal={Physical Review D},
   publisher={American Physical Society (APS)},
   author={Moore, Blake and Favata, Marc and Arun, K. G. and Mishra, Chandra Kant},
   year={2016},
   month=jun }

@article{Klein_2018,
   title={Fourier domain gravitational waveforms for precessing eccentric binaries},
   volume={98},
   ISSN={2470-0029},
   url={http://dx.doi.org/10.1103/PhysRevD.98.104043},
   DOI={10.1103/physrevd.98.104043},
   number={10},
   journal={Physical Review D},
   publisher={American Physical Society (APS)},
   author={Klein, Antoine and Boetzel, Yannick and Gopakumar, Achamveedu and Jetzer, Philippe and de Vittori, Lorenzo},
   year={2018},
   month=nov }

@article{Huerta_2014,
   title={Accurate and efficient waveforms for compact binaries on eccentric orbits},
   volume={90},
   ISSN={1550-2368},
   url={http://dx.doi.org/10.1103/PhysRevD.90.084016},
   DOI={10.1103/physrevd.90.084016},
   number={8},
   journal={Physical Review D},
   publisher={American Physical Society (APS)},
   author={Huerta, E. A. and Kumar, Prayush and McWilliams, Sean T. and O’Shaughnessy, Richard and Yunes, Nicolás},
   year={2014},
   month=oct }

@misc{IMRPhenomEcc,
      title={Time-domain phenomenological multipolar waveforms for aligned-spin binary black holes in elliptical orbits}, 
      author={Maria de Lluc Planas and Antoni Ramos-Buades and Cecilio García-Quirós and Héctor Estellés and Sascha Husa and Maria Haney},
      year={2025},
      eprint={2503.13062},
      archivePrefix={arXiv},
      primaryClass={gr-qc},
      url={https://arxiv.org/abs/2503.13062}, 
}

@article{Manna_IMREcc_2025,
   title={Improved inspiral-merger-ringdown model for BBHs on elliptical orbits},
   volume={111},
   ISSN={2470-0029},
   url={http://dx.doi.org/10.1103/849s-3zy8},
   DOI={10.1103/849s-3zy8},
   number={12},
   journal={Physical Review D},
   publisher={American Physical Society (APS)},
   author={Manna, Pratul and RoyChowdhury, Tamal and Mishra, Chandra Kant},
   year={2025},
   month=jun }

@misc{EOBv5EHM,
      title={Third post-Newtonian dynamics for eccentric orbits and aligned spins in the effective-one-body waveform model SEOBNRv5EHM}, 
      author={Aldo Gamboa and Mohammed Khalil and Alessandra Buonanno},
      year={2024},
      eprint={2412.12831},
      archivePrefix={arXiv},
      primaryClass={gr-qc},
      url={https://arxiv.org/abs/2412.12831}, 
}

@misc{EOBv5PHM,
      title={Adding equatorial-asymmetric effects for spin-precessing binaries into the SEOBNRv5PHM waveform model}, 
      author={Héctor Estellés and Alessandra Buonanno and Raffi Enficiaud and Cheng Foo and Lorenzo Pompili},
      year={2025},
      eprint={2506.19911},
      archivePrefix={arXiv},
      primaryClass={gr-qc},
      url={https://arxiv.org/abs/2506.19911}, 
}

@article{ajith2007IMR,
  title = {Template bank for gravitational waveforms from coalescing binary black holes: Nonspinning binaries},
  author = {Ajith, P. and Babak, S. and Chen, Y. and Hewitson, M. and Krishnan, B. and Sintes, A. M. and Whelan, J. T. and Br\"ugmann, B. and Diener, P. and Dorband, N. and Gonzalez, J. and Hannam, M. and Husa, S. and Pollney, D. and Rezzolla, L. and Santamar\'{\i}a, L. and Sperhake, U. and Thornburg, J.},
  journal = {Phys. Rev. D},
  volume = {77},
  issue = {10},
  pages = {104017},
  numpages = {22},
  year = {2008},
  month = {May},
  publisher = {American Physical Society},
  doi = {10.1103/PhysRevD.77.104017},
  url = {https://link.aps.org/doi/10.1103/PhysRevD.77.104017}
}

@article{Cao_2017,
  title = {Waveform model for an eccentric binary black hole based on the effective-one-body-numerical-relativity formalism},
  author = {Cao, Zhoujian and Han, Wen-Biao},
  journal = {Phys. Rev. D},
  volume = {96},
  issue = {4},
  pages = {044028},
  numpages = {23},
  year = {2017},
  month = {Aug},
  publisher = {American Physical Society},
  doi = {10.1103/PhysRevD.96.044028},
  url = {https://link.aps.org/doi/10.1103/PhysRevD.96.044028}
}

@article{liu2024effective,
  title={Effective-one-body numerical-relativity waveform model for eccentric spin-precessing binary black hole coalescence},
  author={Liu, Xiaolin and Cao, Zhoujian and Zhu, Zong-Hong},
  journal={Classical and Quantum Gravity},
  volume={41},
  number={19},
  pages={195019},
  year={2024},
  publisher={IOP Publishing}
}

@article{Nagar_2018_TEOB,
   title={Time-domain effective-one-body gravitational waveforms for coalescing compact binaries with nonprecessing spins, tides, and self-spin effects},
   volume={98},
   ISSN={2470-0029},
   url={http://dx.doi.org/10.1103/PhysRevD.98.104052},
   DOI={10.1103/physrevd.98.104052},
   number={10},
   journal={Physical Review D},
   publisher={American Physical Society (APS)},
   author={Nagar, Alessandro and Bernuzzi, Sebastiano and Del Pozzo, Walter and Riemenschneider, Gunnar and Akcay, Sarp and Carullo, Gregorio and Fleig, Philipp and Babak, Stanislav and Tsang, Ka Wa and Colleoni, Marta and Messina, Francesco and Pratten, Geraint and Radice, David and Rettegno, Piero and Agathos, Michalis and Fauchon-Jones, Edward and Hannam, Mark and Husa, Sascha and Dietrich, Tim and Cerdá-Duran, Pablo and Font, José A. and Pannarale, Francesco and Schmidt, Patricia and Damour, Thibault},
   year={2018},
   month=nov }

@misc{nagar2024TEOBGO,
      title={Effective-one-body waveform model for non-circularized, planar, coalescing black hole binaries: the importance of radiation reaction}, 
      author={Alessandro Nagar and Rossella Gamba and Piero Rettegno and Veronica Fantini and Sebastiano Bernuzzi},
      year={2024},
      eprint={2404.05288},
      archivePrefix={arXiv},
      primaryClass={gr-qc},
      url={https://arxiv.org/abs/2404.05288}, 
}

@article{Ajith_2011,
   title={Inspiral-Merger-Ringdown Waveforms for Black-Hole Binaries with Nonprecessing Spins},
   volume={106},
   ISSN={1079-7114},
   url={http://dx.doi.org/10.1103/PhysRevLett.106.241101},
   DOI={10.1103/physrevlett.106.241101},
   number={24},
   journal={Physical Review Letters},
   publisher={American Physical Society (APS)},
   author={Ajith, P. and Hannam, M. and Husa, S. and Chen, Y. and Brügmann, B. and Dorband, N. and Müller, D. and Ohme, F. and Pollney, D. and Reisswig, C. and Santamaría, L. and Seiler, J.},
   year={2011},
   month=jun }

@article{Khan_2016,
   title={Frequency-domain gravitational waves from nonprecessing black-hole binaries. II. A phenomenological model for the advanced detector era},
   volume={93},
   ISSN={2470-0029},
   url={http://dx.doi.org/10.1103/PhysRevD.93.044007},
   DOI={10.1103/physrevd.93.044007},
   number={4},
   journal={Physical Review D},
   publisher={American Physical Society (APS)},
   author={Khan, Sebastian and Husa, Sascha and Hannam, Mark and Ohme, Frank and Pürrer, Michael and Forteza, Xisco Jiménez and Bohé, Alejandro},
   year={2016},
   month=feb }

@article{Cho_2022,
   title={Generalized quasi-Keplerian solution for eccentric, nonspinning compact binaries at 4PN order and the associated inspiral-merger-ringdown waveform},
   volume={105},
   ISSN={2470-0029},
   url={http://dx.doi.org/10.1103/PhysRevD.105.064010},
   DOI={10.1103/physrevd.105.064010},
   number={6},
   journal={Physical Review D},
   publisher={American Physical Society (APS)},
   author={Cho, Gihyuk and Tanay, Sashwat and Gopakumar, Achamveedu and Lee, Hyung Mok},
   year={2022},
   month=mar }

@article{blanchet1986radiative,
  title={Radiative gravitational fields in general relativity I. General structure of the field outside the source},
  author={Blanchet, Luc and Damour, Thibault},
  journal={Philosophical transactions of the royal society of London. Series a, mathematical and physical sciences},
  volume={320},
  number={1555},
  pages={379--430},
  year={1986},
  publisher={The Royal Society London}
}

@article{Bini_2012,
   title={Gravitational radiation reaction along general orbits in the effective one-body formalism},
   volume={86},
   ISSN={1550-2368},
   url={http://dx.doi.org/10.1103/PhysRevD.86.124012},
   DOI={10.1103/physrevd.86.124012},
   number={12},
   journal={Physical Review D},
   publisher={American Physical Society (APS)},
   author={Bini, Donato and Damour, Thibault},
   year={2012},
   month=dec }

@article{Fumagalli_2025,
   title={Nonadiabatic dynamics of eccentric black-hole binaries in post-Newtonian theory},
   volume={112},
   ISSN={2470-0029},
   url={http://dx.doi.org/10.1103/znmj-6wvt},
   DOI={10.1103/znmj-6wvt},
   number={2},
   journal={Physical Review D},
   publisher={American Physical Society (APS)},
   author={Fumagalli, Giulia and Loutrel, Nicholas and Gerosa, Davide and Boschini, Matteo},
   year={2025},
   month=jul }

@article{Leibovich_2023,
   title={Radiation reaction for nonspinning bodies at 4.5PN in the effective field theory approach},
   volume={108},
   ISSN={2470-0029},
   url={http://dx.doi.org/10.1103/PhysRevD.108.024017},
   DOI={10.1103/physrevd.108.024017},
   number={2},
   journal={Physical Review D},
   publisher={American Physical Society (APS)},
   author={Leibovich, Adam K. and Pardo, Brian A. and Yang, Zixin},
   year={2023},
   month=jul }

@misc{blanchet_2024_v9RR,
      title={Gravitational radiation reaction for compact binary systems at the fourth-and-a-half post-Newtonian order}, 
      author={Luc Blanchet and Guillaume Faye and David Trestini},
      year={2024},
      eprint={2407.18295},
      archivePrefix={arXiv},
      primaryClass={gr-qc},
      url={https://arxiv.org/abs/2407.18295}, 
}

@article{Pound_2008,
   title={Multiscale analysis of the electromagnetic self-force in a weak gravitational field},
   volume={77},
   ISSN={1550-2368},
   url={http://dx.doi.org/10.1103/PhysRevD.77.044012},
   DOI={10.1103/physrevd.77.044012},
   number={4},
   journal={Physical Review D},
   publisher={American Physical Society (APS)},
   author={Pound, Adam and Poisson, Eric},
   year={2008},
   month=feb }

@article{Miller_2021,
   title={Two-timescale evolution of extreme-mass-ratio inspirals: Waveform generation scheme for quasicircular orbits in Schwarzschild spacetime},
   volume={103},
   ISSN={2470-0029},
   url={http://dx.doi.org/10.1103/PhysRevD.103.064048},
   DOI={10.1103/physrevd.103.064048},
   number={6},
   journal={Physical Review D},
   publisher={American Physical Society (APS)},
   author={Miller, Jeremy and Pound, Adam},
   year={2021},
   month=mar }

@article{Yunes_2009,
   title={Post-circular expansion of eccentric binary inspirals: Fourier-domain waveforms in the stationary phase approximation},
   volume={80},
   ISSN={1550-2368},
   url={http://dx.doi.org/10.1103/PhysRevD.80.084001},
   DOI={10.1103/physrevd.80.084001},
   number={8},
   journal={Physical Review D},
   publisher={American Physical Society (APS)},
   author={Yunes, Nicolas and Arun, K. G. and Berti, Emanuele and Will, Clifford M.},
   year={2009},
   month=oct }

@article{K_nigsd_rffer_2006,
   title={Phasing of gravitational waves from inspiralling eccentric binaries at the third-and-a-half post-Newtonian order},
   volume={73},
   ISSN={1550-2368},
   url={http://dx.doi.org/10.1103/PhysRevD.73.124012},
   DOI={10.1103/physrevd.73.124012},
   number={12},
   journal={Physical Review D},
   publisher={American Physical Society (APS)},
   author={Königsdörffer, Christian and Gopakumar, Achamveedu},
   year={2006},
   month=jun }

@article{Tiwari_2020,
   title={Combining post-circular and Padé approximations to compute Fourier domain templates for eccentric inspirals},
   volume={102},
   ISSN={2470-0029},
   url={http://dx.doi.org/10.1103/PhysRevD.102.084042},
   DOI={10.1103/physrevd.102.084042},
   number={8},
   journal={Physical Review D},
   publisher={American Physical Society (APS)},
   author={Tiwari, Srishti and Gopakumar, Achamveedu},
   year={2020},
   month=oct }

@article{Morras_2025,
   title={Improved post-Newtonian waveform model for inspiralling precessing-eccentric compact binaries},
   volume={111},
   ISSN={2470-0029},
   url={http://dx.doi.org/10.1103/PhysRevD.111.084052},
   DOI={10.1103/physrevd.111.084052},
   number={8},
   journal={Physical Review D},
   publisher={American Physical Society (APS)},
   author={Morras, Gonzalo and Pratten, Geraint and Schmidt, Patricia},
   year={2025},
   month=apr }

@misc{Morras_2025_1PN,
      title={Modeling Gravitational Wave Modes from Binaries with Arbitrary Eccentricity}, 
      author={Gonzalo Morras},
      year={2025},
      eprint={2507.00169},
      archivePrefix={arXiv},
      primaryClass={gr-qc},
      url={https://arxiv.org/abs/2507.00169}, 
}

@article{Agazie_2023,
   title={The NANOGrav 15 yr Data Set: Evidence for a Gravitational-wave Background},
   volume={951},
   ISSN={2041-8213},
   url={http://dx.doi.org/10.3847/2041-8213/acdac6},
   DOI={10.3847/2041-8213/acdac6},
   number={1},
   journal={The Astrophysical Journal Letters},
   publisher={American Astronomical Society},
   author={Agazie, Gabriella and Anumarlapudi, Akash and Archibald, Anne M. and Arzoumanian, Zaven and Baker, Paul T. and Bécsy, Bence and Blecha, Laura and Brazier, Adam and Brook, Paul R. and Burke-Spolaor, Sarah and Burnette, Rand and Case, Robin and Charisi, Maria and Chatterjee, Shami and Chatziioannou, Katerina and Cheeseboro, Belinda D. and Chen, Siyuan and Cohen, Tyler and Cordes, James M. and Cornish, Neil J. and Crawford, Fronefield and Cromartie, H. Thankful and Crowter, Kathryn and Cutler, Curt J. and DeCesar, Megan E. and DeGan, Dallas and Demorest, Paul B. and Deng, Heling and Dolch, Timothy and Drachler, Brendan and Ellis, Justin A. and Ferrara, Elizabeth C. and Fiore, William and Fonseca, Emmanuel and Freedman, Gabriel E. and Garver-Daniels, Nate and Gentile, Peter A. and Gersbach, Kyle A. and Glaser, Joseph and Good, Deborah C. and Gültekin, Kayhan and Hazboun, Jeffrey S. and Hourihane, Sophie and Islo, Kristina and Jennings, Ross J. and Johnson, Aaron D. and Jones, Megan L. and Kaiser, Andrew R. and Kaplan, David L. and Kelley, Luke Zoltan and Kerr, Matthew and Key, Joey S. and Klein, Tonia C. and Laal, Nima and Lam, Michael T. and Lamb, William G. and W. Lazio, T. Joseph and Lewandowska, Natalia and Littenberg, Tyson B. and Liu, Tingting and Lommen, Andrea and Lorimer, Duncan R. and Luo, Jing and Lynch, Ryan S. and Ma, Chung-Pei and Madison, Dustin R. and Mattson, Margaret A. and McEwen, Alexander and McKee, James W. and McLaughlin, Maura A. and McMann, Natasha and Meyers, Bradley W. and Meyers, Patrick M. and Mingarelli, Chiara M. F. and Mitridate, Andrea and Natarajan, Priyamvada and Ng, Cherry and Nice, David J. and Ocker, Stella Koch and Olum, Ken D. and Pennucci, Timothy T. and Perera, Benetge B. P. and Petrov, Polina and Pol, Nihan S. and Radovan, Henri A. and Ransom, Scott M. and Ray, Paul S. and Romano, Joseph D. and Sardesai, Shashwat C. and Schmiedekamp, Ann and Schmiedekamp, Carl and Schmitz, Kai and Schult, Levi and Shapiro-Albert, Brent J. and Siemens, Xavier and Simon, Joseph and Siwek, Magdalena S. and Stairs, Ingrid H. and Stinebring, Daniel R. and Stovall, Kevin and Sun, Jerry P. and Susobhanan, Abhimanyu and Swiggum, Joseph K. and Taylor, Jacob and Taylor, Stephen R. and Turner, Jacob E. and Unal, Caner and Vallisneri, Michele and van Haasteren, Rutger and Vigeland, Sarah J. and Wahl, Haley M. and Wang, Qiaohong and Witt, Caitlin A. and Young, Olivia},
   year={2023},
   month=jun, pages={L8} }

@article{Reardon_2023,
   title={Search for an Isotropic Gravitational-wave Background with the Parkes Pulsar Timing Array},
   volume={951},
   ISSN={2041-8213},
   url={http://dx.doi.org/10.3847/2041-8213/acdd02},
   DOI={10.3847/2041-8213/acdd02},
   number={1},
   journal={The Astrophysical Journal Letters},
   publisher={American Astronomical Society},
   author={Reardon, Daniel J. and Zic, Andrew and Shannon, Ryan M. and Hobbs, George B. and Bailes, Matthew and Di Marco, Valentina and Kapur, Agastya and Rogers, Axl F. and Thrane, Eric and Askew, Jacob and Bhat, N. D. Ramesh and Cameron, Andrew and Curyło, Małgorzata and Coles, William A. and Dai, Shi and Goncharov, Boris and Kerr, Matthew and Kulkarni, Atharva and Levin, Yuri and Lower, Marcus E. and Manchester, Richard N. and Mandow, Rami and Miles, Matthew T. and Nathan, Rowina S. and Osłowski, Stefan and Russell, Christopher J. and Spiewak, Renée and Zhang, Songbo and Zhu, Xing-Jiang},
   year={2023},
   month=jun, pages={L6} }

@article{EPTA2023,
   title={The second data release from the European Pulsar Timing Array: III. Search for gravitational wave signals},
   volume={678},
   ISSN={1432-0746},
   url={http://dx.doi.org/10.1051/0004-6361/202346844},
   DOI={10.1051/0004-6361/202346844},
   journal={Astronomy amp; Astrophysics},
   publisher={EDP Sciences},
   author={Antoniadis, J. and Arumugam, P. and Arumugam, S. and Babak, S. and Bagchi, M. and Bak Nielsen, A.-S. and Bassa, C. G. and Bathula, A. and Berthereau, A. and Bonetti, M. and Bortolas, E. and Brook, P. R. and Burgay, M. and Caballero, R. N. and Chalumeau, A. and Champion, D. J. and Chanlaridis, S. and Chen, S. and Cognard, I. and Dandapat, S. and Deb, D. and Desai, S. and Desvignes, G. and Dhanda-Batra, N. and Dwivedi, C. and Falxa, M. and Ferdman, R. D. and Franchini, A. and Gair, J. R. and Goncharov, B. and Gopakumar, A. and Graikou, E. and Grießmeier, J.-M. and Guillemot, L. and Guo, Y. J. and Gupta, Y. and Hisano, S. and Hu, H. and Iraci, F. and Izquierdo-Villalba, D. and Jang, J. and Jawor, J. and Janssen, G. H. and Jessner, A. and Joshi, B. C. and Kareem, F. and Karuppusamy, R. and Keane, E. F. and Keith, M. J. and Kharbanda, D. and Kikunaga, T. and Kolhe, N. and Kramer, M. and Krishnakumar, M. A. and Lackeos, K. and Lee, K. J. and Liu, K. and Liu, Y. and Lyne, A. G. and McKee, J. W. and Maan, Y. and Main, R. A. and Mickaliger, M. B. and Niţu, I. C. and Nobleson, K. and Paladi, A. K. and Parthasarathy, A. and Perera, B. B. P. and Perrodin, D. and Petiteau, A. and Porayko, N. K. and Possenti, A. and Prabu, T. and Quelquejay Leclere, H. and Rana, P. and Samajdar, A. and Sanidas, S. A. and Sesana, A. and Shaifullah, G. and Singha, J. and Speri, L. and Spiewak, R. and Srivastava, A. and Stappers, B. W. and Surnis, M. and Susarla, S. C. and Susobhanan, A. and Takahashi, K. and Tarafdar, P. and Theureau, G. and Tiburzi, C. and van der Wateren, E. and Vecchio, A. and Venkatraman Krishnan, V. and Verbiest, J. P. W. and Wang, J. and Wang, L. and Wu, Z.},
   year={2023},
   month=oct, pages={A50} }

@article{Xu_2023,
   title={Searching for the Nano-Hertz Stochastic Gravitational Wave Background with the Chinese Pulsar Timing Array Data Release I},
   volume={23},
   ISSN={1674-4527},
   url={http://dx.doi.org/10.1088/1674-4527/acdfa5},
   DOI={10.1088/1674-4527/acdfa5},
   number={7},
   journal={Research in Astronomy and Astrophysics},
   publisher={IOP Publishing},
   author={Xu, Heng and Chen, Siyuan and Guo, Yanjun and Jiang, Jinchen and Wang, Bojun and Xu, Jiangwei and Xue, Zihan and Nicolas Caballero, R. and Yuan, Jianping and Xu, Yonghua and Wang, Jingbo and Hao, Longfei and Luo, Jingtao and Lee, Kejia and Han, Jinlin and Jiang, Peng and Shen, Zhiqiang and Wang, Min and Wang, Na and Xu, Renxin and Wu, Xiangping and Manchester, Richard and Qian, Lei and Guan, Xin and Huang, Menglin and Sun, Chun and Zhu, Yan},
   year={2023},
   month=jun, pages={075024} }

@article{George_2018,
   title={Deep neural networks to enable real-time multimessenger astrophysics},
   volume={97},
   ISSN={2470-0029},
   url={http://dx.doi.org/10.1103/PhysRevD.97.044039},
   DOI={10.1103/physrevd.97.044039},
   number={4},
   journal={Physical Review D},
   publisher={American Physical Society (APS)},
   author={George, Daniel and Huerta, E. A.},
   year={2018},
   month=feb }

@article{SXS_2019,
   title={The SXS collaboration catalog of binary black hole simulations},
   volume={36},
   ISSN={1361-6382},
   url={http://dx.doi.org/10.1088/1361-6382/ab34e2},
   DOI={10.1088/1361-6382/ab34e2},
   number={19},
   journal={Classical and Quantum Gravity},
   publisher={IOP Publishing},
   author={Boyle, Michael and Hemberger, Daniel and Iozzo, Dante A B and Lovelace, Geoffrey and Ossokine, Serguei and Pfeiffer, Harald P and Scheel, Mark A and Stein, Leo C and Woodford, Charles J and Zimmerman, Aaron B and Afshari, Nousha and Barkett, Kevin and Blackman, Jonathan and Chatziioannou, Katerina and Chu, Tony and Demos, Nicholas and Deppe, Nils and Field, Scott E and Fischer, Nils L and Foley, Evan and Fong, Heather and Garcia, Alyssa and Giesler, Matthew and Hebert, Francois and Hinder, Ian and Katebi, Reza and Khan, Haroon and Kidder, Lawrence E and Kumar, Prayush and Kuper, Kevin and Lim, Halston and Okounkova, Maria and Ramirez, Teresita and Rodriguez, Samuel and Rüter, Hannes R and Schmidt, Patricia and Szilagyi, Bela and Teukolsky, Saul A and Varma, Vijay and Walker, Marissa},
   year={2019},
   month=sep, pages={195006} }

@misc{SXS_2025,
      title={The SXS Collaboration's third catalog of binary black hole simulations}, 
      author={Mark A. Scheel and Michael Boyle and Keefe Mitman and Nils Deppe and Leo C. Stein and Cristóbal Armaza and Marceline S. Bonilla and Luisa T. Buchman and Andrea Ceja and Himanshu Chaudhary and Yitian Chen and Maxence Corman and Károly Zoltán Csukás and C. Melize Ferrus and Scott E. Field and Matthew Giesler and Sarah Habib and François Hébert and Daniel A. Hemberger and Dante A. B. Iozzo and Tousif Islam and Ken Z. Jones and Aniket Khairnar and Lawrence E. Kidder and Taylor Knapp and Prayush Kumar and Guillermo Lara and Oliver Long and Geoffrey Lovelace and Sizheng Ma and Denyz Melchor and Marlo Morales and Jordan Moxon and Peter James Nee and Kyle C. Nelli and Eamonn O'Shea and Serguei Ossokine and Robert Owen and Harald P. Pfeiffer and Isabella G. Pretto and Teresita Ramirez-Aguilar and Antoni Ramos-Buades and Adhrit Ravichandran and Abhishek Ravishankar and Samuel Rodriguez and Hannes R. Rüter and Jennifer Sanchez and Md Arif Shaikh and Dongze Sun and Béla Szilágyi and Daniel Tellez and Saul A. Teukolsky and Sierra Thomas and William Throwe and Vijay Varma and Nils L. Vu and Marissa Walker and Nikolas A. Wittek and Jooheon Yoo},
      year={2025},
      eprint={2505.13378},
      archivePrefix={arXiv},
      primaryClass={gr-qc},
      url={https://arxiv.org/abs/2505.13378}, 
}

@misc{CIT_2023,
      title={A catalogue of precessing black-hole-binary numerical-relativity simulations}, 
      author={Eleanor Hamilton and Edward Fauchon-Jones and Mark Hannam and Charlie Hoy and Chinmay Kalaghatgi and Lionel London and Jonathan E. Thompson and Dave Yeeles and Shrobana Ghosh and Sebastian Khan and Panagiota Kolitsidou and Alex Vano-Vinuales},
      year={2023},
      eprint={2303.05419},
      archivePrefix={arXiv},
      primaryClass={gr-qc},
      url={https://arxiv.org/abs/2303.05419}, 
}

@article{Mino_1997,
   title={Gravitational radiation reaction to a particle motion},
   volume={55},
   ISSN={1089-4918},
   url={http://dx.doi.org/10.1103/PhysRevD.55.3457},
   DOI={10.1103/physrevd.55.3457},
   number={6},
   journal={Physical Review D},
   publisher={American Physical Society (APS)},
   author={Mino, Yasushi and Sasaki, Misao and Tanaka, Takahiro},
   year={1997},
   month=mar, pages={3457–3476} }

@article{Pound_2012,
   title={Second-Order Gravitational Self-Force},
   volume={109},
   ISSN={1079-7114},
   url={http://dx.doi.org/10.1103/PhysRevLett.109.051101},
   DOI={10.1103/physrevlett.109.051101},
   number={5},
   journal={Physical Review Letters},
   publisher={American Physical Society (APS)},
   author={Pound, Adam},
   year={2012},
   month=jul }

@misc{landini2025opticalgravitationalwavessignals,
      title={Optical gravitational waves as signals of Gravitationally-Decaying Particles}, 
      author={Giacomo Landini and Alessandro Strumia},
      year={2025},
      eprint={2501.09794},
      archivePrefix={arXiv},
      primaryClass={hep-ph},
      url={https://arxiv.org/abs/2501.09794}, 
}

@article{London_2018,
   title={First Higher-Multipole Model of Gravitational Waves from Spinning and Coalescing Black-Hole Binaries},
   volume={120},
   ISSN={1079-7114},
   url={http://dx.doi.org/10.1103/PhysRevLett.120.161102},
   DOI={10.1103/physrevlett.120.161102},
   number={16},
   journal={Physical Review Letters},
   publisher={American Physical Society (APS)},
   author={London, Lionel and Khan, Sebastian and Fauchon-Jones, Edward and García, Cecilio and Hannam, Mark and Husa, Sascha and Jiménez-Forteza, Xisco and Kalaghatgi, Chinmay and Ohme, Frank and Pannarale, Francesco},
   year={2018},
   month=apr }

@article{Garc_a_Quir_s_2020,
   title={Multimode frequency-domain model for the gravitational wave signal from nonprecessing black-hole binaries},
   volume={102},
   ISSN={2470-0029},
   url={http://dx.doi.org/10.1103/PhysRevD.102.064002},
   DOI={10.1103/physrevd.102.064002},
   number={6},
   journal={Physical Review D},
   publisher={American Physical Society (APS)},
   author={García-Quirós, Cecilio and Colleoni, Marta and Husa, Sascha and Estellés, Héctor and Pratten, Geraint and Ramos-Buades, Antoni and Mateu-Lucena, Maite and Jaume, Rafel},
   year={2020},
   month=sep }

@article{Akcay_2021,
   title={Hybrid post-Newtonian effective-one-body scheme for spin-precessing compact-binary waveforms up to merger},
   volume={103},
   ISSN={2470-0029},
   url={http://dx.doi.org/10.1103/PhysRevD.103.024014},
   DOI={10.1103/physrevd.103.024014},
   number={2},
   journal={Physical Review D},
   publisher={American Physical Society (APS)},
   author={Akcay, Sarp and Gamba, Rossella and Bernuzzi, Sebastiano},
   year={2021},
   month=jan }

@article{Klein_2013,
   title={Gravitational waveforms for precessing, quasicircular binaries via multiple scale analysis and uniform asymptotics: The near spin alignment case},
   volume={88},
   ISSN={1550-2368},
   url={http://dx.doi.org/10.1103/PhysRevD.88.124015},
   DOI={10.1103/physrevd.88.124015},
   number={12},
   journal={Physical Review D},
   publisher={American Physical Society (APS)},
   author={Klein, Antoine and Cornish, Neil and Yunes, Nicolás},
   year={2013},
   month=dec }

@article{Trestini_2023,
   title={Gravitational-wave tails of memory},
   volume={107},
   ISSN={2470-0029},
   url={http://dx.doi.org/10.1103/PhysRevD.107.104048},
   DOI={10.1103/physrevd.107.104048},
   number={10},
   journal={Physical Review D},
   publisher={American Physical Society (APS)},
   author={Trestini, David and Blanchet, Luc},
   year={2023},
   month=may }

@article{Loutrel_2017,
   title={Hereditary effects in eccentric compact binary inspirals to third post-Newtonian order},
   volume={34},
   ISSN={1361-6382},
   url={http://dx.doi.org/10.1088/1361-6382/aa59c3},
   DOI={10.1088/1361-6382/aa59c3},
   number={4},
   journal={Classical and Quantum Gravity},
   publisher={IOP Publishing},
   author={Loutrel, Nicholas and Yunes, Nicolás},
   year={2017},
   month=feb, pages={044003} }

@article{Trestini_2025_ST,
   title={Gravitational waves from quasielliptic compact binaries in scalar-tensor theory to one-and-a-half post-Newtonian order},
   volume={42},
   ISSN={1361-6382},
   url={http://dx.doi.org/10.1088/1361-6382/adedf5},
   DOI={10.1088/1361-6382/adedf5},
   number={15},
   journal={Classical and Quantum Gravity},
   publisher={IOP Publishing},
   author={Trestini, David},
   year={2025},
   month=jul, pages={155016} }

@article{Trestini_2025_Schott,
   title={Schott term in the binding energy for compact binaries on circular orbits at fourth post-Newtonian order},
   volume={112},
   ISSN={2470-0029},
   url={http://dx.doi.org/10.1103/lsbb-sv45},
   DOI={10.1103/lsbb-sv45},
   number={2},
   journal={Physical Review D},
   publisher={American Physical Society (APS)},
   author={Trestini, David},
   year={2025},
   month=jul }

@article{Iyer_1995,
  title = {Post-Newtonian gravitational radiation reaction for two-body systems: Nonspinning bodies},
  author = {Iyer, Bala R. and Will, Clifford M.},
  journal = {Phys. Rev. D},
  volume = {52},
  issue = {12},
  pages = {6882--6893},
  numpages = {0},
  year = {1995},
  month = {Dec},
  publisher = {American Physical Society},
  doi = {10.1103/PhysRevD.52.6882},
  url = {https://link.aps.org/doi/10.1103/PhysRevD.52.6882}
}

@article{Gopakumar_1997,
  title = {Second post-Newtonian gravitational radiation reaction for two-body systems: Nonspinning bodies},
  author = {Gopakumar, A. and Iyer, Bala R. and Iyer, Sai},
  journal = {Phys. Rev. D},
  volume = {55},
  issue = {10},
  pages = {6030--6053},
  numpages = {0},
  year = {1997},
  month = {May},
  publisher = {American Physical Society},
  doi = {10.1103/PhysRevD.55.6030},
  url = {https://link.aps.org/doi/10.1103/PhysRevD.55.6030}
}

@misc{blanchet_2025_ISCO,
      title={Innermost stable circular orbit of arbitrary-mass compact binaries at fourth post-Newtonian order}, 
      author={Luc Blanchet and David Langlois and Etienne Ligout},
      year={2025},
      eprint={2505.01278},
      archivePrefix={arXiv},
      primaryClass={gr-qc},
      url={https://arxiv.org/abs/2505.01278}, 
}

\end{document}